\documentclass[12pt]{article}
 
\usepackage{graphicx}
\usepackage{amsmath}
\usepackage{amssymb}
\usepackage{geometry}
\usepackage{setspace}
\usepackage{hyperref}
\usepackage{natbib}
\usepackage{booktabs}
\usepackage{float}
\usepackage{caption}
\usepackage{array}
\usepackage{multirow}

\title{Recovering Direct Price Effects of Environmental Amenities in Housing Markets: Regression and Causal Machine Learning Model Assessment with Empirical Monte Carlo Simulation}
\author{}
\date{}
\begin{document}
\maketitle
 
\begin{abstract}
Hedonic price models are widely used to value environmental amenities, yet practical guidance for estimating direct unconfounded price effects remains sparse. We conduct an empirical Monte Carlo simulation to evaluate traditional regression and causal machine learning (CML) approaches for estimating the direct unmediated price effect of spatially delineated amenities on treated properties (DUET), a lower-bound approximation of welfare change with direct applications to benefit-cost analysis. Whereas previous simulations rely on parametric assumptions, we preserve the actual data-generating process underlying over 1 million property transactions from upstate New York (1990--2024) and randomly assign treatment locations across iterations, establishing a ground truth against which estimation error can be measured precisely. Generalized difference-in-differences (DID) regression consistently outperforms baseline DID and two-way fixed effects models across all scenarios. Causal forest DID performs comparably to generalized DID in most scenarios and, in larger samples (at least 3,000 treated observations) increasingly common in contemporary hedonic studies, achieves significantly lower estimation error and better confidence interval coverage. Its advantage widens further when the treatment induces a nonlinear shift in the hedonic price function. The performance of causal forest DID, however, hinges on specification choices that differ from regression practice, such as including geographic attributes only in the price models and avoiding high-dimensional fixed effects. We translate these findings into method-specific best-practice recommendations for both regression and CML approaches.



\end{abstract}
\doublespacing
\newpage

\section{Introduction}
\label{sec:intro}

The hedonic property value model has been one of the most widely used tools in environmental economics since its introduction in the 1970s, with applications accelerating alongside growth in data accessibility and advances in econometrics and computing \citep{bishop2020}. A key methodological advance in this literature is the generalized difference-in-differences (GDID) framework\footnote{GDID extends standard DID by allowing the covariate coefficients to vary across periods, thereby accommodating equilibrium shift and price function change over time. Note that we use ``GDID" to represent the general framework, whereas specific estimators are denoted by other acronyms.} introduced in \citet{kuminoff2010} (hereafter KPP) with its advantage demonstrated via Monte Carlo simulations.\footnote{Following the  Monte Carlo framework developed in \citet{cropper1988}, KPP simulate housing market equilibria under parametric assumptions to evaluate the ability of hedonic models to recover welfare measures from amenity shocks. Specifically, their simulations use structural equilibrium models with known utility, bidding, and cost functions, generate synthetic housing price data from nonmarginal amenity shocks, and assess how well alternative hedonic specifications can recover willingness to pay.} \citet{banzhaf2021} demonstrates that the GDID approach identifies the direct unmediated price effect on treated (DUET), the price effect of treatment evaluated along the post-treatment price function, isolated from indirect effects arising from equilibrium shift and mediated effects through treatment-associated amenities. Further, \citet{banzhaf2021} shows that DUET provides a lower-bound approximation of the welfare change measured by Hicksian equivalent surplus (ES).\footnote{The equivalent surplus measures the change needed in wealth that holds utility constant at ex post levels to forgo a change in amenity. While the more widely recognized concept, equivalent variation, is more appropriate for price changes, ES is often used for changes in quantities or qualities where individuals cannot adjust quantity in a public good context. \citet{banzhaf2021} shows that DUET consistently serves as a lower-bound for ES, which is comparable to the role of compensating surplus (CS).} This theoretical development not only clarifies the welfare interpretation of the widely used hedonic DID estimator, but also motivates a fundamental reconceptualization of the simulation approach pioneered by \citet{cropper1988} (hereafter CDM) and KPP.

We undertake a new DUET-based simulation analysis for three reasons. First, the hedonic literature has advanced considerably over the past fifteen years, with two particularly notable developments that call for updated methodological guidance: the substantial expansion in dataset sizes and the potential incorporation of machine-learning-based modeling techniques \citep[e.g.,][]{chen2025heterogeneous}. Second, providing a consistent welfare interpretation, DUET offers a natural target for estimator assessment, one that is more appropriate than evaluating models by their ability to recover welfare measures directly, as we demonstrate below. Third, unlike prior simulation frameworks that required constructing synthetic datasets, the target of DUET allows researchers to simulate price changes along the post-treatment hedonic price function using real transaction data, improving the generalizability of the evaluation results.

Accurately recovering DUET makes sense from both a policy and a research-design perspective. First, DUET is easy to interpret. Employing a logarithmic price outcome enables a straightforward interpretation of DUET as a percentage change in post-treatment housing prices, thereby offering an intuitive and valuable metric for policy evaluation \citep[e.g.,][]{keiser2019, hu2025, mamun2023, guo2024}. Second, as proved in \citet{banzhaf2021}, DUET consistently serves as a lower bound approximation for the ES\footnote{As shown in \citet{banzhaf2021}, DUET serves as a lower bound for ES even in the presence of endogenous residential sorting and treatment-induced general equilibrium effects. The conditions that may invalidate this relationship are twofold: 1. The violation of local noninterference, where the treatment of one house directly leads to the change in prices of other houses. 2. The violation of nonnegative profits, where landlords make aggregate net losses when adjusting housing features (other than the treatment) in response to the treatment.}, which offers practical advantages for typical policy contexts. For a favorable amenity change, DUET provides a conservative (lower bound) measure of societal benefit, which is desirable in benefit-cost analyses as it bolsters regulatory credibility. For an undesirable amenity change, DUET provides an overestimate (upper bound) of the damage and hence a conservative compensation estimate, which aligns with regulatory principles for protecting victims and mitigating risk.\footnote{For instance, EPA generally adopts ``default assumptions'', which are standard scientific assumptions erring toward overestimation of damages to protect public health.} More importantly, targeting precise recovery of ES or other welfare measures is typically infeasible, as the gap between DUET and ES is determined by structural economic factors that are intrinsic to the policy context rather than amenable to econometric refinement. \footnote{These factors include, but are not limited to, the extent of residential mobility, the magnitude of general equilibrium price adjustments, and the shape of household preference distributions} Theoretical guidelines exist for accurately estimating DUET (e.g., using DID to mitigate bias from time-invariant unobservables), yet no analytical technique exists to systematically narrow the DUET-ES gap. Finally, as a result of the theoretical clarity provided by \citet{banzhaf2021}, the DUET target allows us to set up the empirical simulation conveniently with real transaction data.

While the theoretical framework for DUET estimation is well established, its empirical implementation faces two practical challenges that can violate the core identifying assumption of \textit{conditional parallel trends}. First, property transaction data exhibit \textit{spatial autocorrelation} in both observed and unobserved characteristics, which can make proximity-based treatment and control groups systematically different even when treatment sites are randomly located. Second, the pre-post \textit{equilibrium shift} in the hedonic price function turns the effects of time-invariant factors into time-varying effects that survive first differencing or spatial fixed effects, so DID alone cannot eliminate the resulting bias \citep{klaiber2013, kuminoff2014, banzhaf2021}. Strategies that may mitigate these challenges include, but are not limited to, the following: allowing the price function to shift over time (GDID),\footnote{Although GDID offers demonstrated advantages and has been adopted in some hedonic studies \citep[e.g.,][]{keiser2019, cheng2024, guignet2024, hu2025}, its adoption in the literature remains far from widespread and many recent hedonic studies have not adopted it \citep[e.g.,][]{mamun2023, lang2023, kang2024, guo2024, jarvis2025}} modeling the price function more flexibly and improving treatment-control balance with causal machine learning (CML) \citep{chernozhukov2016, athey2019}, incorporating a richer set of observables, and restricting the sample to improve balance. Section~\ref{sec:IdProblems} discusses these problems and strategies in detail.

How reliably these strategies and the associated estimators recover DUET is ultimately an empirical question. We address it with an empirical Monte Carlo simulation framework whose design and choice of methods draw on the prior simulation literature. CDM introduces the Monte Carlo framework to assess which hedonic models can recover willingness to pay (WTP) for nonmarginal amenity shocks, and KPP updates the sample size and the evaluated models to align with the concurrent literature. Later simulation studies closely follow KPP's design in sample size, equilibrium formulation, and simulation target \citep{gopalakrishnan2011, klaiber2013, banzhaf2021}, and several of them clarify that capitalization estimates from quasi-experimental methods conflate direct price effects with price function changes driven by equilibrium shift \citep{klaiber2013, kuminoff2014, banzhaf2021}.\footnote{When analysts assume a stable price function and use reduced-form methods to infer policy impacts, this concern is basically the housing market version of the Lucas critique \citep{lucas1976}.} KPP recommends a generalized DID approach, which, under correct functional form assumptions, deals with the equilibrium shift, whether induced by the treatment shock or otherwise. \citet{banzhaf2021} formally shows that this approach recovers DUET, a lower-bound approximation (75\% to 92\%) of ES. Complementing these studies, \citet{bishop2020} offers empirical best-practice recommendations, including the specification of spatial fixed effects and balance-improving preprocessing techniques, several of which we evaluate.

\begin{figure}[htbp]
  \centering
  \includegraphics[width=.79\textwidth]{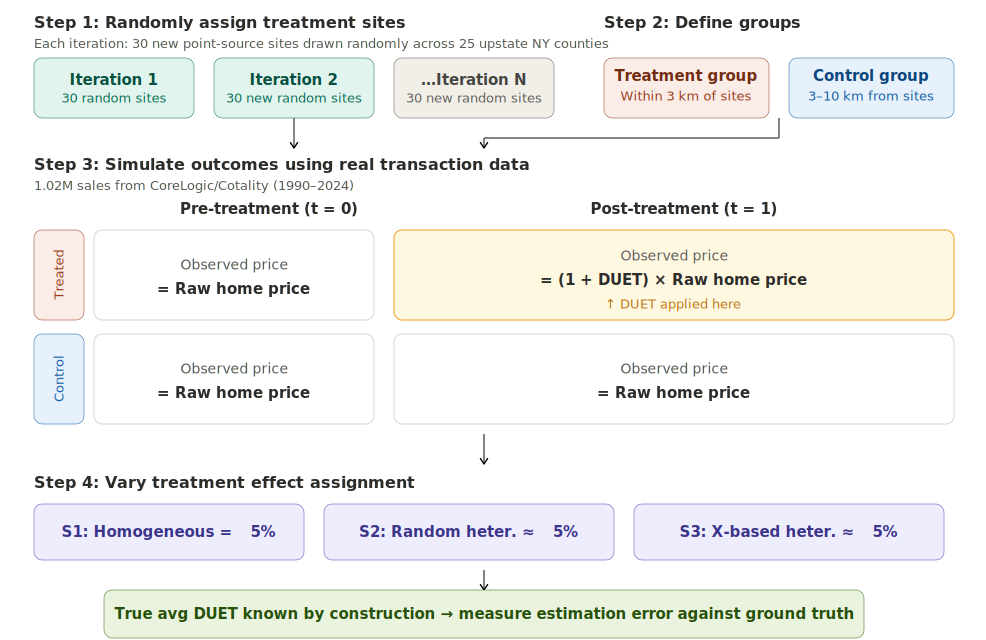}
  \caption{Empirical Simulation Design. Except for DUET-related price changes, all variables are from the raw transaction data. The original data generation process, including natural equilibrium shift, actual price functions, and actual spatial autocorrelation, is preserved. S2 and S3 in step 4 assign heterogeneous DUET across observations within each iteration. The average DUET of 5\% is from the main analysis, with variations in robustness checks.}
  \label{fig:design}
\end{figure}

While our design shares the core logic of KPP, re-estimating a known effect repeatedly in a Monte Carlo simulation and evaluating competing estimators, including GDID, by their estimation error, it departs from KPP in four ways. We focus exclusively on DUET rather than on recovering WTP from complete housing market equilibria. We preserve the real data-generating process rather than imposing parametric assumptions: we take observed pre-treatment prices and simulate post-treatment prices by applying the DUET directly to real transaction prices (Figure~\ref{fig:design}), whereas KPP generate pseudo-data from specified utility functions, preference distributions, housing supply assumptions, and bidding algorithms. We randomly reassign treatment locations across iterations, which improves the generalizability of the results across spatial configurations of treatment, whereas KPP use a fixed treatment.\footnote{KPP use a fixed treatment in a local market with randomness coming from subsampling and equilibrium formation. This design yields results specific to that unique treatment assignment and likely lacks generalizability to other treatments or locations.} Finally, we simulate a broader regional market with far more housing units than KPP to reflect contemporary sample sizes.\footnote{The sample size of 2,000 in KPP reflects data constraints in earlier hedonic studies.}

The simulation study spans 25 counties in upstate New York (Figure~\ref{fig:region}). In each iteration, we randomly place 30 point-source amenity sites, each inducing a percentage price change for homes within a 3~km radius, and apply that change to actual post-treatment transaction prices. We then estimate the DUET with each estimator under investigation. Scenarios vary by how the treatment effect is assigned: homogeneous, randomly heterogeneous, or heterogeneous in observable characteristics. The baseline design does not generate a treatment-induced equilibrium shift, which we relax in Section~\ref{sec:indshift}, but the naturally occurring equilibrium shift remains present in the data, making the generalized DID framework necessary for consistent estimation. Because the true average DUET is known by construction, we can assess estimation accuracy directly. We evaluate three traditional regression estimators, baseline DID (BDID), generalized DID with coarse spatial fixed effects (T-GDID), and generalized DID with property fixed effects (T-GDIDP), together with a set of causal forest estimators that embed the generalized DID framework in a two-stage machine learning procedure. Section~\ref{sec:estimationapproaches} describes each estimator.

\begin{figure}[htbp]
  \centering
  \includegraphics[width=0.75\textwidth]{Figure_2.jpg}
  \caption{Study Region. Map lines delineate study areas and do not necessarily depict accepted national boundaries.}
  \label{fig:region}
\end{figure}

Our simulation results yield several key findings. First, traditional generalized difference-in-differences regression (T-GDID) consistently outperforms other traditional methods including baseline DID, two-way fixed effects panel models (T-GDIDP), and baseline post-period OLS across all scenarios, confirming the importance of allowing price function shifts over time. Second, causal machine learning methods, particularly causal forest DID on adjacent pairs (CF2), achieve performance comparable to T-GDID in general, and demonstrate superior performance in larger samples. Third, the optimal sample frame choices seem to vary across methods and differ from widely adopted practices: specifying a too-narrow (e.g., with a 1~km width) proximity bin for control observations may severely diminish traditional or causal machine learning estimator performance; shortening the pre-treatment period does improve traditional T-GDID performance, but this improvement tends to degrade in CF2, demonstrating the flexibility of causal machine learning to extract relevant information from less relevant samples. Fourth, when geographic attributes are appropriately incorporated (i.e., in the price model only, not the propensity score model or causal forest algorithm) and sample size is sufficiently large (e.g., at least 3,000 treated observations in our simulation context), CF2 achieves the lowest estimation errors with near-nominal CI coverage rate among all tested estimators. T-GDID reaches nominal coverage only when inference is clustered at the treatment-site level with a wild cluster bootstrap, at the cost of intervals far wider than CF2's. Fifth, when we additionally impose a treatment-induced equilibrium shift, its functional form determines which estimator gains. A shift that is nonlinear in property characteristics considerably widens CF2's advantage over T-GDID as the shift intensifies (in large samples, T-GDID's RMSE more than triples while CF2's rises only modestly). To the extent that real amenity shocks reshape the price function nonlinearly, our main comparison, which omits the induced shift, likely understates CF2's advantage over T-GDID. These findings suggest that the choice between traditional and machine learning methods should be guided by sample size and price function complexity, with CF2 offering substantial advantages in large-sample contexts common in contemporary hedonic studies.

This study makes four contributions to the hedonic price literature. First, to our knowledge it is the first simulation study that targets the direct unmediated price effect on treated (DUET) as the estimation objective, providing guidance aligned with the welfare interpretation established in \citet{banzhaf2021}. Second, by adapting causal machine learning methods to the generalized DID framework, it offers the first systematic assessment of how double machine learning could improve hedonic price effect estimation for spatially delineated treatments. Third, unlike previous simulation studies that rely on parametric assumptions about utility functions and market equilibria, our empirical simulation preserves the real data-generating process, enhancing the external validity of the resulting recommendations. Finally, we provide method-specific best practice recommendations that reflect the different specification requirements of regression and machine learning approaches.

Our findings matter for policy because hedonic estimates of the price effects of spatially delineated amenity shocks feed directly into policy analyses that support consequential decisions, from national and regional regulation changes to the local siting and compensation decisions for new infrastructure (e.g., factories, large-scale solar/wind sites, and data centers). If adopted in practice, our empirical simulation framework would serve as a transparent, data-based way to evaluate estimators and specifications, boosting the credibility of benefit estimation and related policymaking efforts. Furthermore, as demonstrated in this study, the improvement in average effect estimation accuracy from better models (e.g., GDID estimators over BDID) could translate into considerable changes in dollar amount of estimated benefit (e.g., a seven percentage point improvement in MAPE means an expected \$1.968 million\footnote{Note that this total benefit change is based on the relatively low average property value of \$134,207.4 in our study region (as shown in Appendix Table D.1), a treatment effect size of 5\%, and an average number of 4189 treated properties for 30 random treatment sites in our simulation sample with $\geq$3k treated.} change in estimated benefit for 30 random sites in our study region), potentially shifting the underlying benefit-cost analysis. Lastly, as large datasets become increasingly available for environmental policymaking, we show how causal machine learning can be applied to estimate DUET in a hedonic DID setting, contributing to an emerging literature that brings modern machine learning tools to policy-relevant benefit-cost analysis \citep[e.g.,][]{chen2025heterogeneous, faccioli2026random, johnston2026random}.

The remainder of this paper proceeds as follows. Section~\ref{sec:method} lays out the conceptual framework for DUET, the identification problems and strategies, and the estimators we compare. Section~\ref{sec:simdata} describes the simulation design, its maintained assumptions, and the property transaction data. Section~\ref{sec:results} presents the simulation results. Section~\ref{sec:discussion} discusses best practice recommendations and caveats, and Section~\ref{sec:conclusion} concludes.

\section{Methodology}
 \label{sec:method}
\subsection{Conceptual Framework}
 \label{sec:conceptframe}
We illustrate the concept and estimation of the direct unmediated price effect on treated, or DUET, in a framework closely following \citet{banzhaf2021}.
 
Let $g$ denote the environmental amenity of interest, $p$ denote the property price function, and $\mathbf{X}$ denote a comprehensive set of property features excluding $g$. If we let $a$ denote the state of the world and abstract the timeline into a binary variable $t$, with $t = 0$ representing the pre-treatment period and $t = 1$ the post-treatment period, we can write the original state as $a^{0}$, the observed post-treatment state as $a^{*}$, and the counterfactual state at $t = 1$ without treatment as $a'$. The corresponding price functions are then represented by $p^{a^{0}}$, $p^{a'}$, and $p^{a^{*}}$, and other symbols share the same superscript convention.
 
The natural equilibrium shift in the absence of treatment from $t = 0$ to $t = 1$ can then be described by $p^{a'}(\cdot) - p^{a^{0}}(\cdot)$, and the treatment-induced equilibrium shift is $p^{a^{*}}(\cdot) - p^{a'}(\cdot)$. At $t = 1$, the treatment group will have $g^{a^{*}} = 1$ and $g^{a'} = 0$, and controls will have $g^{a^{*}} = 0$ and $g^{a'} = 0$. The total effect (TE) of the treatment on the price for an individual house $i$ in the treatment group can be calculated as:
 
\begin{equation}\label{eq:TE}
\begin{split}
{TE}_{i} &= \left( p^{a^{*}}\left( g^{a^{*}},\mathbf{X}^{a^{*}} \right) - p^{a^{0}}\left( g^{a^{0}},\mathbf{X}^{a^{0}} \right) \right) - \left( p^{a'}\left( g^{a'},\mathbf{X}^{a'} \right) - p^{a^{0}}\left( g^{a^{0}},\mathbf{X}^{a^{0}} \right) \right)\\
&= p^{a^{*}}\left( g^{a^{*}},\mathbf{X}^{a^{*}} \right) - p^{a'}\left( g^{a'},\mathbf{X}^{a'} \right) = p^{a^{*}}\left( 1,\mathbf{X}^{a^{*}} \right) - p^{a'}\left( 0,\mathbf{X}^{a'} \right).
\end{split}
\end{equation}
 
However, $TE$ is not a good value measure. First, $TE$ includes both the price change from the treatment-induced amenity improvement and the price change from the equilibrium shift in the hedonic price function (i.e., the function change from $p^{a'}(\cdot)$ to $p^{a^{*}}(\cdot)$), with the latter affecting both treatment and control groups. This equilibrium-shift-induced change is called the indirect price effect ($IE$) of the treatment in \citet{banzhaf2021}, and can be expressed as ${IE}_{i} = p^{a^{*}}\left( 0,\mathbf{X}_{g = 0}^{a^{*}} \right) - p^{a'}\left( 0,\mathbf{X}^{a'} \right)$. Second, since we cannot observe the counterfactual state $a'$, neither $TE$ nor $IE$ can be conveniently estimated in practice. In contrast, the direct price effect of an amenity change along the post-treatment price function is a more viable estimation target, which can be expressed as:
 
\begin{equation}\label{eq:DE}
{DE}_{i} = p^{a^{*}}\left( 1,\mathbf{X}_{g = 1}^{a^{*}} \right) - p^{a^{*}}\left( 0,\mathbf{X}_{g = 0}^{a^{*}} \right).
\end{equation}
This $DE$ measure can also be expressed as $TE - IE$.\footnote{Equation~(\ref{eq:DE}) is introduced only to explain the concept of $DE$. Our main simulations preserve the natural equilibrium shift in the data but set the treatment-induced shift to zero ($IE_i = 0$); Section~\ref{sec:assumptions} discusses this assumption and Section~\ref{sec:indshift} relaxes it by adding an induced shift.}
 
In practice, when amenity $g$ changes, not only can the equilibrium shift, but the observable covariates $\mathbf{X}$ may also change. Therefore, $TE$, $IE$, and $DE$ all involve the price effects mediated through changes in $\mathbf{X}$. If we could effectively hold $\mathbf{X}$ constant (e.g., at $\widetilde{\mathbf{X}}$) while changing $g$, we can measure the direct unmediated price change:
 
\begin{equation}\label{eq:DUE}
{DUE}_{i} = p^{a^{*}}\left( 1,\widetilde{\mathbf{X}} \right) - p^{a^{*}}\left( 0,\widetilde{\mathbf{X}} \right)
\end{equation}
\citet{banzhaf2021} showed that a generalized DID estimator recovers the average $DUE$ on the treated ($DUET$ hereafter). Also, $DUET$ is regarded as a proper value measure of the amenity change in \citet{banzhaf2021}, as it is shown to be a lower-bound approximation of the welfare change measured by Hicksian ES.
 
From equation~(\ref{eq:DUE}), it appears that the analyst can estimate $DUET$ easily just from data in $t = 1$, given that $p^{a^{*}}(\cdot)$ is typically observable. However, spatial autocorrelation induced endogeneity (clarified in the following subsection) will make the treated ($g^{a^{*}} = 1$) and controls ($g^{a^{*}} = 0$) systematically different in $\mathbf{X}^{a^{*}}$, so that it is difficult to hold $\mathbf{X}^{a^{*}}$ constant across the two groups for unbiased DUET estimation.
 
This endogeneity problem can be mitigated by investigating pre-post price changes (i.e., in a DID framework) since the changes in $\mathbf{X}^{a^{*}}$ across time are less likely systematically different between the treatment and control groups. The generalized DID approach in KPP essentially recommends estimating separate price functions for the pre- and post-treatment periods (to accommodate price function shifts) and using the resulting price residuals in a DID framework. With linear-in-$\mathbf{X}$ price modeling as an example, \citet{banzhaf2021} showed that this generalized DID estimation recovers the average DUET:
\begin{equation}\label{eq:GDID}
\begin{split}
\overline{DUET\left( a^{*} \right)} &= E\left[ \left( p^{a^{*}}\left( g^{a^{*}} = 1 \right) - \widehat{\mathbf{\gamma}_{g = 1}^{a^{*}}}\mathbf{X}_{g = 1}^{a^{*}} \right) - \left( p^{a^{0}} - \widehat{\mathbf{\gamma}_{g = 0}^{a^{0}}}\mathbf{X}^{a^{0}} \right)\mid g^{a^{*}} = 1 \right]\\
&\quad - E\left[ \left( p^{a^{*}}\left( g^{a^{*}} = 0 \right) - \widehat{\mathbf{\gamma}_{g = 0}^{a^{*}}}\mathbf{X}_{g = 0}^{a^{*}} \right) - \left( p^{a^{0}} - \widehat{\mathbf{\gamma}_{g = 0}^{a^{0}}}\mathbf{X}^{a^{0}} \right)\mid g^{a^{*}} = 0 \right],
\end{split}
\end{equation}
where $\widehat{\mathbf{\gamma}_{g = 1}^{a^{*}}} = \widehat{\mathbf{\gamma}_{g = 0}^{a^{*}}}$, with $\widehat{\mathbf{\gamma}}$ representing the price coefficients related to $\mathbf{X}$. The right-hand side of equation~(\ref{eq:GDID}) is the generalized DID estimator. While more detailed derivation and proof related to DUET can be found in \citet{banzhaf2021}, we briefly show why equation~(\ref{eq:GDID}) holds. Starting from equation~(\ref{eq:DUE}), let $\mathbf{X}_{g = 1}^{a^{*}}$ represent the actual treated group feature vector while $\widetilde{\mathbf{X}}$ represents the feature vector the treatment units would have exhibited had they been assigned the control status in $t = 1$, we can write $\overline{DUET\left( a^{*} \right)}$ as:
\begin{equation}\label{eq:DUETexp}
\begin{split}
\overline{DUET\left( a^{*} \right)} &= E\left[ p^{a^{*}}\left( 1,\widetilde{\mathbf{X}} \right) - p^{a^{*}}\left( 0,\widetilde{\mathbf{X}} \right) \mid g^{a^{*}} = 1 \right]\\
&= E\left[ p^{a^{*}}\left( 1,\mathbf{X}_{g = 1}^{a^{*}} \right) + \widehat{\mathbf{\gamma}_{g = 1}^{a^{*}}}(\widetilde{\mathbf{X}} - \mathbf{X}_{g = 1}^{a^{*}}) - p^{a^{*}}\left( 0,\widetilde{\mathbf{X}} \right) \mid g^{a^{*}} = 1 \right]\\
&= E\left[ \left( p^{a^{*}}\left( 1,\mathbf{X}_{g = 1}^{a^{*}} \right) - \widehat{\mathbf{\gamma}_{g = 1}^{a^{*}}}\mathbf{X}_{g = 1}^{a^{*}} \right) - \left( p^{a^{0}} - \widehat{\mathbf{\gamma}_{g = 0}^{a^{0}}}\mathbf{X}^{a^{0}} \right)\mid g^{a^{*}} = 1 \right]\\
&\quad - E\left[ \left( p^{a^{*}}\left( 0,\widetilde{\mathbf{X}} \right) - \widehat{\mathbf{\gamma}_{g = 1}^{a^{*}}}\widetilde{\mathbf{X}} \right) - \left( p^{a^{0}} - \widehat{\mathbf{\gamma}_{g = 0}^{a^{0}}}\mathbf{X}^{a^{0}} \right) \mid g^{a^{*}} = 1 \right].
\end{split}
\end{equation}
Applying the conditional parallel trends condition, \footnote{This conditional parallel trends condition suggests that, in the absence of treatment, the treatment and control groups would exhibit parallel trends conditional on $\mathbf{X}$. Note that this is a variant of the conditional parallel trends condition applied on residual prices, which mirrors Assumption 2 (Conditional mean independence in differences for the treated) in \citet{banzhaf2021}.}
\begin{equation}\label{eq:CPT}
\begin{split}
E\left[ \left( p^{a^{*}}\left( 0,\widetilde{\mathbf{X}} \right) - \widehat{\mathbf{\gamma}_{g = 1}^{a^{*}}}\widetilde{\mathbf{X}} \right) - \left( p^{a^{0}} - \widehat{\mathbf{\gamma}_{g = 0}^{a^{0}}}\mathbf{X}^{a^{0}} \right) \mid g^{a^{*}} = 1 \right] \\
\quad = E\left[ \left( p^{a^{*}}\left( 0,\mathbf{X}_{g = 0}^{a^{*}} \right) - \widehat{\mathbf{\gamma}_{g = 0}^{a^{*}}}\mathbf{X}_{g = 0}^{a^{*}} \right) - \left( p^{a^{0}} - \widehat{\mathbf{\gamma}_{g = 0}^{a^{0}}}\mathbf{X}^{a^{0}} \right) \mid g^{a^{*}} = 0 \right],
\end{split}
\end{equation}
it is immediately clear that equation~(\ref{eq:GDID}) holds.

\subsection{Identification Problems and Strategies}
 \label{sec:IdProblems}
While the theoretical framework for DUET estimation is clarified above, its empirical implementation presents several challenges, which we discuss below. 

The \textit{Conditional Parallel Trends} assumption, as presented in equation~(\ref{eq:CPT}), is the core identifying assumption underlying the above conceptual framework and all methods employed in this study. In practice, however, it may be violated due to the following challenges.

\textit{Spatial autocorrelation} is typically exhibited in property transaction data for both observed and unobserved characteristics. Even if treatment is randomly assigned across locations, the treatment indicator might be endogenous as spatial autocorrelation may induce systematic differences between treated and control groups. The associated confounding bias from time-invariant unobservables (and observables) can be mitigated via first differencing (or spatial fixed effects) but cannot be eliminated because of the equilibrium shift problem.
 
The pre-post \textit{equilibrium or price function shift} \citep{klaiber2013, kuminoff2014, banzhaf2021} exacerbates the spatial-autocorrelation-induced confounding from time-invariant factors, even after first differencing. Intuitively, the pre-post shift in the price function transforms the effects of time-invariant factors into time-varying effects that survive first differencing. Combined with confounding from genuinely time-variant factors, this equilibrium-shift-induced confounding can severely bias baseline DID estimation when spatial autocorrelation is strong, as systematic treatment-control differences in covariate-related terms persist and contaminate identification. 

To address these issues, we adopt several general strategies as listed below, and the simulation results allow us to evaluate their effectiveness.
 
First, allowing the price function to shift over time (i.e., the generalized DID framework proposed in KPP) can mitigate this problem as shown in the literature (e.g., KPP and \citealt{banzhaf2021}).
 
Second, better price modeling that reduces specification bias related to $\mathbf{X}$ will further mitigate the confounding effect from (time-variant and time-invariant) observables and reduce bias in treatment effect estimation \citep{caetano2024difference}. Causal machine learning estimators utilize flexible price modeling with machine learning (e.g., addressing complex interactions and nonlinearity), improve sample balance via adaptive matching\footnote{These estimators essentially create matched samples (i.e., treatment-control pairs) to improve treatment-control balance — in causal forests, this is achieved within the leaves generated by the honest tree-splitting process.}, and thus may recover ATT more precisely \citep{chernozhukov2016}.
 
Third, with large-scale modern datasets and machine learning methods capable of managing high-dimensional covariates, acquiring more observables can help researchers reduce bias by better accounting for (or proxying) confounding unobservables in practice.
 
Moreover, sample-balancing strategies that discard a portion of the data upfront, such as selecting control observations within a narrow proximity bin close to treated units or restricting the analysis to a shorter time window, can enhance sample balance and thereby improve estimation precision. However, these approaches may backfire in causal machine learning models, which are typically data-hungry and can suffer from reduced performance when the effective sample size is substantially decreased.
 
\subsection{Estimation Approaches}
\label{sec:estimationapproaches}
 
Based on the strategies stated above, we put to the test several widely used hedonic models as well as a few novel machine learning approaches. We compare these estimators with three general objectives in mind: 

\begin{enumerate}
    \item Investigate whether traditional generalized DID estimators, which are capable of addressing equilibrium shift and recovering DUET, perform better than baseline models;
    \item Evaluate whether machine-learning-based estimators, which improve price function modeling and mitigate spatial autocorrelation, improve estimation performance over traditional generalized DID estimators;
    \item Assess how different specification choices, including variable specifications and sample frame selections, affect the performance of each estimator.
\end{enumerate}

The estimators we investigate (enumerated below) fall into two groups: traditional regression-based methods (BDID, T-GDID, T-GDIDP), which serve as benchmarks, and causal machine learning methods, which adapt causal forest algorithms to the generalized DID framework. In the main analysis, we focus on the best-performing causal machine learning method, CF2. Appendix A provides further details on all estimators, including formal descriptions and simulation results for additional causal machine learning approaches.
 
\textit{Benchmark DID (BDID)} is the traditional difference-in-differences regression with a comprehensive set of covariates and zip-code-by-year fixed effects. This approach does not allow the price function (specifically, the $\mathbf{X}$ coefficients) to shift over time, and hence leave more $\mathbf{X}$-related confounding in the estimation.\footnote{We summarize the specifications for standard error estimation here. For all OLS regression approaches, standard errors are two-way clustered by zip code and year. For causal forest approaches including CF2, the standard error of the average effect is the heteroskedasticity-robust standard error of the residual-on-residual regression computed by the \texttt{average\_treatment\_effect} function of the \texttt{grf} package with the overlap option, which combines out-of-bag estimates of the propensity score and of the expected price change \citep{athey2019}. Neither procedure fully reflects that treatment is assigned through a limited number of spatial sites, so both may understate sampling variability. Section~\ref{sec:efficiency} and Appendix Table D.5 examine this limitation, including site-level clustering with a wild cluster bootstrap for T-GDID.}
 
\textit{Traditional Generalized DID (T-GDID)} is the generalized difference-in-differences regression (recommended in KPP), involving a comprehensive set of covariates, interactions between these covariates with a time indicator (allowing the price function to shift), and zip-code-by-year fixed effects.
 
\textit{Traditional Generalized Panel DID (T-GDIDP)} is identical to T-GDID except that it uses parcel-level fixed effects and year-fixed effects (instead of zip-code-by-year fixed effects), so the effect identification relies only on repeated sales. T-GDIDP is also similar to the commonly known two-way fixed effects model, except that T-GDIDP allows the price function to shift over time.
 
\textit{Causal Forests DID with Adjacent-pairs (CF2)} runs causal forest on the first difference of prices, applying only to the closest pre- and post-treatment transaction pairs (adjacent pairs hereafter). The causal forest \citep{athey2019} estimation involves two stages: it first models the outcome (i.e., first-differences of prices in this context) and the treatment with random forests, then passes the residuals to the causal forest algorithm in the second step, which produces the GDID estimates. Formally, let $\Delta \ln P_i$ denote the first difference in log price for adjacent pair $i$ ($\Delta \ln P_i = \ln P_{i,\text{post}} - \ln P_{i,\text{pre}}$), the first stage of CF2 estimates the outcome and treatment models (i.e., Robinson's transformation in causal forests\footnote{For further details on Robinson's transformation and the causal forest algorithm, see \citet{athey2019} and the documentation for the \texttt{grf} R package: \url{https://grf-labs.github.io/grf/REFERENCE.html\#orthogonalization-the-r-learner}.}):

\begin{equation}\label{eq:Ytilde}
\tilde{Y}_{it} = Y_{it} - \hat{E}\left[Y_{it} \mid \mathbf{X}_{it}\right],
\end{equation}

\begin{equation}\label{eq:Dtilde}
\tilde{D}_{i} = D_{i} - \hat{E}\left[D_{i} \mid \mathbf{X}_{it}\right] = D_i - \hat{e}(\mathbf{X}_{it}),
\end{equation}
where $Y_{it} = \Delta \ln P_{it}$,  $\hat{E}\left[Y_{it} \mid \mathbf{X}_{it}\right]$ is a random forest model of outcome given covariates, and $\hat{e}(\mathbf{X}_{it})$ is the estimated propensity score from a random forest on the treatment indicator. To minimize overfitting, these first-stage random forest models are estimated using a cross-fitting technique\footnote{In the \texttt{grf} package, this is implemented through out-of-bag prediction. Since each tree is grown on a subsample of the data, each observation's prediction averages only over trees that were not trained on it. The second-stage causal forest uses the same procedure.}, where each observation's residual is essentially computed using a held-out sample that excludes that observation. Note that the feature vector $\mathbf{X}_{it}$, compared to $\mathbf{X}$ used in traditional GDID estimators, additionally includes the post-transaction year and the within-pair difference in transaction years, providing the forest with sufficient time information to internally model and accommodate price function shifts in equation~(\ref{eq:Ytilde}).\footnote{Rearranging the right-hand side of equation~(\ref{eq:GDID}), we can find that, compared to baseline DID, the GDID estimator essentially just further models the first-difference of price with time-appropriate $\mathbf{X}$. CF2 does exactly this: it models the first-differenced-prices with random forests and $\mathbf{X}_{it}$, then passes the residuals to the causal forest algorithm, which conducts the second difference and gives the GDID estimate.} The resulting residuals $\tilde{Y}_{i}$ and $\tilde{D}_i$ are then passed to the second stage (i.e., the causal forest).

The second stage of CF2, the causal forest algorithm, estimates a conditional treatment effect $\hat{\tau}(\mathbf{X}_i)$ for each observation:

\begin{equation}\label{eq:CF}
\tilde{Y}_i = \tau(\mathbf{X}_i)\tilde{D}_i + \varepsilon_i,
\end{equation}
where $\tau(\mathbf{X}_i)$ varies flexibly with covariates, as a result of the causal forest's adaptive matching algorithm. In general, these conditional average treatment estimates are aggregated across observations with the doubly robust AIPW approach to estimate an average effect. Because overlap is limited in our setting, we use the overlap-weighted variant of this aggregation to estimate the average DUET (as detailed in Appendix A).

\section{Simulation and Data}
\label{sec:simdata}

This section presents the empirical simulation framework used to evaluate the two families of estimators introduced in Section \ref{sec:estimationapproaches}: traditional regression-based methods (BDID, T-GDID, T-GDIDP), which serve as the benchmark, and the causal machine learning method (CF2), which adapts the causal forest algorithm to the generalized DID framework. 
 
\subsection{Simulation Design}
\label{sec:simulationdesign}
 
This subsection presents details of the empirical simulation design used to evaluate DUET estimators. To facilitate reading, we present a summary of all simulation parameters and settings in Table~\ref{tab:design}.
 
\textit{Outcome} The treatment effect (i.e., DUET) on home prices is specified as percentage changes on real post-treated prices. We do not generate artificial transactions; the price modification is based on real transaction prices in the post-treated group. To accommodate for situations where price changes might be heterogeneous, we investigate the following three scenarios for the price change. Although the assigned effects are heterogeneous across observations in Scenario 2 and Scenario 3, and causal machine learning estimators are designed to estimate such heterogeneous treatment effect, we target the average effect as our primary estimand to maintain comparability with traditional approaches in this study.
 
Scenario 1. The price effects are homogeneous;
 
Scenario 2. The price effects are heterogeneous, where the heterogeneity is randomly drawn from a list of potential values unrelated to observable features;
 
Scenario 3. The price effects are heterogeneous, where the heterogeneity is related to observable features.
 
\textit{Treatment} Spatially delineated treatments of interest are assumed to have proximity-based effects, with the treatment sites abstracted to points on the map.\footnote{This point specification intends to accelerate the simulation process without loss of generality.} To enhance generalizability across varying treatment locations, the simulation randomly assigns treatment points across repeated iterations. The treatment effect is assumed to be imposed on treated properties within a certain radius of treatment points (i.e., 3~km in the simulation). The outcome and treatment design in this simulation are further illustrated in Figure~\ref{fig:design} and Table~\ref{tab:design}. 

\textit{Estimation Sample} The analysis is initially conducted using transactions located within a 10~km radius of the treatment sites spanning 1990 through 2024. The sample radius and time frame are subsequently varied to examine how they influence the estimation performance.

\textit{Estimator Evaluation Criteria} Since our aim is to assess the performance in estimating average DUET across estimators, we must first establish the true average DUET --- calculated as the average of assigned individual DUET across all treated observations per iteration. In each simulation iteration, the DUET estimate from a certain estimator will be compared against the true average DUET, generating error measures for that estimator. Error measures in this study include Root Mean Squared Error (RMSE) and Mean Absolute Percentage Error (MAPE) measured across all iterations.\footnote{RMSE (Root Mean Squared Error) measures the typical estimation error: $RMSE = \sqrt{\frac{1}{n}\sum_{i = 1}^{n}({\widehat{\tau}}_{i} - \tau_{i})^{2}}$, where ${\widehat{\tau}}_{i}$ is the estimated effect and $\tau_{i}$ is the true effect in iteration $i$. MAPE (Mean Absolute Percentage Error) expresses average error as a percentage of the true effect: $MAPE = \frac{100\%}{n}\sum_{i = 1}^{n} \mid \frac{{\widehat{\tau}}_{i} - \tau_{i}}{\tau_{i}} \mid$. RMSE penalizes large errors more heavily, while MAPE provides scale-free comparison across scenarios.} When two estimation approaches achieve very similar error measures, we will further examine the width of the confidence intervals (measured by the average standard error) and the empirical coverage rates (i.e., the empirical probability that the confidence intervals cover the true DUET). A narrower interval is desirable only if it still covers the true DUET at close to the nominal rate, so we read the two measures together.

\begin{table}[H]
\centering
\caption{Simulation Design Details}
\label{tab:design}
\renewcommand{\arraystretch}{0.75}
\begin{tabular}{@{}p{5cm}p{11cm}@{}}
\toprule
\textbf{Feature} & \textbf{Specification} \\
\midrule
Empirical Data Location & Twenty-five counties in upstate NY (Figure~\ref{fig:region}) \\
Effect Distance & 3~km \\
Sample Inclusion Criterion & 10~km from the treatment site; vary to show sensitivity \\
Treatment Site Assignment & Randomly assigned to coordinates \\
Treatment Site Number & 30 \\
Treatment Effect & \textit{Scenario 1}: Homogeneous DUET (.05); \newline
\textit{Scenario 2}: Heterogeneous DUET randomly drawn from (.01, .02, .04, .05, .06, .08, .09), with average approximating .05; \newline
\textit{Scenario 3}: Heterogeneous DUET related to observables (i.e., a function of building age, living space, number of bathrooms, and sewer type), with average approximating .05. \\
Treatment Year & 2010, uniform across sites and iterations \\
Estimation Methods & Traditional generalized DID (T-GDID); Traditional generalized DID on repeated sales (T-GDIDP); Benchmark DID (BDID); Causal forest GDID on repeated sales (CF2) \\
Sample Size & Varies based on treatment site location and associated property/transaction density \\
\bottomrule
\end{tabular}
\end{table}

To best preserve the original data generation process (including the natural equilibrium shift, actual price functions, and actual spatial autocorrelation), the main simulations do not impose a treatment-induced equilibrium shift ($IE_i = 0$). Section~\ref{sec:assumptions} explains why this does not affect the theoretical conclusions or any empirical procedure, and Section~\ref{sec:indshift} relaxes it by adding induced shifts of increasing intensity.

\subsection{What Our Simulation Does and Does Not Assume}
\label{sec:assumptions}

Because our empirical Monte Carlo design draws its data-generating process from real transactions rather than from a fully specified structural model, the assumptions it relies on are easy to leave implicit. We therefore state them explicitly, give a formal statement of the conditions under which the causal machine learning estimator (CF2) identifies the average DUET, and explain why these conditions alone do not determine which estimator performs best.

We list all \textit{assumptions} adopted in our simulation design here. 

First, \textit{the real market is the data-generating process}: every price, covariate, transaction date, and location is taken from the observed 1990--2024 transaction record (the CoreLogic data detailed in Section~\ref{sec:data}), so the natural equilibrium shift, the true (unknown) hedonic price surface, and the empirical spatial autocorrelation are inherited rather than parameterized. The simulation manipulates only the treatment assignment and the treatment-induced price change, so the ground-truth effect is known by construction while the rest of the environment remains empirically realistic. 

Second, \textit{the baseline design imposes no treatment-induced equilibrium shift} ($IE_i = 0$): in the main simulations the treatment moves only the treated units' prices, not the shape of the price function, and the surrounding price surface is shaped by the natural equilibrium shift. This does not affect the theoretical conclusions of \citet{banzhaf2021} or any empirical procedure, because all estimators target $DE$ rather than the split between $DE$ and $IE$, and the natural shift already present in the data functions in the same way as the induced shift and gives generalized DID its advantage over baseline DID. Its only consequence is that the post-period price function the estimators must approximate may be simpler than in settings where the treatment-induced shift presents. Whether this favors traditional or CML estimators depends on the form of the induced shift (Section~\ref{sec:indshift}). Section~\ref{sec:indshift} and Appendix E relax this assumption by adding an explicit induced shift. 

Third, \textit{conditional parallel trends}: because treated and control price changes are generated from the same underlying surface (absent the imposed effect), the identifying condition of equation~(\ref{eq:CPT}) --- that treated and control units share the same expected period-to-period change in price given $\mathbf{X}$ --- is assumed to hold in the design. In an empirical setting, conditional parallel trends might be violated as stated in Section~\ref{sec:IdProblems}), leading to final estimation bias. 

Fourth, \textit{local non-interference (SUTVA)} holds: changing one house's treatment status does not appreciably move any other house's price, so there is no tipping or general-equilibrium feedback in the baseline. this is Assumption A1 of \citet{banzhaf2021}, and it is what makes the unit-level potential outcomes well defined.

Finally, causal forests approaches also rely on \textit{overlap}: which requires that every covariate value has a positive probability of being observed in both the treatment and control groups, that is, $0 < e(\mathbf{X}_i) \equiv P(D_i = 1 \mid \mathbf{X}_i) < 1$. In our context, however, (A2) fails in covariate cells at both extremes. Close to treatment sites, nearly all observations are treated and comparable controls rarely exist, whereas far from them, nearly all observations are controls and comparable treated units are rare. We therefore target the overlap-weighted average effect \citep{li2018}, which assigns vanishing weight to both types of cells and avoids dividing by estimated propensity terms close to zero.\footnote{This is implemented with the \texttt{target.sample = "overlap"} option of \texttt{average\_treatment\_effect()} in the \texttt{grf} package, whose target is $E[e(\mathbf{X})(1-e(\mathbf{X}))\,\tau(\mathbf{X})]\,/\,E[e(\mathbf{X})(1-e(\mathbf{X}))]$. A few things are worth noting when this overlap-weighted average effect is compared to the ATT. The two estimands coincide when the treatment effect does not vary systematically with $\mathbf{X}$, as in Scenarios 1 and 2. In Scenario 3, the overlap weights discount covariate cells in which nearly all units are treated, so the two estimands can differ, and any gap is counted as CF2 estimation error because all estimators are evaluated against the average DUET on all treated units. This gap appears to be minor, as CF2's error measures in Scenario 3 are nearly identical to those in Scenarios 1 and 2 (Table~\ref{tab:efficiency}). For ease of exposition, we continue to refer to the CF2 estimates as the average DUET.}

\textit{What the design does not assume.} The exercise does not posit a parametric structural model of utility, bidding, or sorting. Therefore, it speaks to the estimation accuracy of the average DUET but not to the tightness of the welfare bound (e.g., the Hicksian equivalent-surplus slack that \citet{banzhaf2021} characterizes, or the WTP slack that KPP characterizes). It also does not assume that any estimator's functional form is correctly specified, since assessing that is the objective of the study. Finally, it does not assume treatment is unconditionally random: site coordinates are drawn at random, but a property's treatment status depends on its distance to a site, and hence on the spatially autocorrelated covariates, so the design contains genuine confounding rather than assuming it away.

The conditions adopted, however, are asymptotic and conditional, and they do not settle which estimator performs better in practice, which is why the empirical simulation is needed. First, identification conditions are established under independent sampling, whereas our simulation preserves real data that are spatially correlated and involve unobservables and complex relationships (especially between the outcome and other factors). A simulation is necessary to show how well point identification and the associated standard-error theory apply empirically. Second, choosing estimators and specifications is important in empirical settings, but identification conditions alone do not rank them. Estimators that are consistent under the same assumptions can differ widely in finite samples, depending on how well each approximates the price function and how much variance its flexibility costs. Part of this trade-off can be examined in real data (e.g., out-of-sample fit in price models), but the resulting estimation error in the causal effect cannot, which requires a simulation design in which the effect is known by construction \citep{huber2013}. Third, the asymptotic normality of CF2 requires the forest nuisance estimates to converge faster than $n^{-1/4}$ \citep{athey2019application}. This rate cannot be verified directly, and whether the resulting normal approximation is adequate in a typical hedonic study is therefore an empirical question, which our simulation addresses by checking empirical confidence-interval coverage. 

\subsection{Data}
\label{sec:data}
 
This study utilizes comprehensive property transaction and characteristic data from CoreLogic (now known as Cotality), covering single-family residential properties in 25 counties across upstate New York from 1990 to 2024. CoreLogic maintains one of the most extensive databases of property characteristics and historical transaction records in the United States, providing detailed information on property attributes, sale prices, and transaction dates.
 
We focus exclusively on single-family residential homes for our analysis. To ensure that transaction prices reflect fair market values rather than distressed sales or non-arm's-length transfers, we implement a rigorous screening process generally following \citet{hu2025}. First, we utilize CoreLogic's proprietary indicator of arm's-length transactions, excluding properties that lack this designation. We further exclude transactions flagged as short sales, foreclosures, sales between relatives, bulk investor purchases, or sales for new construction. To prevent price distortions from speculative ``flipping,'' we remove any transaction occurring within 120 days of a previous sale of the same property. Finally, we exclude transactions in the bottom 1st percentile and top 99th percentile of the price distribution to diminish the influence of extreme outliers \citep{johnston2019}.
 
Geographic attributes augment the CoreLogic data and include distance measures to key infrastructure and amenities: transmission lines, primary roads, airports, railroads, water bodies, and conservation areas. These spatial variables are derived from multiple sources including U.S.\ Census Bureau TIGER/Line files, NOAA databases, the New York Protected Areas Database (NYPAD), and OurAirports.com. Additionally, we incorporate census tract identifiers and metropolitan area designations to enable spatial fixed effects and control for time-invariant location characteristics.
 
The final dataset encompasses over 1.02 million property transactions corresponding to 594,802 single-family residences across the study region. This rich dataset provides the empirical foundation for our Monte Carlo simulations, preserving real-world housing market structure, price distributions and trends, and spatial correlations in property features.

\section{Results}
 \label{sec:results}
 
\subsection{Can Generalized DID Improve Average DUET Estimation?}
 
KPP (via simulation) and \citet{banzhaf2021} (theoretically) proved that the generalized DID framework, allowing the price function to shift over time, outperforms the baseline DID models in recovering welfare measures. In this subsection, we demonstrate that the generalized DID framework improves the estimation accuracy of average DUET, compared to baseline DID models. We also investigate the CF variation of generalized DID to show the potential of causal machine learning models in hedonic studies.
 
In Table~\ref{tab:baseline}, we show the results for the first three scenarios, where the siting location is random. Table~\ref{tab:baseline} Panel A shows the estimation accuracy measures (the root mean square error, RMSE, and mean absolute percentage error, MAPE) for models based on the generalized DID framework, while Panel B shows corresponding error measures for the baseline models. These error measures confirm two best practice recommendations in the traditional hedonic literature but not always followed by practitioners: First, allowing price function shift over time and adopting a generalized DID approach improves (i.e., T-GDID vs. BDID) estimation accuracy \citep{kuminoff2010}; Second, compared to panel DID models that use property fixed effects (e.g., T-GDIDP), regression models that combine cross-sectional and within-panel variation through coarser geographic fixed effects (e.g., T-GDID) improve estimation accuracy in typical housing market datasets \citep{bishop2020}.

The causal machine learning approach, CF2,\footnote{Results for other causal machine learning approaches are available in Appendix A.} performs comparably to T-GDID, which outperforms other conventional estimators. Visual inspection of error patterns (Appendix Figures B.1-B.3) suggests estimation errors increase markedly in smaller samples and the relative performance of estimators seems to change with sample size. To formally identify the sample size threshold, we employ Kolmogorov-Smirnov tests \citep{smirnov1948, conover1999} comparing error distributions between CF2 and T-GDID across sample size regimes (Appendix Table B.1). Results indicate that 3,000 treated observations represent the lowest threshold at which CF2 begins to reliably outperform T-GDID, with statistically significant distributional differences across the threshold (KS statistic = 0.162, p $<$ 0.01) and CF2 achieving lower errors in 51.7\% of iterations in samples at or above the threshold.

\begin{table}[H]
\centering
\caption{Error Measures for Baseline Models}
\label{tab:baseline}
\small
\resizebox{\textwidth}{!}{%
\renewcommand{\arraystretch}{0.90}
\begin{tabular}{@{}llcccccc@{}}
\toprule
& & \multicolumn{3}{c}{\textbf{All iterations}} & \multicolumn{3}{c}{\textbf{No. Treated $\geq$3k}} \\
\cmidrule(lr){3-5} \cmidrule(lr){6-8}
& & \textbf{Scenario 1} & \textbf{Scenario 2} & \textbf{Scenario 3} & \textbf{Scenario 1} & \textbf{Scenario 2} & \textbf{Scenario 3} \\
& & Homo. & Rand. Heter.  & X-Heter., & Homo. & Rand. Heter.  & X-Heter. \\
\midrule
\multicolumn{8}{l}{\textbf{Panel A: Allowing Price Function Shift}} \\
CF2 & RMSE & 0.0366 & 0.0364 & 0.0365 & \textbf{0.0282} & \textbf{0.0281} & \textbf{0.0282} \\
& MAPE & 54.04 & 53.69 & 52.96 & \textbf{43.40} & \textbf{43.14} & \textbf{45.17} \\
T-GDID & RMSE & \textbf{0.0347} & \textbf{0.0346} & \textbf{0.0347} & 0.0287 & 0.0287 & 0.0286 \\
& MAPE & \textbf{53.37} & \textbf{53.30} & \textbf{52.87} & 44.64 & 44.65 & 46.68 \\
T-GDIDP & RMSE & 0.0442 & 0.0442 & 0.0441 & 0.0354 & 0.0353 & 0.0353 \\
& MAPE & 66.76 & 66.71 & 65.70 & 54.49 & 54.49 & 56.98 \\
\midrule
\multicolumn{8}{l}{\textbf{Panel B: Benchmark Models - No Price Function Shift}} \\
BDID & RMSE & 0.0413 & 0.0413 & 0.0412 & 0.0338 & 0.0338 & 0.0337 \\
& MAPE & 61.91 & 61.77 & 61.48 & 52.07 & 52.02 & 54.45 \\
\bottomrule
\end{tabular}
}
\begin{flushleft}
{\footnotesize
\noindent \textit{Note}: Errors are measured against the assigned average direct unmediated price effect on treated (or average DUET). Monte Carlo simulation results are based on 500 iterations for each scenario-method combination. Results for more estimators are available in Appendix Table A.3. To ensure fair comparisons, all methods across all scenarios use identical data generation processes, with treatment randomly assigned in each iteration using seeds equal to the loop counter. In the homogeneous treatment effect scenario, effect size is set to 5 percent on prices of treated transactions. In the heterogeneous effect scenarios, effect size is set so that the average effect is approximately 5 percent. Bold text denotes the top-performing approaches for each scenario.
}
\end{flushleft}
\end{table}

Therefore, the results suggest when treated observations (i.e., post-treatment transactions in the treatment group) number at least 3,000,\footnote{This threshold is estimated from our upstate New York simulation (3~km treatment radius, 5\% treatment effect, random siting) and should be treated as context-specific rather than universal. Performance thresholds are likely to vary with market density, transaction volume, treatment radius, and effect size. The directional pattern, however, should be consistent: as treated sample size grows, flexible machine learning models increasingly outperform methods that rely on stronger parametric assumptions.} CF2 seems to produce lower RMSE and MAPE compared to T-GDID. To further investigate whether the performance difference in larger samples is statistically significant, we use bootstrapped error distributions to conduct pair-wise statistical tests across methods as described in Appendix C. The difference between the better performing models (T-GDID and CF2) and the other traditional models (BDID and T-GDIDP) are statistically significant at 5\% level across most scenarios, and this advantage generally increases with sample size. With the current specifications, however, CF2 does not significantly outperform T-GDID, as shown in Appendix Tables C.1--C.4, the last of which adds 500 further iterations and yields similar results for these comparisons.
 
With these initial lessons in mind, we devote the following subsections to investigate specification changes that may help further improve the estimation accuracy of the two top-performing models in Table~\ref{tab:baseline}, CF2 and T-GDID.

\subsection{Adding Geographic Attributes}
 
The specifications above deliberately omit geographic attributes to enable a direct test of whether their inclusion enhances estimation precision. Theoretically, a rich set of geographic controls should reduce the influence of unobserved factors, addressing potential endogeneity and reducing estimation bias.  Empirically, however, the question is more complicated, as traditional linear regression models may struggle to exploit the additional information in geographic controls, which likely involve nonlinearities and complicated interactions. We formally test the impact of adding geographic attributes\footnote{This set of geographic attributes includes the natural log of nearest distance to transmission lines, primary roads, airports, railroads, water bodies, conservation areas, and whether the property is in a metropolitan area.} on estimation accuracy across the better-performing methods above.

The results are presented in Table~\ref{tab:geo}. We observe that, when the number of treated is at least 3,000, adding geographic attributes has different impacts on different methods. CF2 can utilize these geographic attributes well, performing considerably better across all scenarios. It is worth noting that the performance improvement in CF2 depends on the correct inclusion strategy: We can only incorporate these spatial attributes into the price models rather than the propensity score model or the causal forest algorithm. Because these proximity-based spatial attributes are highly correlated with the proximity-based treatment, including them elsewhere would destabilize propensity score estimation and prevent the causal forest from effectively separating confounding from treatment effect heterogeneity \citep{athey2019}.\footnote{If these geographic features are added to the causal forest algorithm, the forest over-splits on them, creating treatment effect estimates that vary heavily by geography (as shown by very high variable importance scores in certain geographic attributes). Specifically, different geographic areas have different treatment intensities, and spatial autocorrelation creates conditional price differences. The algorithm attributes these differences to heterogeneous effects rather than confounding even after residualization (i.e., overfitting to residual spatial patterns), and will heavily use these geographical features to create very pure treatment/control leaves. \citet{athey2019} suggests: When some covariates are very strong predictors of the treatment, including them in all steps of causal forests can lead to bias. In such cases, they recommend the two-step orthogonalized procedure (i.e., residualizing outcome and treatment, then using residuals in the final causal forest), excluding these strong predictor features from the second step and the propensity model.}

\begin{table}[H]
\centering
\caption{Error Measures with Geographic Attributes}
\label{tab:geo}
\small
\resizebox{\textwidth}{!}{%
\renewcommand{\arraystretch}{0.80}
\begin{tabular}{@{}llcccccc@{}}
\toprule
& & \multicolumn{3}{c}{\textbf{With Geographic Attributes}} & \multicolumn{3}{c}{\textbf{Without Geographic Attributes}} \\
\cmidrule(lr){3-5} \cmidrule(lr){6-8}
& & \textbf{Scenario 1} & \textbf{Scenario 2} & \textbf{Scenario 3} & \textbf{Scenario 1} & \textbf{Scenario 2} & \textbf{Scenario 3} \\
& & No. Treated & No. Treated & No. Treated & No. Treated & No. Treated & No. Treated \\
& & $\geq$3k & $\geq$3k & $\geq$3k & $\geq$3k & $\geq$3k & $\geq$3k \\
\midrule
CF2 & RMSE & \textbf{0.0246} & \textbf{0.0247} & \textbf{0.0246} & 0.0282 & 0.0281 & 0.0282 \\
& MAPE & \textbf{38.95} & \textbf{39.09} & \textbf{40.47} & 43.40 & 43.14 & 45.17 \\
T-GDID & RMSE & 0.0279 & 0.0279 & 0.0278 & 0.0287 & 0.0287 & 0.0286 \\
& MAPE & 43.89 & 43.88 & 45.58 & 44.64 & 44.65 & 46.68 \\
\bottomrule
\end{tabular}
}
\begin{flushleft}
{\footnotesize
\noindent \textit{Note}: Errors are measured against the assigned average direct unmediated price effect on treated (or average DUET). Monte Carlo simulation results are based on 500 iterations for each scenario-method combination. To ensure fair comparisons, all methods across all scenarios use identical data generation processes. Bold text denotes the top-performing approaches for each scenario.
}
\end{flushleft}
\end{table}

In contrast, these geographic attributes, entering in their natural logarithm forms, do not really change the performance of T-GDID (i.e., a slight improvement but the change is not way statistically significant).  This result seems to suggest that adding these geographic attributes does not provide enough additional information for the linear regression model to perform better in estimation accuracy, while the same action considerably boosts the performance of CF2.

\subsection{Control Group Radius}

Choosing control observations that are closer to the treatment observations will generally increase sample balance, which should benefit traditional estimation methods (analogous to matching or boundary strategies). However, at the same time, this approach drops a big proportion of the control observations, potentially limiting information used to train machine learning models. We investigate limiting the control group radius to 5~km instead of 10~km, and assess the estimation performance of models that have shown advantages across different scenarios so far (i.e., CF2 and T-GDID as a baseline).

Results are presented in Appendix Table D.2, indicating that specifying a control group that is geographically closer only slightly improves the performance of T-GDID and CF2. Although utilizing control groups that are geographically closer to the treated is a balance improving best practice in the literature \citep{pope2008, muehlenbachs2015, bishop2020, hu2025}, the observed performance improvement in T-GDID in Appendix Table D.2 is close to zero and statistically insignificant. Moreover, for CF2, the performance change is also approximately zero, suggesting that a tighter control group proximity criterion offers no gain for CF2, potentially because it restricts the information to train the machine learning based price models. Since the results thus far offer no definitive guidance on optimal control radius selection, we conduct a more comprehensive investigation in Section~\ref{sec:efficiency}.

\subsection{Time Frame}

In this subsection, we show how time frame selection affects the performance of CF2 and T-GDID. In practice, hedonic DID studies typically select a time window close to the treatment event while retaining enough pre-treatment periods for pre-trend testing and dynamic effect analysis. The exact choice of time frame, however, remains largely ad hoc -- a gap simulation work like ours can help inform. For our 2010 treatment, common practice would suggest dropping transactions before 2000. Our earlier analyses deliberately use the full 1990–2024 window so that we can directly test whether such a restriction improves estimation accuracy.

The results are presented in Table~\ref{tab:C1}. Unsurprisingly, excluding older transactions (ten years or more prior to the treatment) considerably improves T-GDID performance. However, the same practice only slightly improves the performance of CF2. These results, in combination with control-radius results in Appendix Table D.2, seem to suggest that T-GDID, with a set of rigid modeling assumptions, needs more data pre-processing to improve treatment-control balance. In contrast, the causal machine learning approach, with its flexible algorithm, benefits from additional data and is more robust to the specification of the pre-treatment period. 

\begin{table}[H]
\centering
\caption{Error Measures with a Shorter Pre-period}
\small
\setlength{\tabcolsep}{3pt}
\renewcommand{\arraystretch}{0.85}
\begin{tabular}{@{}llcccccc@{}}
\toprule
& & \multicolumn{3}{c}{\textbf{3--10~km Control}} & \multicolumn{3}{c}{\textbf{3--5~km Control}} \\
\cmidrule(lr){3-5} \cmidrule(lr){6-8}
& & \textbf{Scenario 1} & \textbf{Scenario 2} & \textbf{Scenario 3} & \textbf{Scenario 1} & \textbf{Scenario 2} & \textbf{Scenario 3} \\
& & No. Treated & No. Treated & No. Treated & No. Treated & No. Treated & No. Treated \\
& & $\geq$3k & $\geq$3k & $\geq$3k & $\geq$3k & $\geq$3k & $\geq$3k \\
\midrule
\multicolumn{8}{l}{\textbf{Panel A: Shorter pre-period from 2000 to 2009}} \\
CF2 & RMSE & \textbf{0.0234} & \textbf{0.0234} & \textbf{0.0236} & \textbf{0.0238} & \textbf{0.0242} & \textbf{0.0242} \\
& MAPE & \textbf{37.27} & \textbf{37.31} & \textbf{39.05} & \textbf{37.51} & \textbf{37.97} & \textbf{39.33} \\
T-GDID & RMSE & 0.0249 & 0.0249 & 0.0249 & 0.0253 & 0.0253 & 0.0255 \\
& MAPE & 38.80 & 38.83 & 39.99 & 38.14 & 38.17 & 40.00 \\
\midrule
\multicolumn{8}{l}{\textbf{Panel B: Benchmark - pre-period from 1990 to 2009}} \\
CF2 & RMSE & 0.0246 & 0.0247 & 0.0246 & 0.0244 & 0.0244 & 0.0245 \\
& MAPE & 38.95 & 39.09 & 40.47 & 38.19 & 37.98 & 38.92 \\
T-GDID & RMSE & 0.0279 & 0.0279 & 0.0278 & 0.0279 & 0.0279 & 0.0280 \\
& MAPE & 43.89 & 43.88 & 45.58 & 43.26 & 43.31 & 45.60 \\
\bottomrule
\end{tabular}
\label{tab:C1}
\end{table}

{\footnotesize
\noindent \textit{Note}: Errors are measured against the assigned average direct unmediated price effect on treated or average DUET. Monte Carlo simulation results are based on 500 iterations for each scenario-method combination. To ensure fair comparisons, all methods across all scenarios use identical data generation processes, with treatment randomly assigned in each iteration using seeds equal to the loop counter. Geographic attributes are included in all specifications and modeled consistently with the approach presented in Table~\ref{tab:geo}. Bold text denotes the top-performing approaches for each scenario.
}

\subsection{Location and Year Encoding}
\label{sec:fixedeffects}

The regression estimators absorb location-by-time price variation through zip-code-by-year fixed effects (Section~\ref{sec:estimationapproaches}), which raises the question of whether CF2 should mirror this specification. Our main CF2 specification instead enters zip code and sale year as two ordered numerical features, and a forest has no direct analogue of a fixed effect: an indicator variable is simply one more candidate on which a tree may split. To test how the encoding matters, we re-estimate the preferred CF2 specifications (3--10~km control, Scenario 1) under two alternative encodings, each entering all three forests (the propensity, outcome, and causal forests). The first replaces zip code and sale year with separate sets of binary indicators (i.e., analog to zip and year fixed effects). The second uses one binary indicator for each zip-code-by-year cell (i.e., the zip-by-year fixed effects of T-GDID). 

The results are reported in Appendix Table D.4. Separate zip and year one-hot-encoded indicators slightly increase the error measures of CF2 across the board. With the short pre-period (i.e., the best performing specification), RMSE differs from the main specification by less than 0.0001 across all iterations and by 0.0006 in the large-sample regime (No.\ Treated $\geq$ 3k), the only difference that is statistically significant. Zip-by-year indicators, by contrast, substantially degrade CF2. In the large-sample regime, RMSE rises from 0.0241 to 0.0355 with the short pre-period and from 0.0248 to 0.0393 with the long pre-period (both increases are significant under paired bootstrap tests), leaving CF2 well behind T-GDID (RMSE of 0.0258 on the same iterations, Table~\ref{tab:efficiency}). At the same time, the reported standard errors shrink, and the coverage rate of the 95\% confidence interval falls to 67\% and 60\%, respectively.

The mechanism lies in how the forest selects splitting variables. At each split, the generalized random forest evaluates only a random subset of the $p$ features, whose size is drawn from a Poisson distribution with a mean of $\min\{\sqrt{p}+20,\ p\}$ \citep{athey2019}. With separate zip and year indicators, the causal forest sees roughly 120 total features, and about four of the 17 structural and timing features enter each candidate subset. With zip-by-year indicators, the feature count rises to about 850, about 98\% of which are binary indicators, so each candidate subset contains on average only one informative feature. Each positive zip-by-year indicator represents only a limited number of repeat-sale pairs, so the indicators themselves are rarely used (they account for about 1\% of the causal forest's variable importance). On the other hand, they crowd the informative features out of the candidate subsets and the forest loses much of its ability to choose splits that fit the data. Therefore, fixed-effect specifications cannot be carried over mechanically from regression to forest. A low-dimensional encoding of location and time preserves CF2's performance, whereas high-cardinality interaction fixed effects dominate the feature space and render it largely uninformative.

\subsection{Interval Width, CI Coverage Rate, and More Iterations}
\label{sec:efficiency}

To resolve the open questions on sample time frame and control radius, this section makes three changes to the design. First, in addition to RMSE and MAPE, Table~\ref{tab:efficiency} reports the interval width (``Ave.\ SE'') and empirical coverage rate (``True eff.\ in CI'') defined in Section~\ref{sec:simulationdesign} for the leading T-GDID and CF2 specifications. Because a standard error that understates sampling variability yields narrow intervals that fail to cover the true DUET, we judge precision by RMSE and read interval width together with coverage. Second, we add a 3--4~km control group to examine whether an even narrower control bin improves performance. Finally, we increase the number of simulation iterations to 2000 to enhance the reliability of the results.

\begin{table}[H]
\centering
\caption{Interval Width and CI Coverage Rate for Best-performing Models}
\label{tab:efficiency}
\small
\resizebox{\textwidth}{!}{%
\renewcommand{\arraystretch}{0.70}
\begin{tabular}{@{}llccc@{}}
\toprule
& & \textbf{Scenario 1} & \textbf{Scenario 2} & \textbf{Scenario 3} \\
& & No. Treated $\geq$3k & No. Treated $\geq$3k & No. Treated $\geq$3k \\
\midrule
\multicolumn{5}{l}{\textbf{Panel A: T-GDID with Short Pre-period (from 2000 to 2009)}} \\
T-GDID 3--4~km Control & RMSE & 0.0422 & 0.0421 & 0.0281 \\
& MAPE & 45.268 & 45.326 & 45.068 \\
& Ave. SE & 1.6155 & 12.8687 & 0.0238 \\
& True eff. in CI (\%) & 88.4\% & 88.1\% & 90.5\% \\
T-GDID 3--5~km Control & RMSE & 0.0262 & 0.0262 & 0.0261 \\
& MAPE & 40.407 & 40.386 & 42.000 \\
& Ave. SE & 0.0200 & 0.0201 & 0.0216 \\
& True eff. in CI (\%) & 84.4\% & 84.2\% & 90.3\% \\
T-GDID 3--10~km Control & RMSE & 0.0258 & 0.0258 & 0.0257 \\
& MAPE & 39.329 & 39.325 & 40.733 \\
& Ave. SE & 0.0169 & 0.0171 & 0.0188 \\
& True eff. in CI (\%) & 81.7\% & 81.7\% & 87.0\% \\
\midrule
\multicolumn{5}{l}{\textbf{Panel B: CF2 with Short Pre-period (from 2000 to 2009)}} \\
CF2 3--4~km Control & RMSE & 0.0264 & 0.0263 & 0.0265 \\
& MAPE & 41.516 & 41.381 & 43.354 \\
& Ave. SE & 0.0245 & 0.0246 & 0.0243 \\
& True eff. in CI (\%) & 94.1\% & 95.0\% & 94.7\% \\
CF2 3--5~km Control & RMSE & 0.0244 & 0.0245 & 0.0243 \\
& MAPE & 38.039 & 38.384 & 39.643 \\
& Ave. SE & 0.0216 & 0.0217 & 0.0214 \\
& True eff. in CI (\%) & 92.9\% & 93.2\% & 93.1\% \\
CF2 3--10~km Control & RMSE & \textbf{0.0241} & \textbf{0.0241} & \textbf{0.0242} \\
& MAPE & \textbf{37.021} & \textbf{37.271} & \textbf{38.483} \\
& Ave. SE & 0.0196 & 0.0197 & 0.0195 \\
& True eff. in CI (\%) & 90.1\% & 91.0\% & 90.3\% \\
\midrule
\multicolumn{5}{l}{\textbf{Panel C: CF2 with Long Pre-period (from 1990 to 2009)}} \\
CF2 3--4~km Control & RMSE & 0.0265 & 0.0267 & 0.0267 \\
& MAPE & 41.004 & 41.223 & 42.919 \\
& Ave. SE & 0.0225 & 0.0226 & 0.0223 \\
& True eff. in CI (\%) & 89.8\% & 89.6\% & 90.5\% \\
CF2 3--5~km Control & RMSE & 0.0253 & 0.0254 & 0.0253 \\
& MAPE & 38.936 & 39.273 & 40.568 \\
& Ave. SE & 0.0198 & 0.0199 & 0.0197 \\
& True eff. in CI (\%) & 88.4\% & 87.8\% & 88.8\% \\
CF2 3--10~km Control & RMSE & 0.0248 & 0.0249 & 0.0249 \\
& MAPE & 38.686 & 38.776 & 39.927 \\
& Ave. SE & 0.0179 & 0.0180 & 0.0178 \\
& True eff. in CI (\%) & 86.5\% & 85.6\% & 86.3\% \\
\bottomrule
\end{tabular}
}
\end{table}
\begin{flushleft}
{\footnotesize
\noindent \textit{Note}: Errors are measured against the assigned average direct unmediated price effect on treated (or average DUET). Results are based on 2000 Monte Carlo iterations for each scenario-method combination, restricted to iterations with No.\ Treated $\geq$3k. To ensure fair comparisons, all methods across all scenarios use identical data generation processes, with treatment randomly assigned in each iteration using seeds equal to the loop counter. Ave.\ SE measures the width of the confidence interval and is calculated as the average of the per-iteration standard errors. True eff.\ in CI (\%) is the empirical coverage rate of the 95\% confidence interval, calculated as the proportion of iterations in which the iteration's CI contains the true DUET value. The large Ave.\ SE of T-GDID with a 3--4~km control in Scenarios 1 and 2 is driven by a few near-degenerate iterations. Geographic attributes are included in all specifications and modeled consistently with the approach presented in Table~\ref{tab:geo}. Bold text indicates the top-performing approaches overall.
}
\end{flushleft}

The results show that CF2 with a shorter pre-period and a larger control radius (10~km) is the best performing estimator overall, achieving the lowest error measures and high CI coverage rates. For T-GDID, the 10~km control radius (with a 2000–2009 pre-period) slightly outperforms the 5~km specification in error metrics and reports smaller standard errors, although its CI coverage rate is lower. The 4~km specification in T-GDID exhibits the worst accuracy and severe instability — with inflated standard errors in Scenarios 1 and 2 — suggesting that an overly narrow spatial bin undermines reliable inference. Similarly, while CF2 with a 4~km control radius maintains stable standard errors, it consistently underperforms the 5~km and 10~km specifications in both RMSE and MAPE, confirming that restricting the control sample to a very narrow proximity bin does not improve estimation performance for either method. When sample sizes are sufficiently large,\footnote{We have also investigated estimator performance in small samples among the 2000 simulation iterations, as shown in Appendix Table B.2. The results suggest that CF2 becomes increasingly unlikely to outperform T-GDID as sample size shrinks.} CF2's performance advantage in accuracy over T-GDID (10~km control) becomes statistically significant, as shown by bootstrap tests in Appendix Table C.5: in samples with at least 3,000 treated observations, both RMSE and MAPE are significantly lower at the .05 significance level. Compared with the best performing CF2 specification, CF2 with the most extensive sample (i.e., a long pre-period and a 10~km control radius) achieves smaller standard errors but has slightly larger error measures and lower CI coverage rates. More generally, the undercoverage of both estimators stems mainly from standard errors that are too small rather than from bias. For T-GDID, two-way clustering on zip code and year understates sampling variability because treatment is assigned through a limited number of spatial sites. Appendix Table D.5 examines this directly by recomputing the inference for the 3--10~km T-GDID specification on the same 2,000 iterations. Clustering two ways on treatment site and year raises coverage only modestly (to 85.1\% with No.\ Treated $\geq 3{,}000$), because only about 18 sites per iteration contribute observations and analytic cluster-robust standard errors remain too small with so few clusters. A wild cluster restricted bootstrap at the treatment-site level \citep{cameron2008}, the level at which treatment is assigned \citep{abadie2023}, restores nominal coverage (95.0\%), but its intervals are wide. Its implied standard error averages 0.037 (median 0.032), compared with a standard deviation of the T-GDID estimates of about 0.025 across iterations. Accurate inference for T-GDID is therefore attainable, but at a considerable cost in interval width. Design-based inference for spatial treatments \citep{pollmann2026} offers a further alternative. The coverage rates in Table~\ref{tab:efficiency} should thus be read as the performance of the standard inference procedures that practitioners most commonly use. 

To show that these results are not specific to the 5\% average DUET used in the main simulation, we conduct a robustness check where the average DUET targets -5\%. As shown in Appendix Table D.3, the result patterns are largely identical to those in Table~\ref{tab:efficiency}: the relative advantages of different specifications remain unchanged under the alternative average DUET. 

\subsection{Simulation with Treatment-Induced Equilibrium Shift}
\label{sec:indshift}

The main simulation preserves the natural pre-post equilibrium shift in the data but does not impose a \textit{treatment-induced} shift in the hedonic price function ($IE_i = 0$, Section~\ref{sec:assumptions}). A natural question is whether our conclusions, in particular the relative standing of CF2 and T-GDID, survive once the treatment itself moves the price function, as it may in some empirical settings. The key conceptual point, derived formally in Appendix E, is that within the generalized DID family the \textit{source} of a price-function shift does not affect estimation: natural and induced shifts both impact every estimator through its ability to approximate the shifted post-period price function. What matters is therefore how \textit{complex} the shift is relative to each estimator's flexibility. A linear-in-covariate shift is absorbed exactly by T-GDID at any magnitude, whereas a shift with nonlinearities and interactions that a linear specification cannot represent is where the flexible CF2 price model should gain ground. This yields a testable prediction. A linear shift should leave T-GDID unchanged at any intensity, while CF2 should pay a minor cost that grows with intensity (i.e., have to approximate the shift). In contrast, a nonlinear shift should considerably widen CF2's advantage as its intensity grows. Because we cannot vary the natural shift, we test this prediction by simulating induced shifts of both forms at several intensities.

In this section, we augment the Scenario 1 data-generating process with a treatment-induced shift applied to \textit{all} post-period transactions in the estimation market (treated and control alike), on top of the direct effect on post-period treated units. The shift is imposed additively in log price (i.e., directly modifying the log-price function) and scaled so that its cross-sectional standard deviation equals a fixed fraction (the ``shift intensity'' $\kappa$) of the post-period log-price standard deviation. $\kappa$ is swept from zero to 0.9 in this simulation. As any level change in average price will be absorbed by the post variable or time fixed effects, we center the mean induced shift at zero so the average post-period price level is unchanged. We consider two forms of the shift: a \textit{linear} form, a randomly weighted combination of four covariates drawn at random from those T-GDID includes, and a \textit{nonlinear} form built from complex interactions between covariates and level indicators that lie outside the linear span (both redrawn in each iteration, see Appendix E). Using four covariates matches the number of variables entering the nonlinear form, so the two forms differ in functional form rather than in dimensionality. Because DUET is defined along the post-treatment price function and imposed only on post-treated units, imposing this shift on both treated and control units leaves the ground-truth DUET benchmark unchanged. Appendix E details the design, the formal justification for why GDID accommodates such shifts, and the expected pattern of results.

The results are reported in Table \ref{tab:indshift}, restricted to the large-sample regime (No.\ Treated $\geq 3{,}000$) in which the main text establishes CF2's edge. The full-sample counterparts, along with the MAPE and coverage measures, appear in Appendix Table E.1. The baseline column ($\kappa = 0$, no induced shift) is the preferred 3--10~km short-pre-period specification from Table~\ref{tab:efficiency}, and every column uses the same 1,107 iterations. The pattern is as predicted. Under the linear shift (Panel A), T-GDID's estimates are numerically identical to those without the shift at every intensity, confirming that its price model absorbs the shift exactly. CF2's RMSE slightly rises from 0.0241 at baseline to 0.0245, 0.0268, and 0.0288 as $\kappa$ increases from 0.1 to 0.9. A forest approximates a linear function with step functions, so a large linear shift adds estimation error that a correctly specified linear model avoids. CF2's edge therefore reverses at higher intensities, with the CF2-minus-T-GDID gap moving from $-0.0016$ to $+0.0030$, but this performance difference is minor compared to the nonlinear results below. The nonlinear shift (Panel B) tells the opposite story. CF2 already edges T-GDID at $\kappa = 0$, and its advantage widens steadily as the nonlinear shift intensifies. T-GDID's RMSE more than triples, from $0.026$ at baseline to $0.085$ at $\kappa = 0.9$, whereas CF2's rises only modestly, from $0.024$ to $0.033$, so the CF2-minus-T-GDID RMSE gap falls monotonically from $-0.002$ to $-0.052$. Comparing the two panels, CF2's error grows by a similar amount under both forms (to 0.029 and 0.033 at $\kappa = 0.9$), whereas T-GDID's error depends entirely on the form, unchanged under the linear shift and more than tripled under the nonlinear one. The relative standing of the two estimators therefore hinges on whether the shift lies within the span of the linear price model. T-GDID is exposed to large approximation error when it does not, while CF2 pays a moderate cost in both cases.

\begin{table}[H]
\centering
\caption{Estimator Performance under a Treatment-Induced Equilibrium Shift (3--10~km Control, No. Treated $\geq$3k, Scenario 1)}
\label{tab:indshift}
\small
\begin{tabular}{@{}llcccc@{}}
\toprule
& & \multicolumn{4}{c}{Shift intensity $\kappa$ (share of post $\ln P$ SD)} \\
\cmidrule(l){3-6}
Shift form & Metric & $0$ (baseline) & $0.1$ & $0.5$ & $0.9$ \\
\midrule
\multicolumn{6}{l}{\textbf{Panel A: Linear shift (placebo; in T-GDID's span)}} \\
T-GDID (3--10~km)      & RMSE                 & 0.0258 & 0.0258 & 0.0258 & 0.0258 \\
CF2 (short pre-period) & RMSE                 & 0.0241 & 0.0245 & 0.0268 & 0.0288 \\
CF2 $-$ T-GDID         & $\Delta$RMSE         & $-0.0016$ & $-0.0012$ & $+0.0010$ & $+0.0030$ \\
\midrule
\multicolumn{6}{l}{\textbf{Panel B: Nonlinear shift (interactions / level indicators)}} \\
T-GDID (3--10~km)      & RMSE                 & 0.0258 & 0.0273 & 0.0517 & 0.0847 \\
CF2 (short pre-period) & RMSE                 & 0.0241 & 0.0246 & 0.0295 & 0.0330 \\
CF2 $-$ T-GDID         & $\Delta$RMSE         & $-0.0016$ & $-0.0027$ & $-0.0222$ & $-0.0517$ \\
\bottomrule
\end{tabular}
\end{table}
\begin{flushleft}
{\footnotesize
\noindent \textit{Note}: Each column corresponds to a treatment-induced shift whose cross-sectional standard deviation equals $\kappa$ times the post-period log-price standard deviation, applied to all post-period transactions on top of the 5\% direct effect on treated units. Panel A imposes a linear shift in a random subset of four covariates that T-GDID includes (absorbed exactly by T-GDID), and Panel B imposes a nonlinear shift outside the linear span. Both shapes are redrawn in each iteration. The row ``CF2 $-$ T-GDID'' reports the difference in RMSE (negative favors CF2). All entries are restricted to iterations with No.\ Treated $\geq 3{,}000$, which leaves 1,107 of 2,000 iterations in every column. The baseline column ($\kappa = 0$, no shift) is the main specification (Table~\ref{tab:efficiency}), computed on the same iterations, and is common to both panels. Full-sample results for all iterations and the corresponding MAPE and CI coverage are reported in Appendix Table E.1.
}
\end{flushleft}

\section{Discussion}
\label{sec:discussion}

\subsection{Best Practice Recommendations based on Simulation Results}

Table \ref{tab:bestpractices} summarizes the best practice recommendations that follow from our simulations, listed separately for regression-based and causal machine learning approaches because the two require different specifications. Among traditional OLS-based methods, T-GDID performs best in problems similar to ours, and its accuracy benefits from a shorter sample time frame and, to a lesser extent, from additional geographic attributes. When the sample is sufficiently large (at least 3,000 treated), CF2 outperforms the traditional methods, but only under specification choices that diverge from standard regression practice. Location and time must enter in a low-dimensional form (as continuous variables, or at most as separate zip and year indicators) rather than as zip-code-by-year fixed effects, whose hundreds of sparse indicators crowd informative features out of the forest's split candidates (Section~\ref{sec:fixedeffects}).\footnote{Within upstate New York, numerically similar zip codes typically correspond to geographic neighbors, particularly within metro areas. By leveraging the spatial structure of the region's zip code system, the causal forest can efficiently partition the data into geographically local neighborhoods, consistent with the adaptive matching logic of the Generalized Random Forest framework \citep{athey2019}.} Spatial attributes that are highly correlated with the treatment belong in the price models only, not in the propensity score model or the causal forest \citep{athey2019}. And simple sample restrictions that tries to improve treatment-control balance offer little to CF2, because the discarded observations help train the price model while the causal forest's adaptive matching already handles balance.

\begin{table}[H]
\centering
\caption{Method-Specific Best Practice Recommendations for DUET Estimation}
\label{tab:bestpractices}
\singlespacing
\small
\setlength{\tabcolsep}{4pt}
\renewcommand{\arraystretch}{1.1}
\begin{tabular}{@{}>{\raggedright\arraybackslash}p{2.9cm}>{\raggedright\arraybackslash}p{5.9cm}>{\raggedright\arraybackslash}p{6.9cm}@{}}
\toprule
\textbf{Specification} & \textbf{Traditional GDID} & \textbf{Causal Machine Learning GDID} \\
\midrule
\textbf{Best Estimators} & 
T-GDID & 
CF2 \\
\addlinespace[2pt]
\textbf{Recommended Sample Size} & 
Suitable for \textit{smaller samples} (e.g., $<$3,000 treated obs.) & 
Performs best in \textit{large samples} (e.g., when treated obs. $\geq$3,000) \\
\addlinespace[2pt]
\textbf{Time Frame} & 
\textit{Shorter pre-period preferred:} Excluding older transactions (e.g., 10 years or older) improves performance. & 
\textit{Relatively flexible with longer pre-periods:} A shorter pre-period (e.g., 2000--2009) still gives the best accuracy, but CF2 is less sensitive to a longer pre-period (e.g., 1990-2009) than T-GDID. \\
\addlinespace[2pt]
\textbf{Control Group Radius} & 
\textit{Avoid a too-narrow bin:} 3--10~km performs similarly to 3--5~km in accuracy, with smaller standard errors but somewhat lower CI coverage; 3--4~km considerably worse. & 
\textit{Use broader control radius:} 3--10~km preferred to 3--5~km, with slightly better accuracy and narrower intervals at a small cost in CI coverage. \\
\addlinespace[2pt]
\textbf{Geographic Attributes} & 
\textit{Minimal impact:} Adding geographic controls provides minimal performance gains when working on a well-balanced sample. & 
\textit{Major benefit when correctly specified:} Include geographic attr. in price models only (not in propensity model or causal forest). Considerably improves performance. \\
\addlinespace[2pt]
\textbf{Variable Encoding} & 
Standard categorical encoding, e.g., one-hot encoding for fixed effects. & 
\textit{Keep location and time low-dimensional:} Enter zip code and year as numerical variables (or as separate zip and year dummies). Zip-by-year dummies crowd out informative features and degrade accuracy and CI coverage. \\
\addlinespace[2pt]
\textbf{Spatial Fixed Effects} & 
\textit{Slightly coarser geographic fixed effects} outperform parcel-level fixed effects. & 
\textit{No fixed effects needed.} \\
\addlinespace[2pt]
\textbf{Price Function complexity} & 
Price function shifts can be examined with control groups, and T-GDID performs well \textit{when the shift is high-intensity and linear}. & 
\textit{Performs better when the price function shift is highly complex} \\ 
\bottomrule
\end{tabular}
\begin{flushleft}
{\footnotesize
\noindent \textit{Note}: Recommendations based on Monte Carlo simulations in this study. T-GDID refers to traditional generalized difference-in-differences with zip-code-by-year fixed effects. CF2 refers to causal forest DID on adjacent transaction pairs. The 3,000 treated observation threshold reflects upstate New York simulation conditions in this study.
}
\end{flushleft}
\end{table}

\subsection{Caveats}

Several important caveats warrant consideration when interpreting our findings.

First, our simulation focuses on average treatment effect estimation for proximity-based treatments, conditional on knowing the spatial extent of treatment influence (i.e., the 3~km radius in our design). Two related questions remain beyond the scope of this analysis: whether these methods can accurately identify the spatial boundaries of treatment effects, and whether they can reliably detect heterogeneous treatment effects across time, property types, and neighborhood profiles. These questions require fundamentally different evaluation criteria that focus on conditional average treatment effect (CATE) estimation rather than average treatment effect on the treated (ATT). These questions could be investigated with modified but feasible empirical Monte Carlo simulations and represent important directions for future research.

Second, while our empirical simulation preserves real data-generating processes rather than relying on parametric assumptions, the housing market structure, transaction patterns, and spatial characteristics reflect upstate New York specifically. Given the general similarity in how housing markets function across the United States, we expect our core findings, particularly the importance of allowing price function shifts and the relative advantages of different methods at varying sample sizes, to generalize to other U.S.\ contexts. However, performance-related thresholds are likely context-dependent. For example, the 3,000 treated observation threshold at which CF2 begins to substantially outperform T-GDID may be different in other contexts. Researchers should view this threshold as indicative rather than definitive and consider pilot analyses to assess method performance in their specific empirical settings.

Third, we do not evaluate performance when treatment location follows specific targeting rules based on observable community or environmental characteristics. In reality, the siting of many amenities and disamenities, including renewable energy facilities (wind farms, solar installations), industrial projects (factories, warehouses, data centers), or environmental improvements (parks, conservation areas), often reflects deliberate consideration of demographic composition, property values, political factors, or geographic features. Targeted site selection based on $\mathbf{X}$ would amplify spatial-autocorrelation-induced bias, likely hurting traditional methods more than CML, whose flexible price modeling and adaptive matching can partially absorb the added confounding.\footnote{The targeting rules are likely nonlinear and involving complex interactions in $X$, which would result in systematic treatment-control price differences that cannot be effectively modeled by traditional linear models (e.g., BDID or T-GDID). This modeling ineffectiveness will leave more $X$-related confounding factors in the causal effect identification and exacerbate the estimation bias of DUET.} Future research could select a specific project type as a prototype, reconstruct its typical targeting criteria from historical siting decisions, and simulate treatment assignment following those rules to assess whether the generalized DID framework and causal machine learning methods maintain their performance advantages.

Finally, our simulation design likely understate the disadvantages of traditional estimators. First, as mentioned above, random treatment site selection omits $\mathbf{X}$-based site selection, which likely benefits traditional methods more than CML approaches. Second, to best preserve the original data generation process, the main simulations do not include a treatment-induced equilibrium shift. Section~\ref{sec:indshift} addresses this omission directly by adding treatment-induced shifts of increasing intensity. When the induced shift is nonlinear in $\mathbf{X}$, T-GDID's estimation error grows rapidly while CF2's grows only modestly, so CF2's advantage widens with the intensity of the shift. Because real amenity shocks may well reshape the price function in nonlinear ways, the main comparison likely understates CF2's advantage over T-GDID. Therefore, our evaluation likely favors traditional methods relative to CML, compared to real-world settings.

\section{Concluding Remarks}
\label{sec:conclusion}
This study evaluates traditional hedonic regression and causal machine learning approaches for estimating the direct unmediated price effect on treated (DUET) of spatially delineated environmental amenities. Using an empirical Monte Carlo simulation framework that preserves real housing transaction patterns and property characteristics from upstate New York, we systematically compare the accuracy of traditional generalized DID methods and causal-machine-learning-based methods across multiple scenarios involving homogeneous and heterogeneous treatment effects.

Our results demonstrate that allowing price function shifts over time substantially improves estimation accuracy, with generalized DID regression (T-GDID) consistently outperforming baseline DID, post-period OLS, and two-way fixed effects panel models. The poor performance of two-way fixed effects panel models (T-GDIDP) highlights the importance of incorporating cross-sectional variation through coarser spatial fixed effects rather than relying solely on repeated sales with property-level fixed effects in traditional OLS based approaches, a finding consistent with \citet{bishop2020} but not universally adopted in practice.

Causal machine learning methods, particularly causal forest DID on adjacent pairs (CF2), demonstrate comparable performance to T-GDID in general and superior performance when the number of treated observations reaches 3,000 or more. In large samples with proper specification, CF2 achieves RMSE as low as 0.0241 and MAPE of 37.02\%, outperforming the best T-GDID specification (i.e., RMSE = 0.0258 and MAPE = 39.3\%). Importantly, the performance of CF2 depends critically on appropriate specification choices that differ from traditional methods (e.g., proximity-based geographic attributes should be excluded from the causal forest algorithm but included in first-stage price models). While traditional methods benefit from techniques that improve treatment-control balance, such as limiting the temporal window (2000-2024 versus 1990-2024), these techniques do not seem to bring considerable benefits to causal machine learning methods. These differences suggest that the flexible machine learning algorithms can effectively extract relevant information from less relevant controls. While sample selection thresholds often rely on subjective judgment of researchers, causal machine learning and simulation techniques seem to enable data-driven approaches to these decisions.

Beyond point estimate accuracy, we examine the width and coverage rates of 95\% confidence intervals. The best-performing CF2 specification (i.e., 10~km control radius, a time frame of 2000-2024, and geographic attributes in price models) reports moderately larger standard errors than T-GDID, but it has lower estimation error and considerably better empirical coverage rates (90.1--91.0\% versus 81.7--87.0\% for the 3--10~km specifications). Its wider intervals therefore reflect a more accurate account of estimation uncertainty rather than a loss of precision, and CF2 offers more reliable statistical inference of the average treatment effect. T-GDID can attain nominal coverage when inference is clustered at the treatment-site level with a wild cluster bootstrap, but only with intervals nearly twice as wide as CF2's on average (an implied standard error of 0.037 versus 0.0196). 

We also relax the simplification that mutes treatment-induced price function shift. Within the generalized DID family, what matters for estimation is not whether a price-function shift is natural or treatment-induced, but how complex the shift is relative to each estimator's flexibility. Consistent with this prediction, adding a treatment-induced shift that is nonlinear in property characteristics widens CF2's advantage as the shift intensifies. In large samples, T-GDID's RMSE rises from 0.026 without the induced shift to 0.085 at the highest intensity, whereas CF2's rises only from 0.024 to 0.033.

Our findings carry direct implications for practitioners conducting hedonic analyses in support of policy evaluation. Regulatory benefit-cost analyses, compensation decisions, and siting assessments for environmental disamenities all depend on credible estimates of localized price effects, precisely what DUET is designed to measure. Method selection for these applications should be guided primarily by sample size. Researchers working with smaller samples (e.g., fewer than 3,000 treated observations) may benefit from employing T-GDID with careful attention to sample balance through temporal and spatial restrictions, inclusion of comprehensive property and geographic controls, and allowance for price function shifts through separate period-specific coefficients. Researchers with larger samples should seriously consider causal machine learning approaches, while recognizing that optimal specification involves: coding location and time in a low-dimensional form (e.g., numerical zip codes and years rather than zip-code-by-year fixed effects) to facilitate the tree-splitting algorithm, excluding spatial features that are highly correlated with the treatment from the treatment model and the final causal forest algorithm while retaining them in price models, maintaining broad inclusion criteria for control observations, and including a comprehensive set of features. Adopting the appropriate method and specification for a given dataset produces DUET estimates with lower error and more reliable confidence intervals, qualities that directly strengthen the credibility of impact assessments and cost-benefit analyses. 

This study provides initial documentation of the new opportunities for improved causal inference provided by the substantial expansion in hedonic dataset sizes over the past two decades and the extensive advances in causal machine learning. As datasets continue to grow and computational methods advance, the relative advantages of flexible machine learning approaches are likely to become even more pronounced. However, the fundamental importance of allowing price function shifts over time, the key insight from KPP and \citet{banzhaf2021}, remains critical regardless of the estimation technique employed. Future research could extend this simulation analysis in several directions, including heterogeneous treatment effect identification, dynamic effect estimation, performance assessment under targeted treatment assignment, and optimal sample inclusion criteria detection. More broadly, our method-selection framework offers a replicable template for credible hedonic-based policy evaluation across a growing set of applications --- land use regulation, environmental hazard exposure, energy infrastructure and AI data center siting, and beyond.

\section*{Declaration of generative AI and AI-assisted technologies in the manuscript preparation process}
During the preparation of this work the authors used Claude (Anthropic) to assist with revising the text and checking analysis code. After using this tool, the authors reviewed and edited the content as needed and take full responsibility for the content of the published article.

\clearpage
\newpage

\bibliographystyle{apalike}
\bibliography{bibliography}
\end{document}


\maketitle

\appendix

\renewcommand{\thefigure}{\Alph{section}.\arabic{figure}}
\renewcommand{\thetable}{\Alph{section}.\arabic{table}}
\numberwithin{equation}{section}

\clearpage
\newpage

\section{Additional Model Information}

\subsection{Glossary}

\begin{table}[htbp]
    \centering
    \caption{Glossary of Acronyms and Technical Terms}
    \label{tab:glossary}
    \small
    \begin{tabularx}{\textwidth}{llX}
        \toprule
        \textbf{Acronym} & \textbf{Full Name} & \textbf{Description} \\
        \midrule
        \multicolumn{3}{l}{\textit{Model Estimators}} \\
        OLS      & Ordinary Least Squares & Benchmark cross-sectional regression using post-treatment data only. \\
        BDID     & Benchmark DID & Traditional DID incorporating pre-treatment data; restricts price function coefficients. \\
        T-GDID   & Trad. Generalized DID & DID model allowing property characteristic coefficients to vary across periods. \\
        T-GDIDP  & Trad. Generalized Panel DID & T-GDID using parcel-level fixed effects, estimated on repeat-sales data. \\
        CF2      & Causal Forest (Adj. Pairs) & CML applied to first-differenced prices of adjacent transaction pairs. \\
        CFRP     & Causal Forest Resid. Prices & CML using OLS residuals of all repeated sales to account for shifts. \\
        CFR2     & Causal Forest Resid. (Adj.) & CML using OLS residuals restricted to adjacent transaction pairs. \\
        \midrule
        \multicolumn{3}{l}{\textit{Technical Terminology}} \\
        AIPW     & Aug. Inverse Prob. Weighting & A doubly robust estimator used to aggregate treatment effects. \\
        CML      & Causal Machine Learning & Estimators identifying treatment effects via algorithms like random forests. \\
        DID      & Difference-in-Differences & Design comparing pre-post changes between treated and control groups. \\
        DUET     & Direct Unmediated Effect on Treated & The target estimand; the direct price \% difference attributable to treatment. \\
        FE       & Fixed Effects & Controls for unobserved factors (e.g., zip-code-by-year or parcel level). \\
        MAPE     & Mean Absolute Percentage Error & Average absolute estimation error relative to the true DUET. \\
        RMSE     & Root Mean Square Error & Square root of the average squared estimation error. \\
        \bottomrule
    \end{tabularx}
\end{table}

\subsection{General Framework and Notation}

All estimators share a common outcome variable and covariate structure. Let $\ln P_{it}$ denote the natural log of the sale price of property $i$ at time $t$. The binary treatment indicator $D_i = 1$ if property $i$ is within 3~km of a treatment site and $D_i = 0$ otherwise. The vector $\mathbf{X}_{it}$ collects property characteristics (e.g., building age, living space, number of bathrooms, and sewer type). The target estimand across all models is the direct unmediated price effect on the treated (DUET), interpreted as the price percentage difference attributable to treatment.

The first category of estimators encompasses traditional ordinary least squares (OLS) regression estimators, differing in their treatment of the time dimension, price function flexibility, and spatial controls. The second category includes causal machine learning (CML) estimators that identify average treatment effects via forests and the Augmented Inverse Probability Weighting (AIPW) approach.

Standard errors for traditional regression estimators are two-way clustered at the zip code and year level. For causal forest estimators, the standard error of the average effect is computed from the doubly robust (AIPW) scores, which combine out-of-bag estimates of the propensity score and of the expected price change. Neither procedure fully reflects the site-level assignment of treatment in our design, a limitation discussed in Section 4.6 of the main text.

\subsection{Traditional Regression Estimators}

The traditional regression estimators follow the general equation:

\begin{equation}
\ln P_{it} = \alpha + \boldsymbol{Treat_i}\boldsymbol{\tau}  + \mathbf{X}_{it}\boldsymbol{\beta}_t + \mu_{zt} + \varepsilon_{it},
\end{equation}
where $ \boldsymbol{Treat_i} $ represents a potential set of variables defining the treatment, $\mu_{zt}$ denotes spatial-by-time fixed effects at certain levels and $\boldsymbol{\beta}_t$ denotes a vector of coefficients on property characteristics that are allowed to vary across time periods. The models below differ in three key dimensions: (1) whether pre-treatment observations are included and over-time first differencing is incorporated, (2) whether $\boldsymbol{\beta}_t$ is restricted to be constant across periods or allowed to vary freely, and (3) whether spatial controls enter as zip-code-by-year fixed effects or parcel-level fixed effects. These differences are summarized in Table \ref{tab:a1}.

\subsubsection{Benchmark OLS (OLS)}

The benchmark OLS estimator is a cross-sectional regression estimated on post-treatment observations only. It imposes $\boldsymbol{\beta}_t = \boldsymbol{\beta}$ (no price function shift) and uses only post-period data:

\begin{equation}
\ln P_{i} = \alpha + \tau D_i + \mathbf{X}_{i}\boldsymbol{\beta} + \mu_{zt} + \varepsilon_{i}, \qquad \forall\ i \in \text{post-period}.
\end{equation}
$D_i$ is the binary treatment group identifier and its coefficient $\tau$ is the estimator of DUET, interpreted as the approximate percentage price difference between treated and control homes in the post-period conditional on observed characteristics. Because OLS uses only post-period data, it cannot difference out time-invariant unobservables correlated with treatment, nor can it account for equilibrium shifts in the price function. It therefore serves as the most restrictive baseline.

\subsubsection{Benchmark DID (BDID)}

The benchmark DID estimator extends OLS by incorporating pre-treatment observations, enabling first differencing to partial out time-invariant unobservables. It still imposes $\boldsymbol{\beta}_t = \boldsymbol{\beta}$, enforcing price function stability across periods:

\begin{equation}
\ln P_{it} = \alpha + \beta_1 D_i + \beta_2 \text{Post}_{t} + \beta_3 \left(D_i \times \text{Post}_{t}\right) + \mathbf{X}_{it}\boldsymbol{\beta} + \mu_{zt} + \varepsilon_{it},
\end{equation}
where $\text{Post}_{t}$ is the binary post-treatment period identifier, $\beta_1$ captures pre-existing price differences between treated and control properties, $\beta_2$ captures the common price trend shared across all properties, and $\beta_3$ is the coefficient of interest, representing the average pre-post price change for treated units relative to controls. While BDID mitigates bias from time-invariant unobservables unlike benchmark OLS, the restriction $\boldsymbol{\beta}_t = \boldsymbol{\beta}$ leaves residual bias in the treatment effect estimate from equilibrium-shift-induced confounding.

\subsubsection{Traditional Generalized DID (T-GDID)}

T-GDID relaxes the restriction on $\boldsymbol{\beta}_t$, allowing the coefficients on property characteristics to vary freely between the pre- and post-treatment periods. This accommodates equilibrium shifts in the hedonic price function as recommended by KPP:

\begin{equation}
\ln P_{it} = \alpha + \beta_1 D_i + \beta_2 \text{Post}_{t} + \beta_3 \left(D_i \times \text{Post}_{t}\right) + \mathbf{X}_{it}\boldsymbol{\beta}_t + \mu_{zt} + \varepsilon_{it},
\end{equation}
where $\boldsymbol{\beta}_t = \boldsymbol{\beta}_0 + \boldsymbol{\beta}_1 \text{Post}_{t}$, so that the price function is estimated separately for the pre- and post-treatment periods---equivalent to including full interactions between $\mathbf{X}_{it}$ and $\text{Post}_t$. By allowing the price function to shift, T-GDID mitigates the equilibrium-shift-induced confounding evident in BDID. The coefficient of interest remains $\beta_3$.

\subsubsection{Traditional Generalized Panel DID (T-GDIDP)}

T-GDIDP is identical to T-GDID in allowing the price function to shift over time, but replaces zip-code-by-year fixed effects with parcel-level fixed effects $\alpha_i$ and year fixed effects $\lambda_t$:

\begin{equation}
\ln P_{it} = \alpha_i + \lambda_t + \beta_1 D_i + \beta_2 \text{Post}_{t} + \beta_3 \left(D_i \times \text{Post}_{t}\right) + \mathbf{X}_{it}\boldsymbol{\beta}_t + \varepsilon_{it},
\end{equation}
where $\boldsymbol{\beta}_t = \boldsymbol{\beta}_0 + \boldsymbol{\beta}_1 \text{Post}_{t}$ and $\beta_3$ is the coefficient of interest. Because identification relies on within-parcel price variation, T-GDIDP is estimated on repeated sales only. The parcel fixed effects absorb all time-invariant property-level unobservables, potentially reducing omitted variable bias relative to T-GDID. However, restricting to repeated sales substantially reduces the effective sample size, which may exacerbate bias induced by outliers. It may also introduce sample selection concerns if the subset of repeat-sale properties is not representative of the broader market, which OLS does not effectively address.

\begin{table}[htbp]
\centering
\caption{Key specification differences across traditional regression estimators.}
\label{tab:a1}
\begin{tabular}{lcccc}
\hline
\textbf{Model} & \textbf{Pre-period} & \textbf{Price function shift} & \textbf{Spatial control} & \textbf{Sample} \\
\hline
OLS     & No  & No  & $\mu_{zt}$           & All post       \\
BDID    & Yes & No  & $\mu_{zt}$           & All            \\
T-GDID  & Yes & Yes & $\mu_{zt}$           & All            \\
T-GDIDP & Yes & Yes & $\alpha_i + \lambda_t$ & Repeated sales \\
\hline
\end{tabular}
\end{table}

\subsection{Causal Machine Learning Estimators}

\subsubsection{Overview of the Two-Stage Logic}
\label{app:twostage}

All causal machine learning estimators follow a two-stage double machine learning framework. In the first stage, separate models partial out the influence of covariates $\mathbf{X}_{it}$ on the outcome and the treatment indicator, producing residualized outcome with mitigated confounding from observables. In the second stage, a causal forest algorithm uses these residualized quantities to estimate heterogeneous treatment effects (conditional on $\mathbf{X}_{it}$), and the average DUET is obtained from an orthogonal moment built on the same residuals. We first describe the AIPW aggregation for the ATT and then the overlap-weighted version used in this study.

Formally, the first stage estimates the outcome and treatment residuals:

\begin{equation}
\label{eq:app-yres}
\tilde{Y}_{it} = Y_{it} - \hat{E}\left[Y_{it} \mid \mathbf{X}_{it}\right],
\end{equation}

\begin{equation}
\tilde{D}_{i} = D_{i} - \hat{E}\left[D_{i} \mid \mathbf{X}_{it}\right] = D_i - \hat{e}(\mathbf{X}_{it}),
\end{equation}
where $\hat{E}\left[Y_{it} \mid \mathbf{X}_{it}\right]$ is a flexible model of outcome given covariates and $\hat{e}(\mathbf{X}_{it})$ is the estimated propensity score from a random forest on the treatment indicator. Within a GDID framework, different CML estimators differ in the choice of $Y_{it}$ and how $\hat{E}\left[Y_{it} \mid \mathbf{X}_{it}\right]$ is estimated: CF2 uses first-difference of price as $Y_{it}$ and handles outcome residualization through the causal forest's internal random forest modeling, while CFRP and CFR2 use explicit period-specific OLS regressions to residualize prices before passing the first difference of price residuals to the causal forest. In each case, these first-stage models are estimated using cross-fitting (i.e., each observation's residual is computed using a model trained on a held-out sample that excludes that observation). This procedure prevents overfitting and is consistent with the out-of-bag prediction procedure in the \texttt{grf} implementation in R. The resulting residuals $\tilde{Y}_{i}$ and $\tilde{D}_i$ are then passed to the second stage - causal forest.

The causal forest then estimates a conditional treatment effect $\hat{\tau}(\mathbf{X}_i)$ for each observation:

\begin{equation}
\label{eq:app-cf}
\tilde{Y}_i = \tau(\mathbf{X}_i)\tilde{D}_i + \varepsilon_i,
\end{equation}
where $\tau(\mathbf{X}_i)$ varies flexibly with covariates, driven by the causal forest's adaptive matching algorithm.

\textit{AIPW aggregation.} To aggregate these conditional estimates into an average effect on the treated, the AIPW estimator constructs a bias-corrected score for each observation:

\begin{equation}
\label{eq:aipw}
\hat{\Gamma}_i = \underbrace{D_i\,\hat{\tau}^{(-i)}(\mathbf{X}_i)}_{\text{1}} + \underbrace{D_i\left(Y_i - \hat{\mu}^{(-i)}(\mathbf{X}_i, 1)\right)}_{\text{2}} - \underbrace{(1 - D_i)\,\frac{\hat{e}^{(-i)}(\mathbf{X}_i)}{1 - \hat{e}^{(-i)}(\mathbf{X}_i)}\left(Y_i - \hat{\mu}^{(-i)}(\mathbf{X}_i, 0)\right)}_{\text{3}},
\end{equation}
where the superscript $(-i)$ denotes that all estimates are computed on a held-out sample excluding observation $i$, and $\hat{\mu}(\mathbf{X}_i, d) = \hat{E}\left[Y_i \mid \mathbf{X}_i\right] + \big(d - \hat{e}(\mathbf{X}_i)\big)\,\hat{\tau}(\mathbf{X}_i)$ is the estimated conditional mean of the outcome given covariates and treatment status $d \in \{0, 1\}$. Term 1 is the treatment effect estimate from the causal forest for treated observations (i.e., the baseline estimate before any correction). Terms 2 and 3 form the augmentation correction. Term 2 adds the residuals of treated observations, so that Terms 1 and 2 together equal the observed outcome of each treated unit minus its predicted outcome without treatment. Term 3 subtracts the residuals of control observations, weighted by the odds of treatment $\hat{e}(\mathbf{X}_i)/(1 - \hat{e}(\mathbf{X}_i))$ so that controls resembling treated units count more. This term corrects for errors in the predicted outcome without treatment, which is the counterfactual that the ATT depends on. If the outcome model were perfectly specified, Terms 2 and 3 would be pure noise with mean zero and the correction would vanish on average.

The ATT is then estimated as:

\begin{equation}
\hat{\tau}_{\text{ATT}} = \frac{1}{N_T}\sum_{i=1}^{N} \hat{\Gamma}_i,
\end{equation}
where $N_T$ is the number of treated observations. The sum runs over all observations because control units supply the correction in Term 3. 

\textit{Identification Assumptions.} Causal forests identify the average DUET under three conditions, which we illustrate using CF2 as an example. The first is \textit{conditional parallel trends} (A1), which requires that, absent treatment, treated and control units share the same expected price change given their covariates, $E[\Delta\ln P_i^{0} \mid \mathbf{X}_i, D_i = 1] = E[\Delta\ln P_i^{0} \mid \mathbf{X}_i, D_i = 0]$ \citep{abadie2005, chang2020, banzhaf2021}. Equation (6) in the main text is its linear form, and the regression benchmarks invoke the same assumption, differing only in how the conditioning on $\mathbf{X}_i$ (i.e., the price modeling) is carried out. The second is \textit{overlap} (A2), which requires that every covariate value has a positive probability of being observed in both groups, $0 < e(\mathbf{X}_i) \equiv P(D_i = 1 \mid \mathbf{X}_i) < 1$. The third is \textit{local non-interference} (A3), the SUTVA condition defined above. Under (A1)--(A3), the conditional difference-in-differences formula of \citet{abadie2005} identifies the ATT.

\textit{Overlap-weighted aggregation.} In our context, however, (A2) fails in covariate cells at both extremes. Close to treatment sites, nearly all observations are treated and comparable controls rarely exist, whereas far from them, nearly all observations are controls and comparable treated units are rare. The first failure is the one that destabilizes the ATT estimator, through the odds weight in Term 3. We therefore target the overlap-weighted average effect \citep{li2018}, which assigns vanishing weight to both types of cells and avoids dividing by estimated propensity terms close to zero.\footnote{This is implemented with the \texttt{target.sample = "overlap"} option of \texttt{average\_treatment\_effect()} in the \texttt{grf} package, whose target is $E[e(\mathbf{X})(1-e(\mathbf{X}))\,\tau(\mathbf{X})]\,/\,E[e(\mathbf{X})(1-e(\mathbf{X}))]$. A few things are worth noting when this average treatment effect on overlapped sample is compared to ATT. The two estimands coincide when the treatment effect does not vary with $\mathbf{X}$, as in Scenarios 1 and 2. In Scenario 3, any remaining gap is counted as CF2 estimation error, because all estimators are evaluated against the average DUET on all treated units. In general, treatment is rare in most of the study area, $1 - e(\mathbf{X}) \approx 1$ for most units, so the overlap weights are close to the ATT weights $e(\mathbf{X})/E[e(\mathbf{X})]$ and the remaining gap in Scenario 3 is minor.  For ease of exposition, we continue to refer to the CF2 estimates as the ATT or average DUET.} With this option, the average DUET is calculated by regressing the outcome residual on the treatment residual:

\begin{equation}
\label{eq:overlap}
\hat{\tau}_{\text{overlap}} = \frac{\sum_{i=1}^{N} \tilde{D}_i\,\tilde{Y}_i}{\sum_{i=1}^{N} \tilde{D}_i^{2}},
\end{equation}
where $\tilde{Y}_i$ and $\tilde{D}_i$ are the out-of-bag residuals from the first stage. Because $E[\tilde{D}_i^{2} \mid \mathbf{X}_i] = e(\mathbf{X}_i)\big(1 - e(\mathbf{X}_i)\big)$, each covariate cell is weighted by its degree of overlap. Equation~(\ref{eq:overlap}) is the overlap-weighted counterpart of the ATT estimator. Multiplying each score in equation~(\ref{eq:aipw}) by $1 - e(\mathbf{X}_i)$ turns the odds weight in Term 3 into $e(\mathbf{X}_i)$, so no division by estimated propensities remains, and changes the weight on each covariate cell from $e(\mathbf{X}_i)$ to $e(\mathbf{X}_i)\big(1 - e(\mathbf{X}_i)\big)$. In expectation, the rescaled moment reduces to the residual-on-residual regression in equation~(\ref{eq:overlap}), in which the conditional effects $\hat{\tau}(\mathbf{X}_i)$ no longer appear. The average DUET therefore depends only on the first-stage propensity and outcome forests, while the causal forest's conditional effects serve the analysis of heterogeneity. The standard error is the heteroskedasticity-robust sandwich standard error of this regression.

Equation~(\ref{eq:overlap}) is the partialling-out estimator of \citet{robinson1988}, and its moment condition is Neyman-orthogonal \citep{chernozhukov2016}, so first-stage estimation errors affect the estimate only to second order. The propensity score and the outcome model (i.e., price residuals and trends) are estimated by the regression forests \citep{athey2019}, and their out-of-bag honest predictions play the role of cross-fitting. In this sense, CF2's average DUET is a double machine learning estimator applied to first-differenced prices, in the spirit of the doubly robust and double machine learning difference-in-differences estimators of \citet{santanna2020} and \citet{chang2020}, but targeting the overlap-weighted effect rather than the ATT. By contrast, simply averaging the forest's CATE estimates, without the residual correction that makes the moment orthogonal, would be a non-orthogonal plug-in of the kind that \citet{chang2020} shows to be first-order biased.

The models below differ in how the first stage is implemented and which sample is used, while all sharing this second-stage aggregation procedure.

\subsubsection{Causal Forest DID with Adjacent Pairs (CF2)}

CF2 applies causal forest directly to first-differenced prices, using only adjacent pairs: the closest pre- and post-treatment transaction pairs for each property. Let $\Delta \ln P_i$ denote the first difference in log price for adjacent pair $i$:

\begin{equation}
\Delta \ln P_i = \ln P_{i,\text{post}} - \ln P_{i,\text{pre}},
\end{equation}
CF2 uses $\Delta \ln P_i$ as the outcome for the causal forest ($Y_{it}$ in equation (6)), with a feature vector $\mathbf{X}_i^{t}$ that, compared to $\mathbf{X}_i$, additionally includes the post-transaction year and the within-pair difference in transaction years, providing the forest with sufficient time information to internally model and accommodate price function shifts. The first-stage residualization of the outcome is therefore handled by the forest's internal random forest modeling rather than by an explicit OLS regression. The propensity score $\hat{e}(\mathbf{X}_i^{t})$ is estimated by a separate random forest on the treatment indicator. The second stage then proceeds as described by equation~(\ref{eq:app-cf}) in Appendix~\ref{app:twostage}, with the AIPW score aggregating the heterogeneous treatment effects into an average DUET.

By restricting to adjacent pairs, CF2 minimizes the time gap between paired transactions, improving the plausibility of the parallel trends assumption. Unlike CFRP and CFR2, CF2 does not rely on an explicit OLS first stage, instead leveraging the causal forest's internal flexibility to model the price function shift directly from the data.

\subsubsection{Causal Forest DID with Residual Prices on Repeated Sales (CFRP)}

CFRP introduces an explicit OLS first stage to residualize prices before passing them to the causal forest. Separate OLS regressions are estimated for the pre- and post-treatment periods:

\begin{equation}
\ln P_{it} = \mathbf{X}_{it}\boldsymbol{\gamma}_t + \mu_{z,t} + \eta_{it} \qquad t \in \{0, 1\},
\end{equation}
where $\boldsymbol{\gamma}_t$ are period-specific coefficients and $\hat{\eta}_{it}$ are the resulting price residuals. By estimating separate regressions for each period, the first stage explicitly accommodates the equilibrium shift in the price function, consistent with the generalized DID framework in A.2.3. For each property transacting in both periods, the first difference in residuals is then formed:

\begin{equation}
\Delta \hat{\eta}_i = \hat{\eta}_{i,\text{post}} - \hat{\eta}_{i,\text{pre}}.
\end{equation}

This first-differenced residual $\Delta \hat{\eta}_i$ is passed as the outcome to the causal forest, replacing ${Y}_i$ in equation~(\ref{eq:app-yres}) of Appendix~\ref{app:twostage}, with the propensity score $\hat{e}(\mathbf{X}_i)$ estimated as described there. This method uses random forests to model the remaining relationships between the feature vector $X$ and first-stage price residuals (i.e., the Orthogonalization or Robinson's transformation in causal forests\footnote{See \citet{athey2019} and the R package page for GRF: \url{https://grf-labs.github.io/grf/REFERENCE.html\#orthogonalization-the-r-learner}.}) and passes the resulting OLS residuals into the causal forest algorithm in equation~(\ref{eq:app-cf}) --- consistent with the idea of boosting \citep{friedman2001, ke2017}. This procedure applies to all repeated sales.
 
\subsubsection{Causal Forest DID with Residual Prices on Adjacent Pairs (CFR2)}

CFR2 mirrors CFRP exactly in its two-stage structure, differing only in the sample used. Rather than all repeated sales, CFR2 restricts to adjacent pairs---the closest pre- and post-treatment transaction pairs for each property:

\begin{equation}
\Delta \hat{\eta}_i = \hat{\eta}_{i,\text{post}} - \hat{\eta}_{i,\text{pre}}, \qquad \forall\ i \in \text{adjacent pairs}.
\end{equation}
This restriction reduces the time gap between paired transactions, improving the plausibility of the parallel trends assumption. However, as with T-GDIDP, restricting to a subset of transactions reduces the effective sample size, which may reduce the performance of the data-hungry causal forest algorithm relative to CFRP.

\subsubsection{Simulation Results for all estimators}

\begin{table}[H]
\centering
\caption{Error Measures for Baseline Models}
\small
\resizebox{\textwidth}{!}{%
\renewcommand{\arraystretch}{0.72}
\begin{tabular}{@{}llcccccc@{}}
\toprule
& & \multicolumn{3}{c}{\textbf{All iterations}} & \multicolumn{3}{c}{\textbf{No. Treated $\geq$3k}} \\
\cmidrule(lr){3-5} \cmidrule(lr){6-8}
& & \textbf{Scenario 1} & \textbf{Scenario 2} & \textbf{Scenario 3} & \textbf{Scenario 1} & \textbf{Scenario 2} & \textbf{Scenario 3} \\
& & Homo. & Rand. Heter.  & X-Heter., & Homo. & Rand. Heter.  & X-Heter. \\
\midrule
\multicolumn{8}{l}{\textbf{Panel A: Allowing Price Function Shift}} \\
CF2 & RMSE & 0.0366 & 0.0364 & 0.0365 & \textbf{0.0282} & \textbf{0.0281} & \textbf{0.0282} \\
& MAPE & 54.04 & 53.69 & 52.96 & \textbf{43.40} & \textbf{43.14} & \textbf{45.17} \\
CFRP & RMSE & 0.0378 & 0.0379 & 0.0381 & 0.0289 & 0.0290 & 0.0290 \\
& MAPE & 57.40 & 57.30 & 56.59 & 44.93 & 44.84 & 46.68 \\
CFR2 & RMSE & 0.0380 & 0.0380 & 0.0379 & 0.0288 & 0.0288 & 0.0287 \\
& MAPE & 56.88 & 56.77 & 55.85 & 45.02 & 44.88 & 46.52 \\
T-GDID & RMSE & \textbf{0.0347} & \textbf{0.0346} & \textbf{0.0347} & 0.0287 & 0.0287 & 0.0286 \\
& MAPE & \textbf{53.37} & \textbf{53.30} & \textbf{52.87} & 44.64 & 44.65 & 46.68 \\
T-GDIDP & RMSE & 0.0442 & 0.0442 & 0.0441 & 0.0354 & 0.0353 & 0.0353 \\
& MAPE & 66.76 & 66.71 & 65.70 & 54.49 & 54.49 & 56.98 \\
\midrule
\multicolumn{8}{l}{\textbf{Panel B: Benchmark Models - No Price Function Shift}} \\
OLS & RMSE & 0.0609 & 0.0608 & 0.0612 & 0.0624 & 0.0624 & 0.0627 \\
& MAPE & 90.07 & 89.99 & 88.66 & 90.35 & 90.31 & 92.43 \\
BDID & RMSE & 0.0413 & 0.0413 & 0.0412 & 0.0338 & 0.0338 & 0.0337 \\
& MAPE & 61.91 & 61.77 & 61.48 & 52.07 & 52.02 & 54.45 \\
\bottomrule
\end{tabular}
}

\begin{flushleft}
{\footnotesize
\noindent \textit{Note}: This table is the full version of Table 2 in the main text. Monte Carlo simulation results are based on 500 iterations for each scenario-method combination. All methods across all scenarios use identical data generation processes. In the homogeneous treatment effect scenario, effect size is set to 5 percent on prices of treated transactions. In the heterogeneous effect scenarios, effect size is set so that the average effect is approximately 5 percent. Bold text denotes the top-performing approaches for each scenario.
}
\end{flushleft}
\end{table}

As shown in Table A.3, causal machine learning methods, CF2, CFR2, and CFRP, perform comparably to T-GDID, which outperforms other conventional estimators. In general, CF2 performs slightly better than CFR2 and significantly better than CFRP. Also, with initial specifications, CF2 and CFR2 do not significantly outperform T-GDID across all scenarios in Table A.3, which is further confirmed by bootstrap tests with additional simulation iterations in Table C.4. With these lessons in mind, we devote the subsequent analyses in the main text to investigate specification changes that may help improve the performance of CF2 and T-GDID. 

\section{Estimation Errors and Tests for Sample Size Threshold}

\setcounter{figure}{0}
\setcounter{table}{0}

In the following figures, for readability reasons, we plot the absolute estimation errors from the first 100 Monte Carlo simulation iterations (out of 500) for each scenario in Table 2. Note that the iterations in Figure B.1 - B.3 are sorted by samples size, in order to intuitively show how sample size affects estimation accuracy.

\begin{figure}[H]
  \centering
  \includegraphics[width=\textwidth]{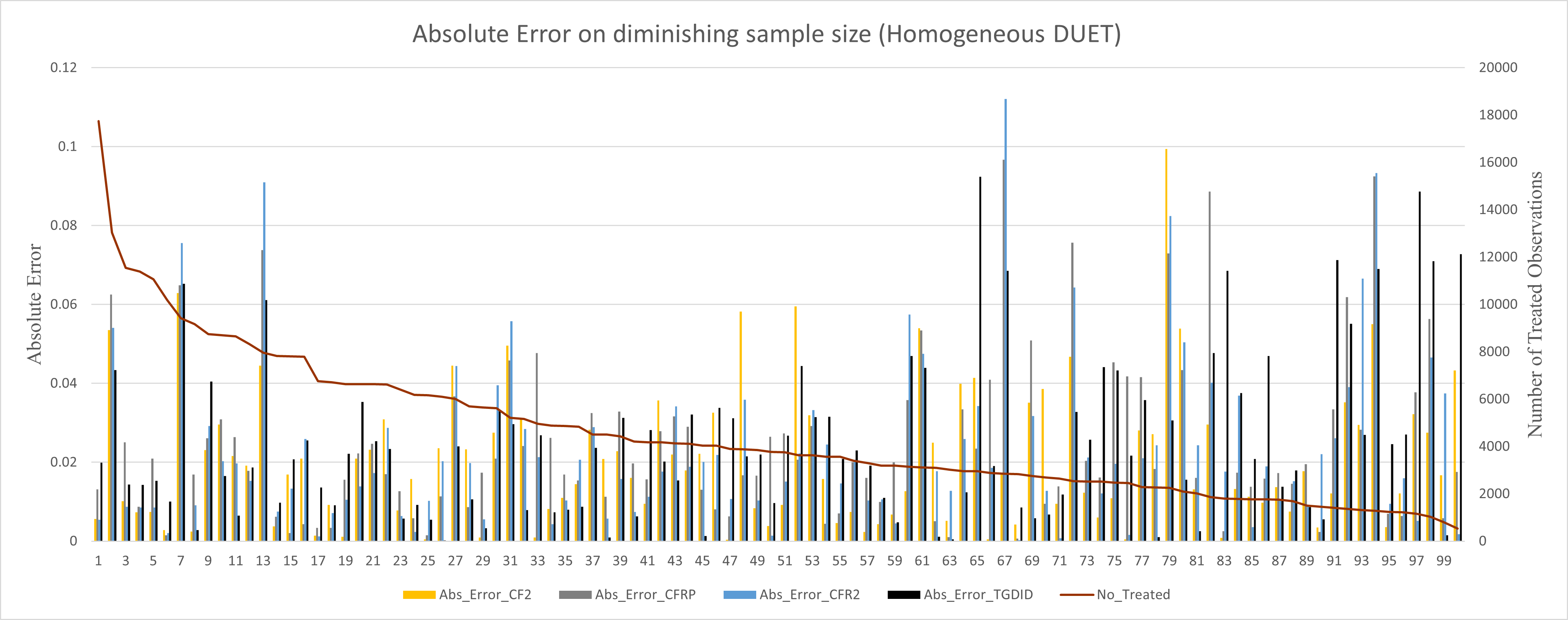}
  \caption{Monte Carlo Simulation Absolute Errors - Iteration Plot (Homogeneous DUET)}
  \label{fig:figA1}
\end{figure}

\begin{figure}[H]
  \centering
  \includegraphics[width=\textwidth]{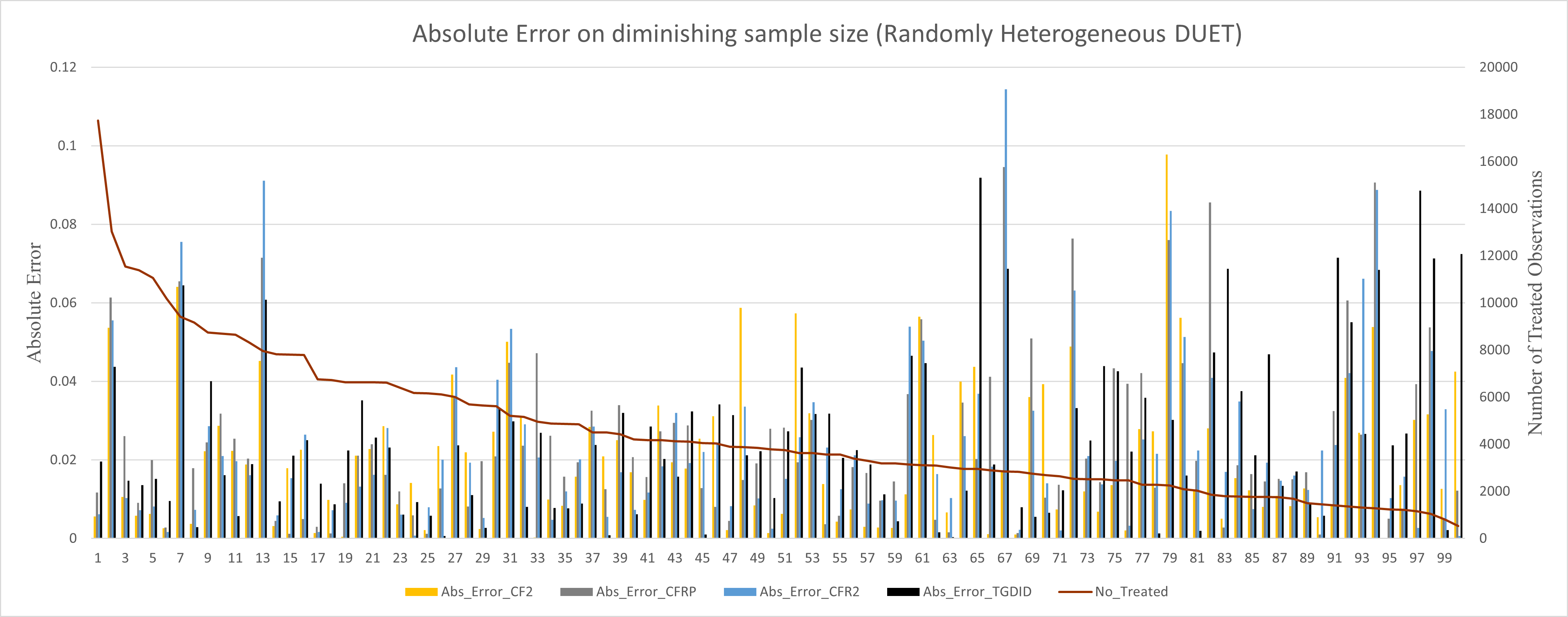}
  \caption{Monte Carlo Simulation Absolute Errors - Iteration Plot (Randomly Heterogeneous DUET)}
  \label{fig:figA2}
\end{figure}

\begin{figure}[H]
  \centering
  \includegraphics[width=\textwidth]{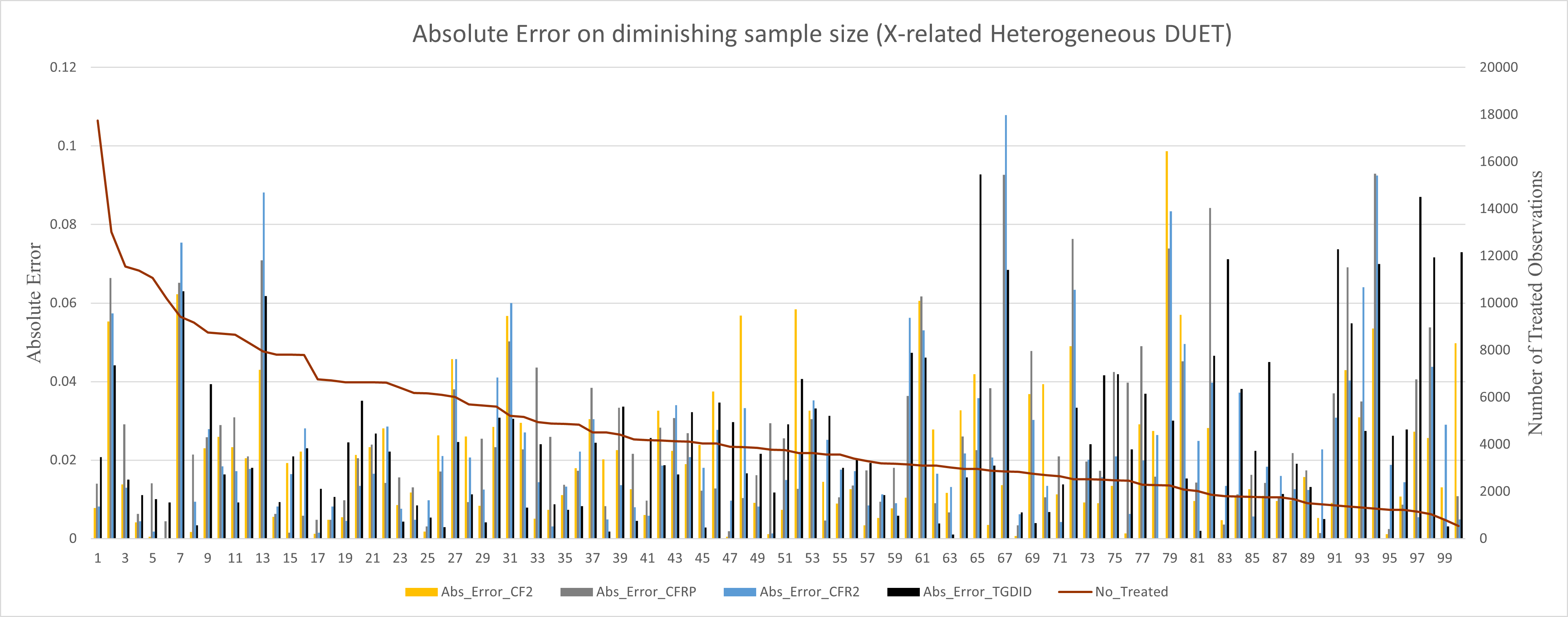}
  \caption{Monte Carlo Simulation Absolute Errors - Iteration Plot (X-related Heterogeneous DUET)}
  \label{fig:figA3}
\end{figure}

Through visual inspection of error patterns across sample sizes in Figure B.1 to B.3, we believe there is a sample size threshold (measured in number of treated observations) below which estimation errors generally exhibit a marked increase from larger samples.

More importantly, we are interested in how the estimation error distribution from different approaches as sample size increases. Since the absolute errors distribute similarly across three scenarios, we focus on scenario 1 (homogeneous DUET) for this analysis. In Table B.1, we divide the samples by different thresholds and calcualte the average mean error difference between CF2 and T-GDID  ($ \frac{1}{n}\sum_{i=1}^{n}(\text{AbsError}_{\text{CF2},i} - \text{AbsError}_{\text{T-GDID},i})$) within each subsample. The results suggest that above a certain threshold, CF2 generally estimates DUET with a lower estimation error. Also, above a certain threshold, CF2 has a higher chance to estimate with a lower error across iterations. This advantage seems to increase as sample size increases.

To formally assess whether estimation accuracy differs meaningfully across sample sizes and to decide the threshold, we employ two-sample (two-sided) Kolmogorov-Smirnov (KS) tests \citep{smirnov1948, conover1999} to compare estimation error distributions between small and large sample regimes. The KS test is a non-parametric test that evaluates whether two samples are drawn from different distributions by measuring the maximum distance between their cumulative distribution functions (with null hypothesis stating both come from the same distribution). We evaluate potential thresholds ranging from 2,000 to 5,000 treated observations to identify where differential error (i.e., $\text{AbsError}_{\text{CF2},i} - \text{AbsError}_{\text{T-GDID},i}$) distributions diverge significantly. Results presented in Table B.1 indicate that all tested thresholds of 3,000 and above yield statistically significant differences in error distribution (mostly p $<$ 0.01), with KS statistics ranging from 0.162 to 0.139. Notably, the threshold of 3,000 represents the first point at which sample sizes become sufficient to reliably favor causal machine learning approaches over traditional methods. At this threshold, the KS statistic of 0.162 (p $<$ 0.01) confirms that error distributions differ significantly, with a negative mean error difference, and an observed chance of 51.7\% for CF2 yielding a lower error. While higher thresholds yield larger KS statistics and increase the probability of CF2 outperforming T-GDID, the threshold of 3,000 represents the first statistically significant turning point in model performance, offering a practical decision rule in our context.

\begin{table}[ht]
\centering
\caption{Kolmogorov-Smirnov Tests for Sample Size Thresholds (Baseline CF2 versus T-GDID)}
\small
\setlength{\tabcolsep}{3pt}
\begin{tabular}{@{}ccccccccc@{}}
\toprule
& \textbf{N Large} & \multicolumn{2}{c}{\textbf{Mean Error Diff.}} & \multicolumn{2}{c}{\textbf{Pr(CF2 Better)}} & & \\
\cmidrule(lr){3-4} \cmidrule(lr){5-6}
\textbf{Threshold} & \textbf{out of 500} & \textbf{Small} & \textbf{Large} & \textbf{Small} & \textbf{Large} & \textbf{KS} & \textbf{P-value} \\
($N_b$) & ($\geq N_b$) & ($<N_b$) & ($\geq N_b$) & ($<N_b$) & ($\geq N_b$) & \textbf{Stat.} & \\
\midrule
2000 & 381 & 0.0013 & 0.0000 & 50.4\% & 50.7\% & 0.127 & 0.109 \\ 
2500 & 337 & 0.0019 & $-$0.0004 & 50.3\% & 50.7\% & 0.128 & 0.054 \\ 
3000 & 288 & 0.0016 & $-$0.0006 & 49.1\% & 51.7\% & 0.162 & 0.003*** \\ 
3500 & 250 & 0.0020 & $-$0.0013 & 47.6\% & 53.6\% & 0.160 & 0.003*** \\ 
4000 & 215 & 0.0018 & $-$0.0015 & 47.7\% & 54.4\% & 0.165 & 0.003*** \\ 
4500 & 183 & 0.0011 & $-$0.0010 & 49.2\% & 53.0\% & 0.155 & 0.008*** \\ 
5000 & 157 & 0.0011 & $-$0.0012 & 49.3\% & 53.5\% & 0.139 & 0.031** \\ 
\bottomrule
\end{tabular}
\vspace{0.5em}

\begin{flushleft}
{\footnotesize
\noindent \textit{Note}: Results are based on 500 simulation iterations under homogeneous DUET as shown in Table 2. Threshold $N_b$ measures the number of treated observations (i.e., post-treated transactions). Mean Error Diff. = $\frac{1}{n}\sum_{i=1}^{n}(\text{AbsError}_{\text{CF2},i} - \text{AbsError}_{\text{T-GDID},i})$ across $n$ iterations. P(CF2 Better) = proportion of iterations where CF2 has lower absolute error than T-GDID. ** $p < 0.05$, *** $p < 0.01$.
}
\end{flushleft}
\end{table}

We also performed the same KS test to the best-performing CF2 and T-GDID specifications in Table 5 (i.e., CF2 with a 10-year pre-period and a 10~km control radius, and T-GDID with a 10-year pre-period and a 10~km control radius). The results are presented in Table A.2. With the Monte Carlo iteration number of 2000, all tested thresholds from 1,000 to 5,000 yield highly significant differences in error distributions (p $<$ 0.001), confirming robust differentiation across sample size regimes. The threshold of 3,000 treated observations remains a critical inflection point: it generates the largest mean error difference in large samples (the lead of CF2 is more pronounced) and the second largest gap in large-sample-chance and small-sample-chance for CF2 to perform better. As sample sizes decrease, CF2 becomes increasingly unlikely to outperform T-GDID. The probability of CF2 achieving lower error than T-GDID falls from 46.7\% at the 5,000-observation threshold to 30.5\% at 1,000, indicating that traditional methods maintain a clear advantage in small-sample settings.

\begin{table}[ht]
\centering
\caption{Kolmogorov-Smirnov Tests for Sample Size Thresholds (Best-performing CF2 versus T-GDID)}
\small
\setlength{\tabcolsep}{3pt}
\begin{tabular}{@{}ccccccccc@{}}
\toprule
& \textbf{N Large} & \multicolumn{2}{c}{\textbf{Mean Error Diff.}} & \multicolumn{2}{c}{\textbf{Pr(CF2 Better)}} & & \\
\cmidrule(lr){3-4} \cmidrule(lr){5-6}
\textbf{Threshold} & \textbf{out of 2000} & \textbf{Small} & \textbf{Large} & \textbf{Small} & \textbf{Large} & \textbf{KS} & \textbf{P-value} \\
($N_b$) & ($\geq N_b$) & ($<N_b$) & ($\geq N_b$) & ($<N_b$) & ($\geq N_b$) & \textbf{Stat.} & \\
\midrule
1000 & 1895 & 0.0200 & 0.0013 & 30.5\% & 49.1\% & 0.367 & 0.0000*** \\ 
1500 & 1657 & 0.0121 & 0.0002 & 39.1\% & 50.0\% & 0.233 & 0.0000*** \\ 
2000 & 1449 & 0.0087 & $-$0.0002 & 41.6\% & 50.7\% & 0.213 & 0.0000*** \\ 
2500 & 1260 & 0.0070 & $-$0.0005 & 41.9\% & 51.8\% & 0.196 & 0.0000*** \\ 
3000 & 1107 & 0.0065 & $-$0.0012 & 42.9\% & 52.4\% & 0.202 & 0.0000*** \\ 
3500 & 958 & 0.0052 & $-$0.0009 & 44.8\% & 51.8\% & 0.183 & 0.0000*** \\ 
4000 & 817 & 0.0043 & $-$0.0007 & 45.5\% & 52.0\% & 0.164 & 0.0000*** \\ 
4500 & 711 & 0.0039 & $-$0.0007 & 45.9\% & 52.2\% & 0.155 & 0.0000*** \\ 
5000 & 609 & 0.0035 & $-$0.0005 & 46.7\% & 51.4\% & 0.150 & 0.0000*** \\ 
\bottomrule
\end{tabular}

\vspace{0.5em}

\begin{flushleft}
{\footnotesize
\noindent \textit{Note}: Results are based on 2000 simulation iterations under homogeneous DUET as shown in Table 5. Threshold $N_b$ measures the number of treated observations (i.e., post-treated transactions). Mean Error Diff. = $\frac{1}{n}\sum_{i=1}^{n}(\text{AbsError}_{\text{CF2},i} - \text{AbsError}_{\text{T-GDID},i})$ across $n$ iterations. Pr(CF2 Better) = proportion of iterations where CF2 has lower absolute error than T-GDID. All p-values are $< 0.01$, denoted by ***.
}
\end{flushleft}
\end{table}

\section{Bootstrapped Standard Errors of RMSE and Statistical Tests}
\setcounter{table}{0}

We construct statistical tests across methods using bootstrapped distributions for results in Table 2.  When sample size (measured by the number of treated observations) is not taken into account, CF2, CFR2, and CFRP demonstrate performance comparable to T-GDID, which substantially outperforms other conventional estimators (Table C.1 to C.3). 

When restricting attention to iterations with at least 3,000 treated observations, CF2, CFR2, and T-GDID outperform all other conventional alternatives---for most scenarios, the bootstrapped distributions reveal that RMSE and MAPE differences are statistically significant at the 5\% level with one-sided tests. The advantage of CF2, CFR2, and T-GDID becomes more pronounced with larger sample sizes. Across most specifications, we find no statistically significant differences in performance between CFRP and the better performing methods, but CFRP shows relatively weaker performance in large samples (e.g., at least 10,000 treated).

To make sure that the comparable performance between CF2, CFR2 and T-GDID is not a coincidence, we run 500 additional Monte Carlo simulations under Scenario 1, with bootstrap tests presented in Table C.4. These tests confirms that, with the configurations in Table 2, the performance of CF2, CFR2, and T-GDID are not statistically different. Moreover, other observed patterns in Table C.4 are also consistent with Table C.1, suggesting 500 Monte Carlo iterations provide stable enough results.

We also run similar bootstrap tests to compare the best-performing specifications in Table 5. The errors of CF2 and T-GDID specifications are compared against T-GDID with a 2000-2009 pre-period and a 10~km control radius. When restricting the analysis to large samples (at least 3,000, 5,000, or 10,000 treated), CF2 with a 2000–2009 pre-period and a 10~km control radius outperforms the reference T-GDID and others in estimation errors. The lead of CF2 over the reference T-GDID becomes statistically significant in a few large-sample cases.

\begin{table}[htbp]
\centering
\caption{Bootstrap Test Results for Method Comparisons Across Sample Sizes (Scenario 1: Homogeneous DUET)}
\small
\setlength{\tabcolsep}{3pt}
\renewcommand{\arraystretch}{0.95}
\begin{tabular}{@{}lccccccl@{}}
\toprule
& \multicolumn{3}{c}{\textbf{RMSE}} & \multicolumn{3}{c}{\textbf{MAPE}} & \textbf{Significantly} \\
\cmidrule(lr){2-4} \cmidrule(lr){5-7}
\textbf{Method} & \textbf{Est.} & \textbf{S.E.} & \textbf{[.05, .95] CI:} & \textbf{Est.} & \textbf{S.E.} & \textbf{[.05, .95] CI:} & \textbf{Better Than} \\
& & & \textbf{Dif vs T-GDID} & & & \textbf{Dif vs T-GDID} & \textbf{(at .05 level)} \\
\midrule
\multicolumn{8}{l}{\textbf{Panel A: Number of Treated $>$ 0 (500 Monte Carlo Iterations)}} \\
CF2 & 0.0366 & 0.0018 & [$-$0.0009, 0.0046] & 54.041 & 2.187 & [$-$3.114, 4.512] & T-GDIDP, BDID, OLS \\
CFR2 & 0.0380 & 0.0016 & [0.0007, 0.0059] & 56.877 & 2.232 & [$-$0.190, 7.346] & T-GDIDP, BDID, OLS \\
CFRP & 0.0378 & 0.0016 & [0.0007, 0.0055] & 57.396 & 2.183 & [0.490, 7.592] & T-GDIDP, BDID, OLS \\
T-GDID & 0.0347 & 0.0013 & [0.000, 0.000] & 53.370 & 1.954 & [0.000, 0.000] & T-GDIDP, BDID, OLS \\
BDID & 0.0413 & 0.0018 & [0.0049, 0.0085] & 61.914 & 2.400 & [6.240, 10.860] & OLS \\
T-GDIDP & 0.0442 & 0.0018 & [0.0070, 0.0119] & 66.757 & 2.546 & [9.908, 16.862] & OLS \\
OLS & 0.0609 & 0.0028 & [0.0215, 0.0309] & 90.069 & 3.626 & [30.697, 42.794] & --- \\
\midrule
\multicolumn{8}{l}{\textbf{Panel B: Number of Treated $\geq$ 3k (288 Monte Carlo Iterations)}} \\
CF2 & 0.0282 & 0.0015 & [$-$0.0028, 0.0018] & 43.396 & 2.119 & [$-$4.798, 2.228] & T-GDIDP, BDID, OLS \\
CFR2 & 0.0288 & 0.0014 & [$-$0.0021, 0.0023] & 45.017 & 2.114 & [$-$3.199, 3.913] & T-GDIDP, BDID, OLS \\
CFRP & 0.0289 & 0.0013 & [$-$0.0019, 0.0023] & 44.932 & 2.125 & [$-$3.287, 3.761] & T-GDIDP, BDID, OLS \\
T-GDID & 0.0287 & 0.0014 & [0.000, 0.000] & 44.639 & 2.137 & [0.000, 0.000] & T-GDIDP, BDID, OLS \\
BDID & 0.0338 & 0.0015 & [0.0034, 0.0067] & 52.073 & 2.527 & [4.775, 10.071] & OLS \\
T-GDIDP & 0.0354 & 0.0017 & [0.0046, 0.0087] & 54.490 & 2.649 & [6.539, 13.099] & OLS \\
OLS & 0.0624 & 0.0042 & [0.0271, 0.0405] & 90.352 & 5.073 & [37.515, 54.290] & --- \\
\midrule
\multicolumn{8}{l}{\textbf{Panel C: Number of Treated $\geq$ 5k (157 Monte Carlo Iterations)}} \\
CF2 & 0.0261 & 0.0017 & [$-$0.0040, 0.0018] & 40.277 & 2.639 & [$-$6.968, 1.834] & T-GDIDP, BDID, OLS \\
CFR2 & 0.0270 & 0.0018 & [$-$0.0030, 0.0026] & 41.662 & 2.719 & [$-$5.406, 3.020] & T-GDIDP, BDID, OLS \\
CFRP & 0.0271 & 0.0017 & [$-$0.0027, 0.0026] & 41.360 & 2.796 & [$-$5.866, 2.923] & T-GDIDP, BDID, OLS \\
T-GDID & 0.0271 & 0.0016 & [0.000, 0.000] & 42.760 & 2.650 & [0.000, 0.000] & T-GDIDP, BDID, OLS \\
BDID & 0.0316 & 0.0020 & [0.0027, 0.0062] & 48.081 & 3.261 & [2.245, 8.414] & OLS \\
T-GDIDP & 0.0314 & 0.0017 & [0.0020, 0.0064] & 49.493 & 3.059 & [3.015, 10.566] & OLS \\
OLS & 0.0678 & 0.0057 & [0.0311, 0.0502] & 97.932 & 7.533 & [43.295, 67.706] & --- \\
\midrule
\multicolumn{8}{l}{\textbf{Panel D: Number of Treated $\geq$ 10k (33 Monte Carlo Iterations)}} \\
CF2 & 0.0234 & 0.0029 & [$-$0.0019, 0.0042] & 37.037 & 5.039 & [$-$7.392, 6.198] & T-GDIDP, BDID, OLS \\
CFR2 & 0.0236 & 0.0028 & [$-$0.0014, 0.0038] & 37.683 & 4.936 & [$-$6.186, 5.928] & T-GDIDP, BDID, OLS \\
CFRP & 0.0291 & 0.0040 & [0.0026, 0.0103] & 42.908 & 6.830 & [$-$3.047, 13.252] & OLS \\
T-GDID & 0.0222 & 0.0023 & [0.000, 0.000] & 37.870 & 4.079 & [0.000, 0.000] & T-GDIDP, OLS \\
BDID & 0.0252 & 0.0026 & [$-$0.0006, 0.0067] & 41.656 & 4.860 & [$-$2.438, 10.431] & OLS \\
T-GDIDP & 0.0307 & 0.0043 & [0.0036, 0.0126] & 45.289 & 7.240 & [$-$1.250, 16.480] & OLS \\
OLS & 0.0653 & 0.0091 & [0.0290, 0.0563] & 100.091 & 14.725 & [41.390, 84.922] & --- \\
\bottomrule
\end{tabular}
\vspace{0.5em}

\begin{flushleft}
{\footnotesize
\textit{Note}: Standard errors and confidence intervals of RMSEs are calculated from 10,000 bootstrap resamples. Because datasets are identical across methods in each Monte Carlo replication, the RMSE across methods are correlated. Independence-based calculation of standard error on RMSE differences will provide overstated standard errors, leading to conservative p-values. We therefore bootstrap the RMSE differences between each competing method and T-GDID, construct [5\%, 95\%] percentile confidence intervals, and formally test for statistical significance. The last column shows the results for one-sided tests. Results are based on 500 simulation iterations as shown in Table 2.
}
\end{flushleft}
\end{table}

\begin{table}[htbp]
\centering
\caption{Bootstrap Test Results for Method Comparisons Across Sample Sizes (Scenario 2: Randomly Heterogeneous DUET)}
\small
\setlength{\tabcolsep}{3pt}
\renewcommand{\arraystretch}{0.95}
\begin{tabular}{@{}lccccccl@{}}
\toprule
& \multicolumn{3}{c}{\textbf{RMSE}} & \multicolumn{3}{c}{\textbf{MAPE}} & \textbf{Significantly} \\
\cmidrule(lr){2-4} \cmidrule(lr){5-7}
\textbf{Method} & \textbf{Est.} & \textbf{S.E.} & \textbf{[.05, .95] CI:} & \textbf{Est.} & \textbf{S.E.} & \textbf{[.05, .95] CI:} & \textbf{Better Than} \\
& & & \textbf{Dif vs T-GDID} & & & \textbf{Dif vs T-GDID} & \textbf{(at .05 level)} \\
\midrule
\multicolumn{8}{l}{\textbf{Panel A: Number of Treated $>$ 0 (500 Monte Carlo Iterations)}} \\
CF2 & 0.0364 & 0.0018 & [$-$0.0010, 0.0046] & 53.688 & 2.185 & [$-$3.393, 4.238] & T-GDIDP, BDID, OLS \\
CFR2 & 0.0380 & 0.0017 & [0.0008, 0.0060] & 56.772 & 2.233 & [$-$0.239, 7.294] & T-GDIDP, BDID, OLS \\
CFRP & 0.0379 & 0.0016 & [0.0007, 0.0056] & 57.301 & 2.184 & [0.424, 7.598] & T-GDIDP, BDID, OLS \\
T-GDID & 0.0346 & 0.0013 & [0.000, 0.000] & 53.303 & 1.950 & [0.000, 0.000] & T-GDIDP, BDID, OLS \\
BDID & 0.0413 & 0.0018 & [0.0049, 0.0085] & 61.770 & 2.396 & [6.163, 10.782] & OLS \\
T-GDIDP & 0.0442 & 0.0019 & [0.0070, 0.0120] & 66.707 & 2.543 & [9.918, 16.878] & OLS \\
OLS & 0.0608 & 0.0028 & [0.0215, 0.0309] & 89.986 & 3.636 & [30.660, 42.776] & --- \\
\midrule
\multicolumn{8}{l}{\textbf{Panel B: Number of Treated $\geq$ 3k (288 Monte Carlo Iterations)}} \\
CF2 & 0.0281 & 0.0015 & [$-$0.0029, 0.0016] & 43.142 & 2.106 & [$-$5.004, 1.945] & T-GDIDP, BDID, OLS \\
CFR2 & 0.0288 & 0.0014 & [$-$0.0021, 0.0024] & 44.885 & 2.132 & [$-$3.342, 3.739] & T-GDIDP, BDID, OLS \\
CFRP & 0.0290 & 0.0013 & [$-$0.0019, 0.0024] & 44.837 & 2.138 & [$-$3.371, 3.638] & T-GDIDP, BDID, OLS \\
T-GDID & 0.0287 & 0.0014 & [0.000, 0.000] & 44.650 & 2.138 & [0.000, 0.000] & T-GDIDP, BDID, OLS \\
BDID & 0.0338 & 0.0015 & [0.0034, 0.0067] & 52.016 & 2.527 & [4.718, 10.000] & OLS \\
T-GDIDP & 0.0353 & 0.0017 & [0.0046, 0.0087] & 54.492 & 2.649 & [6.521, 13.102] & OLS \\
OLS & 0.0624 & 0.0042 & [0.0271, 0.0405] & 90.307 & 5.084 & [37.464, 54.252] & --- \\
\midrule
\multicolumn{8}{l}{\textbf{Panel C: Number of Treated $\geq$ 5k (157 Monte Carlo Iterations)}} \\
CF2 & 0.0259 & 0.0017 & [$-$0.0041, 0.0017] & 40.025 & 2.627 & [$-$7.127, 1.572] & T-GDIDP, BDID, OLS \\
CFR2 & 0.0269 & 0.0018 & [$-$0.0030, 0.0025] & 41.318 & 2.736 & [$-$5.675, 2.714] & T-GDIDP, BDID, OLS \\
CFRP & 0.0270 & 0.0017 & [$-$0.0028, 0.0025] & 41.139 & 2.793 & [$-$6.037, 2.712] & T-GDIDP, BDID, OLS \\
T-GDID & 0.0271 & 0.0016 & [0.000, 0.000] & 42.732 & 2.646 & [0.000, 0.000] & T-GDIDP, BDID, OLS \\
BDID & 0.0316 & 0.0020 & [0.0028, 0.0062] & 48.118 & 3.261 & [2.311, 8.469] & OLS \\
T-GDIDP & 0.0313 & 0.0017 & [0.0020, 0.0064] & 49.449 & 3.065 & [2.962, 10.575] & OLS \\
OLS & 0.0678 & 0.0057 & [0.0311, 0.0503] & 97.968 & 7.548 & [43.362, 67.770] & --- \\
\midrule
\multicolumn{8}{l}{\textbf{Panel D: Number of Treated $\geq$ 10k (33 Monte Carlo Iterations)}} \\
CF2 & 0.0231 & 0.0029 & [$-$0.0021, 0.0040] & 36.076 & 5.100 & [$-$8.276, 5.424] & T-GDIDP, BDID, OLS \\
CFR2 & 0.0235 & 0.0028 & [$-$0.0015, 0.0039] & 37.475 & 4.952 & [$-$6.158, 5.903] & T-GDIDP, BDID, OLS \\
CFRP & 0.0287 & 0.0040 & [0.0025, 0.0100] & 42.723 & 6.740 & [$-$2.883, 13.107] & OLS \\
T-GDID & 0.0221 & 0.0023 & [0.000, 0.000] & 37.704 & 4.080 & [0.000, 0.000] & T-GDIDP, OLS \\
BDID & 0.0251 & 0.0026 & [$-$0.0005, 0.0068] & 41.616 & 4.866 & [$-$2.301, 10.529] & OLS \\
T-GDIDP & 0.0306 & 0.0043 & [0.0036, 0.0126] & 45.162 & 7.285 & [$-$1.186, 16.501] & OLS \\
OLS & 0.0653 & 0.0092 & [0.0290, 0.0565] & 100.092 & 14.771 & [41.445, 85.155] & --- \\
\bottomrule
\end{tabular}
\vspace{0.5em}

\begin{flushleft}
{\footnotesize
\textit{Note}: Standard errors and confidence intervals of RMSEs are calculated from 10,000 bootstrap resamples. Because datasets are identical across methods in each Monte Carlo replication, the RMSE across methods are correlated. Independence-based calculation of standard error on RMSE differences will provide overstated standard errors, leading to conservative p-values. We therefore bootstrap the RMSE differences between each competing method and T-GDID, construct [5\%, 95\%] percentile confidence intervals, and formally test for statistical significance. The last column shows the results for one-sided tests. Results are based on 500 simulation iterations as shown in Table 2.
}
\end{flushleft}
\end{table}

\begin{table}[htbp]
\centering
\caption{Bootstrap Test Results for Method Comparisons Across Sample Sizes (Scenario 3: X-related Heterogeneous DUET)}
\small
\setlength{\tabcolsep}{3pt}
\renewcommand{\arraystretch}{0.95}
\begin{tabular}{@{}lccccccl@{}}
\toprule
& \multicolumn{3}{c}{\textbf{RMSE}} & \multicolumn{3}{c}{\textbf{MAPE}} & \textbf{Significantly} \\
\cmidrule(lr){2-4} \cmidrule(lr){5-7}
\textbf{Method} & \textbf{Est.} & \textbf{S.E.} & \textbf{[.05, .95] CI:} & \textbf{Est.} & \textbf{S.E.} & \textbf{[.05, .95] CI:} & \textbf{Better Than} \\
& & & \textbf{Dif vs T-GDID} & & & \textbf{Dif vs T-GDID} & \textbf{(at .05 level)} \\
\midrule
\multicolumn{8}{l}{\textbf{Panel A: Number of Treated $>$ 0 (500 Monte Carlo Iterations)}} \\
CF2 & 0.0365 & 0.0018 & [$-$0.0010, 0.0046] & 52.960 & 2.067 & [$-$3.646, 3.850] & T-GDIDP, BDID, OLS \\
CFR2 & 0.0379 & 0.0016 & [0.0006, 0.0057] & 55.850 & 2.105 & [$-$0.655, 6.655] & T-GDIDP, BDID, OLS \\
CFRP & 0.0381 & 0.0016 & [0.0008, 0.0057] & 56.593 & 2.093 & [0.224, 7.271] & T-GDIDP, BDID, OLS \\
T-GDID & 0.0347 & 0.0013 & [0.000, 0.000] & 52.874 & 1.916 & [0.000, 0.000] & T-GDIDP, BDID, OLS \\
BDID & 0.0412 & 0.0018 & [0.0047, 0.0083] & 61.477 & 2.421 & [6.203, 11.021] & OLS \\
T-GDIDP & 0.0441 & 0.0019 & [0.0068, 0.0119] & 65.705 & 2.423 & [9.494, 16.144] & OLS \\
OLS & 0.0612 & 0.0029 & [0.0217, 0.0312] & 88.661 & 3.329 & [30.115, 41.523] & --- \\
\midrule
\multicolumn{8}{l}{\textbf{Panel B: Number of Treated $\geq$ 3k (288 Monte Carlo Iterations)}} \\
CF2 & 0.0282 & 0.0015 & [$-$0.0027, 0.0018] & 45.173 & 2.177 & [$-$5.342, 2.293] & T-GDIDP, BDID, OLS \\
CFR2 & 0.0287 & 0.0014 & [$-$0.0021, 0.0023] & 46.523 & 2.152 & [$-$4.008, 3.665] & T-GDIDP, BDID, OLS \\
CFRP & 0.0290 & 0.0013 & [$-$0.0018, 0.0025] & 46.676 & 2.175 & [$-$3.861, 3.738] & T-GDIDP, BDID, OLS \\
T-GDID & 0.0286 & 0.0014 & [0.000, 0.000] & 46.682 & 2.268 & [0.000, 0.000] & T-GDIDP, BDID, OLS \\
BDID & 0.0337 & 0.0015 & [0.0034, 0.0066] & 54.451 & 2.676 & [4.931, 10.578] & OLS \\
T-GDIDP & 0.0353 & 0.0017 & [0.0046, 0.0087] & 56.985 & 2.739 & [6.838, 13.680] & OLS \\
OLS & 0.0627 & 0.0043 & [0.0274, 0.0410] & 92.431 & 4.624 & [37.797, 53.835] & --- \\
\midrule
\multicolumn{8}{l}{\textbf{Panel C: Number of Treated $\geq$ 5k (157 Monte Carlo Iterations)}} \\
CF2 & 0.0261 & 0.0016 & [$-$0.0038, 0.0018] & 41.046 & 2.493 & [$-$7.680, 1.458] & T-GDIDP, BDID, OLS \\
CFR2 & 0.0271 & 0.0018 & [$-$0.0028, 0.0025] & 42.377 & 2.532 & [$-$6.257, 2.657] & T-GDIDP, BDID, OLS \\
CFRP & 0.0273 & 0.0017 & [$-$0.0026, 0.0028] & 42.475 & 2.659 & [$-$6.266, 2.965] & T-GDIDP, BDID, OLS \\
T-GDID & 0.0271 & 0.0016 & [0.000, 0.000] & 44.075 & 2.741 & [0.000, 0.000] & T-GDIDP, BDID, OLS \\
BDID & 0.0315 & 0.0020 & [0.0027, 0.0061] & 49.501 & 3.310 & [2.276, 8.611] & OLS \\
T-GDIDP & 0.0314 & 0.0017 & [0.0021, 0.0065] & 50.728 & 3.099 & [2.655, 10.705] & OLS \\
OLS & 0.0684 & 0.0059 & [0.0314, 0.0510] & 98.362 & 6.690 & [43.327, 65.704] & --- \\
\midrule
\multicolumn{8}{l}{\textbf{Panel D: Number of Treated $\geq$ 10k (33 Monte Carlo Iterations)}} \\
CF2 & 0.0228 & 0.0030 & [$-$0.0021, 0.0042] & 35.273 & 5.066 & [$-$8.332, 5.506] & T-GDIDP, OLS \\
CFR2 & 0.0231 & 0.0029 & [$-$0.0014, 0.0040] & 36.340 & 4.904 & [$-$6.429, 5.512] & T-GDIDP, OLS \\
CFRP & 0.0289 & 0.0041 & [0.0028, 0.0108] & 43.243 & 6.571 & [$-$1.262, 14.281] & OLS \\
T-GDID & 0.0217 & 0.0023 & [0.000, 0.000] & 36.907 & 4.023 & [0.000, 0.000] & T-GDIDP, OLS \\
BDID & 0.0247 & 0.0027 & [$-$0.0005, 0.0067] & 40.632 & 4.611 & [$-$2.360, 10.131] & OLS \\
T-GDIDP & 0.0305 & 0.0044 & [0.0039, 0.0130] & 44.880 & 7.264 & [$-$0.891, 17.250] & OLS \\
OLS & 0.0651 & 0.0094 & [0.0289, 0.0572] & 98.938 & 13.472 & [42.486, 82.815] & --- \\
\bottomrule
\end{tabular}
\vspace{0.5em}

\begin{flushleft}
{\footnotesize
\textit{Note}: Standard errors and confidence intervals of RMSEs are calculated from 10,000 bootstrap resamples. Because datasets are identical across methods in each Monte Carlo replication, the RMSE across methods are correlated. Independence-based calculation of standard error on RMSE differences will provide overstated standard errors, leading to conservative p-values. We therefore bootstrap the RMSE differences between each competing method and T-GDID, construct [5\%, 95\%] percentile confidence intervals, and formally test for statistical significance. The last column shows the results for one-sided tests. Results are based on 500 simulation iterations as shown in Table 2.
}
\end{flushleft}
\end{table}

\begin{table}[htbp]
\centering
\caption{Bootstrap Test Results for Method Comparisons Across Sample Sizes (Scenario 1: Homogeneous DUET, 1000 Iterations)}
\small
\setlength{\tabcolsep}{3pt}
\renewcommand{\arraystretch}{0.95}
\begin{tabular}{@{}lccccccl@{}}
\toprule
& \multicolumn{3}{c}{\textbf{RMSE}} & \multicolumn{3}{c}{\textbf{MAPE}} & \textbf{Significantly} \\
\cmidrule(lr){2-4} \cmidrule(lr){5-7}
\textbf{Method} & \textbf{Est.} & \textbf{S.E.} & \textbf{[.05, .95] CI:} & \textbf{Est.} & \textbf{S.E.} & \textbf{[.05, .95] CI:} & \textbf{Better Than} \\
& & & \textbf{Dif vs T-GDID} & & & \textbf{Dif vs T-GDID} & \textbf{(at .05 level)} \\
\midrule
\multicolumn{8}{l}{\textbf{Panel A: Number of Treated $>$ 0 (1000 Monte Carlo Iterations)}} \\
CF2 & 0.0355 & 0.0012 & [$-$0.0010, 0.0033] & 52.794 & 1.493 & [$-$2.570, 2.945] & T-GDIDP, BDID, OLS \\
CFR2 & 0.0365 & 0.0011 & [0.0000, 0.0043] & 54.922 & 1.516 & [$-$0.486, 5.085] & T-GDIDP, BDID, OLS \\
CFRP & 0.0370 & 0.0011 & [0.0006, 0.0047] & 55.960 & 1.532 & [0.577, 6.042] & T-GDIDP, BDID, OLS \\
T-GDID & 0.0344 & 0.0010 & [0.000, 0.000] & 52.620 & 1.397 & [0.000, 0.000] & T-GDIDP, BDID, OLS \\
BDID & 0.0404 & 0.0012 & [0.0049, 0.0072] & 61.119 & 1.679 & [6.897, 10.117] & OLS \\
T-GDIDP & 0.0438 & 0.0013 & [0.0074, 0.0113] & 66.054 & 1.804 & [10.855, 16.044] & OLS \\
OLS & 0.0612 & 0.0020 & [0.0234, 0.0301] & 90.548 & 2.607 & [33.522, 42.355] & --- \\
\midrule
\multicolumn{8}{l}{\textbf{Panel B: Number of Treated $\geq$ 3k (565 Monte Carlo Iterations)}} \\
CF2 & 0.0285 & 0.0010 & [$-$0.0030, 0.0006] & 43.585 & 1.551 & [$-$5.651, $-$0.010] & T-GDID(P), BDID, OLS \\
CFR2 & 0.0287 & 0.0009 & [$-$0.0026, 0.0006] & 44.828 & 1.513 & [$-$4.303, 1.080] & T-GDIDP, BDID, OLS \\
CFRP & 0.0295 & 0.0010 & [$-$0.0018, 0.0015] & 45.843 & 1.571 & [$-$3.286, 2.129] & T-GDIDP, BDID, OLS \\
T-GDID & 0.0297 & 0.0010 & [0.000, 0.000] & 46.428 & 1.566 & [0.000, 0.000] & T-GDIDP, BDID, OLS \\
BDID & 0.0350 & 0.0011 & [0.0040, 0.0066] & 53.906 & 1.880 & [5.387, 9.534] & OLS \\
T-GDIDP & 0.0362 & 0.0012 & [0.0050, 0.0079] & 55.695 & 1.948 & [6.791, 11.701] & OLS \\
OLS & 0.0630 & 0.0029 & [0.0286, 0.0380] & 90.715 & 3.689 & [38.308, 50.479] & --- \\
\midrule
\multicolumn{8}{l}{\textbf{Panel C: Number of Treated $\geq$ 5k (308 Monte Carlo Iterations)}} \\
CF2 & 0.0265 & 0.0013 & [$-$0.0028, 0.0018] & 39.786 & 2.005 & [$-$6.084, 0.907] & T-GDIDP, BDID, OLS \\
CFR2 & 0.0267 & 0.0012 & [$-$0.0023, 0.0018] & 41.064 & 1.956 & [$-$4.648, 2.091] & T-GDIDP, BDID, OLS \\
CFRP & 0.0284 & 0.0013 & [$-$0.0007, 0.0036] & 43.114 & 2.111 & [$-$2.782, 4.270] & T-GDIDP, BDID, OLS \\
T-GDID & 0.0270 & 0.0012 & [0.000, 0.000] & 42.326 & 1.900 & [0.000, 0.000] & T-GDIDP, BDID, OLS \\
BDID & 0.0302 & 0.0014 & [0.0021, 0.0045] & 46.513 & 2.203 & [1.947, 6.483] & OLS \\
T-GDIDP & 0.0328 & 0.0014 & [0.0039, 0.0077] & 50.344 & 2.396 & [4.933, 11.150] & OLS \\
OLS & 0.0671 & 0.0040 & [0.0337, 0.0466] & 96.015 & 5.359 & [45.188, 62.314] & --- \\
\midrule
\multicolumn{8}{l}{\textbf{Panel D: Number of Treated $\geq$ 10k (63 Monte Carlo Iterations)}} \\
CFR2 & 0.0243 & 0.0024 & [$-$0.0050, 0.0008] & 36.912 & 3.993 & [$-$11.386, $-$0.588] & T-GDID(P), BDID, OLS \\
CF2 & 0.0263 & 0.0030 & [$-$0.0038, 0.0035] & 38.635 & 4.516 & [$-$10.149, 1.755] & T-GDIDP, BDID, OLS \\
CFRP & 0.0286 & 0.0030 & [$-$0.0013, 0.0056] & 41.995 & 4.903 & [$-$6.843, 5.206] & OLS \\
T-GDID & 0.0262 & 0.0021 & [0.000, 0.000] & 42.829 & 3.859 & [0.000, 0.000] & T-GDIDP, OLS \\
BDID & 0.0259 & 0.0020 & [$-$0.0026, 0.0021] & 42.583 & 3.717 & [$-$4.803, 4.350] & OLS \\
T-GDIDP & 0.0359 & 0.0038 & [0.0054, 0.0134] & 52.584 & 6.162 & [2.882, 16.740] & OLS \\
OLS & 0.0632 & 0.0075 & [0.0247, 0.0484] & 89.524 & 11.312 & [29.401, 64.990] & --- \\
\bottomrule
\end{tabular}
\vspace{0.5em}

\begin{flushleft}
{\footnotesize
\textit{Note}: Standard errors and confidence intervals of RMSEs are calculated from 10,000 bootstrap resamples. Because datasets are identical across methods in each Monte Carlo replication, the RMSE across methods are correlated. Independence-based calculation of standard error on RMSE differences will provide overstated standard errors, leading to conservative p-values. We therefore bootstrap the RMSE differences between each competing method and T-GDID, construct [5\%, 95\%] percentile confidence intervals, and formally test for statistical significance. The last column shows the results for one-sided tests. Results are based on 1,000 Monte Carlo iterations (double the 500 iterations in Table 2) to confirm stability of findings.
}
\end{flushleft}
\end{table}

\begin{table}[htbp]
\centering
\caption{Bootstrap Test Results for Comparison across Best-performing Specifications by Sample Sizes (Scenario 1: Homogeneous DUET, 2000 Iterations)}
\small
\setlength{\tabcolsep}{2.5pt}
\renewcommand{\arraystretch}{0.95}
\begin{tabular}{@{}lccccccl@{}}
\toprule
& \multicolumn{3}{c}{\textbf{RMSE}} & \multicolumn{3}{c}{\textbf{MAPE}} & \textbf{Significantly} \\
\cmidrule(lr){2-4} \cmidrule(lr){5-7}
\textbf{Method} & \textbf{Est.} & \textbf{S.E.} & \textbf{[.05, .95] CI:} & \textbf{Est.} & \textbf{S.E.} & \textbf{[.05, .95] CI:} & \textbf{Better Than} \\
& & & \textbf{Dif vs T-GDID} & & & \textbf{Dif vs T-GDID} & \textbf{(at .05 level)} \\
\midrule
\multicolumn{8}{l}{\textbf{Panel A: Number of Treated $>$ 0 (2000 Monte Carlo Iterations)}} \\
CF2 (long pre, 5~km) & 0.0340 & 0.0008 & [0.0009, 0.0042] & 50.756 & 1.016 & [1.000, 5.174] & --- \\
CF2 (long pre, 10~km) & 0.0324 & 0.0007 & [$-$0.0005, 0.0026] & 48.604 & 0.959 & [$-$1.189, 3.011] & --- \\
CF2 (short pre, 5~km) & 0.0367 & 0.0008 & [0.0037, 0.0070] & 54.300 & 1.109 & [4.381, 8.809] & --- \\
CF2 (short pre, 10~km) & 0.0355 & 0.0008 & [0.0025, 0.0058] & 52.227 & 1.074 & [2.301, 6.730] & --- \\
T-GDID (5~km) & 0.0347 & 0.0008 & [0.0023, 0.0044] & 52.147 & 1.031 & [3.030, 5.870] & --- \\
T-GDID (10~km) & 0.0314 & 0.0006 & [0.0000, 0.0000] & 47.685 & 0.913 & [0.000, 0.000] & All but CF2 \\
&&&&&&&(long pre,10~km) \\
\midrule
\multicolumn{8}{l}{\textbf{Panel B: Number of Treated $\geq$ 3k (1107 Monte Carlo Iterations)}} \\
CF2 (long pre, 5~km) & 0.0253 & 0.0006 & [$-$0.0020, 0.0010] & 38.936 & 0.973 & [$-$2.588, 1.735] & --- \\
CF2 (long pre, 10~km) & 0.0248 & 0.0006 & [$-$0.0023, 0.0005] & 38.686 & 0.940 & [$-$2.757, 1.455] & --- \\
CF2 (short pre, 5~km) & 0.0244 & 0.0006 & [$-$0.0028, 0.0001] & 38.039 & 0.919 & [$-$3.483, 0.838] & --- \\
CF2 (short pre, 10~km) & 0.0241 & 0.0006 & [$-$0.0031, $-$0.0002] & 37.021 & 0.923 & [$-$4.512, $-$0.199] & T-GDID (10~km) \\
T-GDID (5~km) & 0.0262 & 0.0007 & [$-$0.0004, 0.0013] & 40.407 & 0.998 & [$-$0.275, 2.476] & --- \\
T-GDID (10~km) & 0.0257 & 0.0007 & [0.0000, 0.0000] & 39.314 & 0.988 & [0.000, 0.000] & --- \\
\midrule
\multicolumn{8}{l}{\textbf{Panel C: Number of Treated $\geq$ 5k (609 Monte Carlo Iterations)}} \\
CF2 (long pre, 5~km) & 0.0233 & 0.0009 & [$-$0.0022, 0.0018] & 36.247 & 1.199 & [$-$2.271, 3.144] & --- \\
CF2 (long pre, 10~km) & 0.0233 & 0.0008 & [$-$0.0021, 0.0017] & 35.859 & 1.214 & [$-$2.760, 2.877] & --- \\
CF2 (short pre, 5~km) & 0.0228 & 0.0008 & [$-$0.0026, 0.0012] & 35.384 & 1.169 & [$-$3.180, 2.258] & --- \\
CF2 (short pre, 10~km) & 0.0226 & 0.0007 & [$-$0.0027, 0.0009] & 34.760 & 1.179 & [$-$3.766, 1.664] & --- \\
T-GDID (5~km) & 0.0243 & 0.0009 & [$-$0.0002, 0.0018] & 37.091 & 1.273 & [$-$0.238, 2.822] & --- \\
T-GDID (10~km) & 0.0235 & 0.0009 & [0.0000, 0.0000] & 35.804 & 1.236 & [0.000, 0.000] & --- \\
\midrule
\multicolumn{8}{l}{\textbf{Panel D: Number of Treated $\geq$ 10k (116 Monte Carlo Iterations)}} \\
CF2 (long pre, 5~km) & 0.0205 & 0.0015 & [$-$0.0036, 0.0018] & 32.272 & 2.383 & [$-$7.253, 2.278] & --- \\
CF2 (long pre, 10~km) & 0.0209 & 0.0015 & [$-$0.0030, 0.0019] & 31.887 & 2.523 & [$-$7.434, 1.706] & --- \\
CF2 (short pre, 5~km) & 0.0192 & 0.0014 & [$-$0.0047, 0.0004] & 29.764 & 2.281 & [$-$9.368, $-$0.571] & T-GDID (10~km) \\
CF2 (short pre, 10~km) & 0.0198 & 0.0015 & [$-$0.0039, 0.0005] & 29.864 & 2.443 & [$-$9.216, $-$0.656] & T-GDID (10~km) \\
T-GDID (5~km) & 0.0198 & 0.0014 & [$-$0.0034, 0.0000] & 31.548 & 2.252 & [$-$6.182, $-$0.299] & T-GDID (10~km) \\
T-GDID (10~km) & 0.0214 & 0.0013 & [0.0000, 0.0000] & 34.772 & 2.366 & [0.000, 0.000] & --- \\
\bottomrule
\end{tabular}
\vspace{0.5em}

\begin{flushleft}
{\footnotesize
\textit{Note}: Standard errors and confidence intervals of RMSEs are calculated from 10,000 bootstrap resamples. Because datasets are identical across methods in each Monte Carlo replication, the RMSE across methods are correlated. Independence-based calculation of standard error on RMSE differences will provide overstated standard errors, leading to conservative p-values. We therefore bootstrap the RMSE differences between each competing method and T-GDID (10~km), construct [5\%, 95\%] percentile confidence intervals, and formally test for statistical significance. The last column shows the results for one-sided tests. Results are based on 2,000 Monte Carlo iterations in Table 5 to confirm stability of findings.
}
\end{flushleft}
\end{table}

\clearpage

\section{More Tables and Figures}
\setcounter{table}{0}

\begin{table}[htbp]
  \centering
  \caption{Summary Statistics}
  \label{tab:sumstats}
  \small
  \begin{tabular}{lrrr}
  \hline\hline
  Variable & N & Mean & Std.\ Dev. \\
  \hline
  \multicolumn{4}{l}{\textit{Price Variables}} \\
  \quad Sales Price (\$) & 1,022,938 & 134,207.4 & 106,568.8 \\
  \quad Log Sales Price & 1,022,938 & 11.6 & 0.7 \\
  \hline
  \multicolumn{4}{l}{\textit{Property Characteristics}} \\
  \quad Building Age (years) & 1,022,938 & 58.9 & 37.0 \\
  \quad Living Area (sq ft) & 1,022,938 & 1,688.7 & 659.5 \\
  \quad Number of Bathrooms & 1,022,938 & 1.9 & 0.9 \\
  \quad Number of Bedrooms & 1,022,938 & 3.2 & 0.8 \\
  \quad Number of Fireplaces & 1,022,938 & 0.5 & 0.6 \\
  \quad Number of Stories & 1,022,938 & 1.6 & 0.5 \\
  \hline
  \multicolumn{4}{l}{\textit{Categorical Features (0/1)}} \\
  \quad Air Conditioning & 1,022,938 & 0.4 & 0.5 \\
  \quad Garage & 1,022,938 & 0.8 & 0.4 \\
  \quad Pool & 1,022,938 & 0.1 & 0.3 \\
  \hline
  \multicolumn{4}{l}{\textit{GIS Distances in miles}} \\
  \quad Dist. to Transmission Line & 1,022,938 & 1.2 & 1.4 \\
  \quad Dist. to Primary Road & 1,022,938 & 3.2 & 4.4 \\
  \quad Dist. to Airport & 1,022,938 & 3.0 & 1.6 \\
  \quad Dist. to Railroad & 1,022,938 & 1.8 & 2.1 \\
  \quad Dist. to Metro Area & 1,022,938 & 1.0 & 2.6 \\
  \quad Dist. to Waterbody & 1,022,938 & 0.4 & 0.3 \\
  \quad Dist. to Conservation Area & 1,022,938 & 0.5 & 0.6 \\
  \hline
\end{tabular}
\begin{flushleft}
{\footnotesize
\noindent \textit{Note}: Total observations = 1,022,938. Except the binary categorical features in the table, the data also include the following categorical property or geographical features: Building Condition, Building Type, Fuel Type, Sewer Type, Water Type, County, City, Zip Code, Sale Year, Census Tract, etc. The geodesic  distance variables are calculated from GIS features coming from the US Census Bureau TIGER/line geodatabase (USCB TIGER),  Homeland Infrastructure Foundation Level Database (HIFLD, involving transmission infrastructure), Ourairports.com (including small airports), U.S. Geological Survey 3D Hydrography Program (including small waterbodies), and New York Protected Areas Database (NYPAD, including geographical spaces managed for conservation, biodiversity, and public recreation).}
\end{flushleft}
\end{table}

\begin{table}[H]
\centering
\caption{Error Measures for Best-performing Models with Closer Control}
\small
\resizebox{\textwidth}{!}{%
\renewcommand{\arraystretch}{0.90}
\begin{tabular}{@{}llcccccc@{}}
\toprule
& & \multicolumn{3}{c}{\textbf{With Geographic Attributes}} & \multicolumn{3}{c}{\textbf{Without Geographic Attributes}} \\
\cmidrule(lr){3-5} \cmidrule(lr){6-8}
& & \textbf{Scenario 1} & \textbf{Scenario 2} & \textbf{Scenario 3} & \textbf{Scenario 1} & \textbf{Scenario 2} & \textbf{Scenario 3} \\
& & No. Treated & No. Treated & No. Treated & No. Treated & No. Treated & No. Treated \\
& & $\geq$3k & $\geq$3k & $\geq$3k & $\geq$3k & $\geq$3k & $\geq$3k \\
\midrule
\multicolumn{8}{l}{\textbf{Panel A: With 3--5~km control}} \\
CF2 & RMSE & \textbf{0.0244} & \textbf{0.0244} & \textbf{0.0245} & 0.0277 & 0.0277 & 0.0277 \\
& MAPE & \textbf{38.19} & \textbf{37.98} & \textbf{38.92} & 42.88 & 42.97 & 44.22 \\
T-GDID & RMSE & 0.0279 & 0.0279 & 0.0280 & 0.0274 & 0.0273 & 0.0274 \\
& MAPE & 43.26 & 43.31 & 45.60 & 42.26 & 42.24 & 44.37 \\
\midrule
\multicolumn{8}{l}{\textbf{Panel B: Benchmark - with 3--10~km control}} \\
CF2 & RMSE & \textbf{0.0246} & \textbf{0.0247} & \textbf{0.0246} & 0.0282 & 0.0281 & 0.0282 \\
& MAPE & \textbf{38.95} & \textbf{39.09} & \textbf{40.47} & 43.40 & 43.14 & 45.17 \\
T-GDID & RMSE & 0.0279 & 0.0279 & 0.0278 & 0.0287 & 0.0287 & 0.0286 \\
& MAPE & 43.89 & 43.88 & 45.58 & 44.64 & 44.65 & 46.68 \\
\bottomrule
\end{tabular}
}
\begin{flushleft}
{\footnotesize
\noindent \textit{Note}: Errors are measured against the assigned average direct unmediated price effect on treated (or average DUET). Monte Carlo simulation results are based on 500 iterations for each scenario-method combination. To ensure fair comparisons, all methods across all scenarios use identical data generation processes, with treatment randomly assigned in each iteration using seeds equal to the loop counter. Bold text denotes the top-performing approaches for each scenario. Geographic attributes for ``w/ Geo'' columns are modeled consistently with the approach presented in Table 3.
}
\end{flushleft}
\end{table}

\begin{table}[H]
\centering
\caption{Interval Width and CI Coverage for Best-performing Models (Ave. DUET=-5\%)}
\small
\renewcommand{\arraystretch}{0.65}
\begin{tabular}{@{}llc@{}}
\toprule
& & \textbf{Scenario 1} \\
& & No. Treated $\geq$3k \\
\midrule
\multicolumn{3}{l}{\textbf{Panel A: T-GDID with Short Pre-period (from 2000 to 2009)}} \\
T-GDID 4~km Control & RMSE & 0.0327 \\
& MAPE & 40.785 \\
& Ave. SE & 1.6503 \\
& True eff. in CI (\%) & 87.8\% \\
T-GDID 5~km Control & RMSE & 0.0237 \\
& MAPE & 36.551 \\
& Ave. SE & 0.0201 \\
& True eff. in CI (\%) & 84.5\% \\
T-GDID 10~km Control & RMSE & {0.0233} \\
& MAPE & {35.583} \\
& Ave. SE & {0.0171} \\
& True eff. in CI (\%) & {80.6\%} \\
\midrule
\multicolumn{3}{l}{\textbf{Panel B: CF2 with Short Pre-period (from 2000 to 2009)}} \\
CF2 4~km Control & RMSE & 0.0244 \\
& MAPE & 38.071 \\
& Ave. SE & 0.0245 \\
& True eff. in CI (\%) & 94.3\% \\
CF2 5~km Control & RMSE & 0.0226 \\
& MAPE & 34.994 \\
& Ave. SE & 0.0216 \\
& True eff. in CI (\%) & 91.9\% \\
CF2 10~km Control & RMSE & \textbf{0.0223} \\
& MAPE & \textbf{34.042} \\
& Ave. SE & 0.0196 \\
& True eff. in CI (\%) & 91.6\% \\
\midrule
\multicolumn{3}{l}{\textbf{Panel C: CF2 with Long Pre-period (from 1990 to 2009)}} \\
CF2 4~km Control & RMSE & 0.0245 \\
& MAPE & 37.856 \\
& Ave. SE & 0.0225 \\
& True eff. in CI (\%) & 90.7\% \\
CF2 5~km Control & RMSE & 0.0233 \\
& MAPE & 35.799 \\
& Ave. SE & 0.0198 \\
& True eff. in CI (\%) & 87.5\% \\
CF2 10~km Control & RMSE & {0.0229} \\
& MAPE & {35.293} \\
& Ave. SE & {0.0179} \\
& True eff. in CI (\%) & {86.6\%} \\
\bottomrule
\end{tabular}
\end{table}
\begin{flushleft}
{\footnotesize
\noindent \textit{Note}: Errors are measured against the assigned average DUET. We set average DUET in this table as - 5\%, in order to check the robustness of main results against changes in average DUET. Results are based on 2000 Monte Carlo iterations, restricted to iterations with No.\ Treated $\geq$3k. To ensure fair comparisons, all methods across all scenarios use identical data generation processes, with treatment randomly assigned in each iteration using seeds equal to the loop counter. Ave.\ SE measures the width of the confidence interval and is calculated as the average of the per-iteration standard errors. True eff.\ in CI (\%) is the empirical coverage rate of the 95\% confidence interval, calculated as the proportion of iterations in which the iteration's own interval (estimate $\pm 1.96$ times its standard error, on the log-price scale) contains the true DUET value; its Monte Carlo standard error is about 1.5--2 percentage points. The large Ave.\ SE of T-GDID with a 4~km control is driven by a few near-degenerate iterations. Geographic attributes are included in all specifications and modeled consistently with the approach presented in Table 3. Bold text indicates the top-performing approaches overall.
}
\end{flushleft}

\begin{table}[H]
\centering
\caption{CF2 under Alternative Zip Code and Year Encodings (3--10~km Control, Scenario 1)}
\label{tab:zipyearfe}
\small
\setlength{\tabcolsep}{4pt}
\renewcommand{\arraystretch}{0.95}
\begin{tabular}{@{}ll ccc @{\hspace{0.8em}} ccc@{}}
\toprule
& & \multicolumn{3}{c}{\textbf{All iterations}} & \multicolumn{3}{c}{\textbf{No.\ Treated $\geq$ 3k}} \\
\cmidrule(lr){3-5}\cmidrule(l){6-8}
& & Continuous & Zip FE $+$ & Zip-by-year & Continuous & Zip FE $+$ & Zip-by-year \\
& & (main spec.) & Year FE & FE & (main spec.) & Year FE & FE \\
\midrule
\multicolumn{8}{l}{\textbf{Panel A: CF2 with Short Pre-period (from 2000 to 2009)}} \\
& RMSE & \textbf{0.0355} & \textbf{0.0355} & 0.0417 & \textbf{0.0241} & 0.0247 & 0.0355 \\
& MAPE & \textbf{52.23} & 52.46 & 61.51 & \textbf{37.02} & 37.82 & 52.27 \\
& Ave.\ SE & 0.0281 & 0.0275 & 0.0226 & 0.0196 & 0.0190 & 0.0148 \\
& True eff.\ in CI (\%) & 91.4 & 89.9 & 74.6 & 90.1 & 87.5 & 66.8 \\
& $\Delta$RMSE vs.\ main & --- & 0.0000 & 0.0062 & --- & 0.0006 & 0.0114 \\
\midrule
\multicolumn{8}{l}{\textbf{Panel B: CF2 with Long Pre-period (from 1990 to 2009)}} \\
& RMSE & \textbf{0.0324} & 0.0327 & 0.0423 & \textbf{0.0248} & 0.0254 & 0.0393 \\
& MAPE & \textbf{48.60} & 48.98 & 63.33 & \textbf{38.69} & 39.30 & 57.94 \\
& Ave.\ SE & 0.0256 & 0.0253 & 0.0212 & 0.0179 & 0.0175 & 0.0140 \\
& True eff.\ in CI (\%) & 89.5 & 89.1 & 70.1 & 86.5 & 85.4 & 60.1 \\
& $\Delta$RMSE vs.\ main & --- & 0.0003 & 0.0100 & --- & 0.0006 & 0.0144 \\
\bottomrule
\end{tabular}
\end{table}
\begin{flushleft}
{\footnotesize
\noindent \textit{Note}: Errors are measured against the assigned average DUET (homogeneous $5\%$ effect, Scenario 1, 3--10~km control, geographic attributes in the price model only). The three encodings differ only in how zip code and sale year enter the propensity model, the outcome model, and causal forests. ``Continuous'' is the main specification (Table 5), in which zip code and sale year enter as ordered numerical variables. ``Zip FE $+$ Year FE'' replaces them with separate sets of zip and year binary indicators, and ``Zip-by-year FE'' replaces them with one binary indicator per zip-code-by-year cell, mirroring the fixed effects of T-GDID. Treatment draws and forest seeds are identical across encodings, and all columns use the same 2,000 iterations (1,107 with No.\ Treated $\geq 3$k), so comparisons are paired and the main-specification RMSE and MAPE coincide with Table 5. $\Delta$RMSE is the RMSE of the alternative encoding minus that of the main specification (positive values favor the main specification). Based on 10,000 paired bootstrap resamples, all $\Delta$RMSE values for the Zip-by-year FE encoding are significant at the 5\% level, while those for the Zip FE $+$ Year FE encoding are not, except the short-pre-period difference with No.\ Treated $\geq 3$k (0.0006, 95\% interval [0.0001, 0.0011]). In the zip-by-year encoding, the causal forest has on average about 850 (short) and 910 (long) features, about 98\% of which are binary indicators (mostly uninformative).
}
\end{flushleft}

\paragraph{Alternative inference for T-GDID.} Table D.5 compares three ways of computing the standard error and 95\% confidence interval of the T-GDID estimate for the main specification (3--10~km control, 2000--2009 pre-period, zip-code-by-year fixed effects, Scenario 1). Treatment draws and seeds are those of Table 5, so the point estimates are identical and only the inference differs. Panel A uses the two-way clustering on zip code and sale year applied throughout the main text. Panel B clusters two ways on treatment site and sale year, where each observation is assigned to its nearest treatment site. Because sites are drawn at random over the study region, only sites with transactions within 10~km contribute observations, about 18 per iteration on average (range 7 to 26). Panel C uses the wild cluster restricted (WCR) bootstrap with clustering at the treatment-site level, the level at which treatment is assigned \citep{cameron2008, abadie2023}. In each of 9,999 bootstrap replications, the null hypothesis is imposed, the restricted residuals of all observations assigned to a site are multiplied by a common Rademacher weight ($\pm 1$ with equal probability), and the site-clustered $t$-statistic is recomputed. The 95\% confidence interval is obtained by test inversion, as the set of effect values not rejected at the 5\% level \citep{roodman2019}. Coverage is the share of iterations whose interval contains the true effect, and the implied standard error is the interval width divided by $2 \times 1.96$, which makes the bootstrap interval comparable to the Ave.\ SE in Table 5.

\begin{table}[H]
\centering
\caption{Alternative Inference for T-GDID (3--10~km Control, Short Pre-period, Scenario 1)}
\label{tab:siteWCB}
\small
\begin{tabular}{@{}llcc@{}}
\toprule
& & All iterations & No.\ Treated $\geq$ 3k \\
\midrule
\multicolumn{4}{l}{\textbf{Point estimates (common to all panels)}} \\
& RMSE                  & 0.0314 & 0.0258 \\
& MAPE (\%)             & 47.69  & 39.33  \\
& SD of estimates       & 0.0298 & 0.0245 \\
\midrule
\multicolumn{4}{l}{\textbf{Panel A: Two-way clustered SE, zip code and year (main text)}} \\
& Ave.\ SE              & 0.0220 & 0.0169 \\
& Median SE             & 0.0198 & 0.0155 \\
& True eff.\ in CI (\%) & 83.5   & 81.7   \\
\midrule
\multicolumn{4}{l}{\textbf{Panel B: Two-way clustered SE, treatment site and year}} \\
& Ave.\ SE              & 0.0239 & 0.0183 \\
& Median SE             & 0.0214 & 0.0170 \\
& True eff.\ in CI (\%) & 86.6   & 85.1   \\
\midrule
\multicolumn{4}{l}{\textbf{Panel C: Wild cluster restricted bootstrap, treatment site}} \\
& Ave.\ implied SE      & 0.0408 & 0.0368 \\
& Median implied SE     & 0.0372 & 0.0324 \\
& True eff.\ in CI (\%) & 94.8   & 95.0   \\
\midrule
& Ave.\ no.\ of site clusters & 17.4 & 17.9 \\
\bottomrule
\end{tabular}
\end{table}
\begin{flushleft}
{\footnotesize
\noindent \textit{Note}: Results are based on 2,000 Monte Carlo iterations (1,107 with No.\ Treated $\geq 3$k), with the homogeneous $5\%$ average DUET. RMSE and MAPE are measured against the true average DUET on the price scale, and the SD of estimates is the standard deviation of the log-scale T-GDID estimates across iterations, the benchmark an accurate standard error should match. True eff.\ in CI (\%) is the share of iterations in which the iteration's own 95\% interval contains the true effect. In Panels A and B the interval is the estimate $\pm 1.96$ times its standard error. In Panel C it is the bootstrap interval, and the implied SE is its width divided by $3.92$. Bootstrap misses are balanced, with the truth below the interval in 2.3\% and above it in 2.6\% of the iterations with No.\ Treated $\geq 3$k. Site clusters are the treatment sites with at least one transaction in the estimation sample.
}
\end{flushleft}

\clearpage
\newpage
\section{Treatment-Induced Equilibrium Shift: Derivation, Design, and Expected Results}
\label{app:indshift}

The main simulation preserves the natural pre-post equilibrium shift present in the real transaction data but sets the treatment-induced shift to zero ($IE_i = 0$) in order to keep the data-generating process fully empirical. This appendix (i) shows formally that the generalized DID (GDID) framework accommodates a treatment-induced shift through the same channel it accommodates the natural shift, so that omitting the induced component does not compromise the interpretation of the exercise; (ii) identifies the feature of a shift that actually drives relative estimator performance --- its \textit{complexity} rather than its magnitude; and (iii) describes the additional simulations reported in Section 4.7, together with the pattern of results.

\subsection{Why GDID Accommodates a Treatment-Induced Shift}
\label{app:indshift-deriv}

\paragraph{Setup.} All estimators compared in this study take as input only the log price $Y=\ln P$, the covariates $\mathbf{X}$, the treatment indicator $D$ ($D=1$ for a property within the treatment radius), and the period $t\in\{0,1\}$. Let
\begin{equation}
m_0(\mathbf{X}) \equiv E[\,Y \mid \mathbf{X},\, t=0\,], \qquad
m_1(\mathbf{X}) \equiv E[\,Y \mid \mathbf{X},\, t=1,\, D=0\,]
\end{equation}
denote the true pre-period price function and the true post-period price function traced out by untreated units (equivalently $p^{a^{0}}(0,\cdot)$ and $p^{a^{*}}(0,\cdot)$ in the notation of Section 2.1). The total pre-post change in the price function faced by control units decomposes into the natural and treatment-induced components introduced in Section 2.1 (where $p^{a'}$ denotes the counterfactual post-period price function that would have prevailed had the treatment not occurred, i.e., the price function in the counterfactual state $a'$):
\begin{equation}
\label{eq:E-decomp}
m_1(\mathbf{X}) - m_0(\mathbf{X})
= \underbrace{\big(p^{a'}-p^{a^{0}}\big)(\mathbf{X})}_{\text{natural shift}}
+ \underbrace{\big(p^{a^{*}}-p^{a'}\big)(\mathbf{X})}_{\text{treatment-induced shift }(IE)}.
\end{equation}

\paragraph{Bias of a GDID estimator.} A GDID estimator forms period-specific fitted price functions $\hat m_0(\mathbf{X})$ and $\hat m_1(\mathbf{X})$ and estimates average DUET as the difference-in-differences of $Y$ net of these fits (this is exactly the structure of equation~(4) in the main text). Define the approximation error of the fitted price function in each period,
\begin{equation}
r_t(\mathbf{X}) \equiv m_t(\mathbf{X}) - \hat m_t(\mathbf{X}), \qquad t\in\{0,1\}.
\end{equation}
Under the conditional parallel trend condition (equation~(6) in the main text, suggesting that, absent treatment, treated and control units share the same period-to-period change in price given $\mathbf{X}$), the bias of the GDID estimator for average DUET is
\begin{equation}
\label{eq:E-bias}
\text{Bias}
= \Big\{ E[\,r_1 \mid D=1\,] - E[\,r_1 \mid D=0\,] \Big\}
- \Big\{ E[\,r_0 \mid D=1\,] - E[\,r_0 \mid D=0\,] \Big\}.
\end{equation}
Two features of \eqref{eq:E-bias} are central. First, the bias depends only on (a) how poorly each period's fitted price function approximates the truth ($r_t$) and (b) how much treated and control units differ in the distribution of $\mathbf{X}$. Second, the decomposition \eqref{eq:E-decomp} of $m_1-m_0$ into natural and treatment-induced components \textit{does not appear at all} in \eqref{eq:E-bias}. The estimator responds to the shifted price function it must fit, not to the reason the function shifted.

\paragraph{The linear case.} Write $m_t(\mathbf{X})=\mathbf{\gamma}^{t}\mathbf{X}$ and let $\Delta\bar{\mathbf{X}} \equiv E[\mathbf{X}\mid D=1]-E[\mathbf{X}\mid D=0]$ denote the spatial-autocorrelation-induced covariate imbalance between treated and control units (nonzero in any single realization even under random siting; see Section 2.2). Then:
\begin{itemize}
\item \textbf{T-GDID} estimates $\mathbf{\gamma}^{0}$ and $\mathbf{\gamma}^{1}$ freely with period-specific coefficients, so $r_0=r_1\approx 0$ and the bias in \eqref{eq:E-bias} vanishes \textit{regardless of the size of} $\mathbf{\gamma}^{1}-\mathbf{\gamma}^{0}$. 
\item \textbf{Baseline DID} constrains $\hat{\mathbf{\gamma}}^{0}=\hat{\mathbf{\gamma}}^{1}=\bar{\mathbf{\gamma}}$, so that $r_1(\mathbf{X})=(\mathbf{\gamma}^{1}-\bar{\mathbf{\gamma}})\mathbf{X}$ and $r_0(\mathbf{X})=(\mathbf{\gamma}^{0}-\bar{\mathbf{\gamma}})\mathbf{X}$, giving
\begin{equation}
\text{Bias} \;\propto\; \Delta\bar{\mathbf{X}}\,'\big(\mathbf{\gamma}^{1}-\mathbf{\gamma}^{0}\big),
\qquad
\mathbf{\gamma}^{1}-\mathbf{\gamma}^{0}
= \underbrace{(\mathbf{\gamma}^{a'}-\mathbf{\gamma}^{a^{0}})}_{\text{natural}}
+ \underbrace{(\mathbf{\gamma}^{a^{*}}-\mathbf{\gamma}^{a'})}_{\text{induced}}.
\end{equation}
The bias is a linear function of the \textit{sum} of the natural and induced coefficient changes, but the baseline DID estimator cannot, and need not, distinguish which component produced it.
\end{itemize}

\paragraph{Observational equivalence.} More generally, the counterfactual state $a'$ enters no estimating equation in Section 2.3 or Appendix A. Consequently, two data-generating processes that induce the same period price functions $m_0$ and $m_1$ --- hence the same joint state of $(\,Y,\mathbf{X},D,t\,)$ on the estimation sample --- are \textit{observationally equivalent} to all estimators. A natural shift and a treatment-induced shift that leave $m_0$ and $m_1$ unchanged are therefore indistinguishable in estimation. This makes precise the claim in the footnote to equation~(2) in the main text that setting $IE_i=0$ ``does not qualitatively affect any theoretical conclusion nor any empirical procedure'': the simulation is not free of a price-function shift at all (the large natural 1990--2024 shift is fully present in the data), and the roughly $16\%$ RMSE advantage of T-GDID over baseline DID in Table 2 is a direct measurement of how much that shift matters --- it simply omits the treatment-induced component of the shift.

\paragraph{What actually differentiates estimators: shift complexity.} Within the GDID family, Eq. \eqref{eq:E-bias} shows that every estimator allows the price function to shift, and what separates them is how well each \textit{approximates} the shifted price functions. T-GDID absorbs any change in linear coefficients exactly, however large, so what survives in $r_t$ is nonlinearity and interaction structure that a period-specific linear model cannot represent. CF2 models $m_t$ with random forests and can represent much more of that structure. The operative feature of a treatment-induced shift is therefore its \textit{complexity} relative to each estimator's flexibility, not its magnitude or its provenance. Because CF2's entire comparative advantage over T-GDID lies in nuisance approximation, omitting a treatment-induced shift, which may add the nonlinear $\mathbf{X}$-interacted structure that CF2 can absorb and T-GDID cannot, likely \textit{understates} CF2's advantage. This is the sense in which the main-text comparison (Sections 4.1--4.6) likely understate CF2's advantage over the traditional estimator.

\subsection{Induced-Shift Simulation Design}
\label{app:indshift-design}

We re-run the Scenario 1 data-generating process (homogeneous $5\%$ direct effect) for the retained main specification (i.e., the 3--10~km control band, with T-GDID at the short pre-period and CF2 at both the short and long pre-periods) and add a treatment-induced shift as follows. In each iteration, together with the direct price effect (i.e., DUET applied to treated post-period units), an induced shift $s(\mathbf{X}_i)$ is added to the log price of \textit{every} post-period transaction $i$ in the estimation market:
\begin{equation}
\ln P_{i,\text{post}} \;\longleftarrow\; \ln P_{i,\text{post}} + s(\mathbf{X}_i),
\qquad
s(\mathbf{X}) = \kappa \cdot \mathrm{sd}\!\left(\ln P_{\text{post}}\right)\cdot \frac{\,\tilde s(\mathbf{X}) - \overline{\tilde s}\,}{\mathrm{sd}(\tilde s)},
\end{equation}
where $\tilde s(\mathbf{X})$ is a raw shift shape (specified below), the shift is centered to be mean-zero over the post-period sample, and it is scaled so that its cross-sectional standard deviation equals the \textit{shift intensity} $\kappa$ times the post-period log-price standard deviation (so that the size of the shift is measured relative to the post-period price standard deviation). Applying the shift to treated and control units alike, and normalizing it, means it represents a pure change in the shape of the post-period price function - the average post-period price level is unchanged as any level change will be absorbed by the post variable or time fixed effects. Because the shift is common to treated and control units conditional on $\mathbf{X}$, the average DUET benchmark is unchanged in the form of $\log(1+0.05)$, so ground truth is preserved and only the estimators' ability to recover it is affected.

We consider two forms of the raw shift shape, holding the intensity $\kappa$ fixed so that the two forms inject shifts of identical magnitude and differ only in functional form:
\begin{itemize}
\item \textbf{Linear.} $\tilde s(\mathbf{X}) = \mathbf{X}_S'\mathbf{w}$, where $\mathbf{X}_S$ is a random subset of four standardized covariates drawn from the 16 structural and geographic covariates that T-GDID includes in its price model, and $\mathbf{w}$ is a vector of independent standard normal weights. Both the subset and the weights are redrawn in each iteration. Using four covariates matches the number of variables entering the nonlinear form below (two interacted continuous covariates, an indicator, and living area), so the two forms differ in functional form rather than in dimensionality. Because the shift is linear in covariates that T-GDID already includes, T-GDID absorbs it exactly at any $\kappa$. The linear form therefore serves as a placebo that isolates whether the presence of an induced shift, absent nonlinearity, affects the estimator comparison.
\item \textbf{Nonlinear.} $\tilde s(\mathbf{X}) = z(V_1)\,z(V_2) + \mathbf{1}\{V_3 = \ell\}\cdot z(\text{LivingArea})$, where $z(\cdot)$ denotes standardization, $V_1$ and $V_2$ are two continuous covariates drawn at random each iteration, and $\mathbf{1}\{V_3 = \ell\}$ is an indicator that a randomly chosen discrete covariate $V_3$ equals a randomly chosen (well-populated) level $\ell$ --- or, with equal probability, a median split of a randomly chosen continuous covariate. The complicated shape of this assigned nonlinear shift lies well outside the linear span of T-GDID's price model. The shape is redrawn each iteration, so the reported results average over a family of nonlinear shifts rather than one hand-picked functional form.
\end{itemize}
We sweep the intensity over $\kappa \in \{0,\, 0.1,\, 0.5,\, 0.9\}$, where $\kappa=0$ reproduces the baseline design of Sections 4.1--4.6. All other simulation settings (2000 iterations, random siting, 30 treatment sites, 3~km treatment radius, seeds equal to the loop counter) are unchanged, so the two arms and the baseline are directly comparable. Because the shift shape for a given iteration is the same (i.e., drawn from the same random seed) across all intensity levels, the intensity sweep varies orthogonally to the randomly varying shape of the induced shift, isolating the effect of intensity.

\subsection{Simulation Results with an Induced Shift}
\label{app:indshift-expected}

The derivation in Appendix \ref{app:indshift-deriv} yields a sharp, falsifiable prediction for each arm:
\begin{itemize}
\item \textbf{Linear arm (Panel A of Table \ref{tab:indshift}).} T-GDID absorbs the shift exactly at every $\kappa$, so its estimates should be unchanged. CF2 must learn the linear shift from data with piecewise-constant trees and therefore pays a minor estimation cost that rises with $\kappa$. The CF2-minus-T-GDID error gap should therefore move only slightly against CF2 as $\kappa$ increases. This arm is the placebo showing that linear treatment-induced shifts do not affect the GDID estimates and do not meaningfully affect the causal forest estimates.
\item \textbf{Nonlinear arm (Panel B of Table \ref{tab:indshift}).} As T-GDID cannot represent the nonlinear, interacted shift, its price modeling error $r_t$ and (through covariate imbalance) estimation bias should grow with $\kappa$. CF2's forest-based price model absorbs much of the shift, so its error should grow far more slowly. The CF2-minus-T-GDID error gap should therefore \textit{widen} in CF2's favor as $\kappa$ increases.
\end{itemize}

Table \ref{tab:indshift} reports both arms, each on all $2000$ iterations ($1107$ in the $\geq 3$k subset). The linear placebo arm (Panel A) behaves as predicted. T-GDID's estimates are numerically identical to the no-shift estimates at every $\kappa$ (RMSE $0.0314$ across all iterations and $0.0258$ in the $\geq 3$k subset), since the shift lies in its linear span. CF2 pays a minor cost for approximating the shift with trees, and the cost rises with $\kappa$. In the $\geq 3$k subset, the RMSE of CF2 with the short pre-period rises from $0.0241$ without the shift to $0.0245$, $0.0268$, and $0.0288$ at $\kappa = 0.1$, $0.5$, and $0.9$, so the CF2-minus-T-GDID gap moves from $-0.0016$ to $-0.0012$, $+0.0010$, and $+0.0030$. Across all iterations, where CF2 already trails T-GDID without the shift ($0.0355$ versus $0.0314$), the gap grows from $+0.0046$ to $+0.0088$. The added error is mostly variance rather than bias, as CF2's mean error stays below $0.005$ at every $\kappa$. A high-intensity induced shift, absent nonlinearity, therefore slightly favors T-GDID. 

The nonlinear runs (Panel B) trace out the predicted pattern, especially in the large-sample regime (No.\ Treated $\geq 3$k, the right block of the table) in which CF2 is supposed to outperform T-GDID. In that subset the CF2-minus-T-GDID RMSE gap is negative at every intensity and widens monotonically in CF2's favor, from $-0.003$ at $\kappa = 0.1$ to $-0.022$ at $\kappa = 0.5$ and $-0.052$ at $\kappa = 0.9$. The driver is T-GDID's inability to address the interacted nonlinear shift. Its RMSE climbs from $0.0273$ at $\kappa = 0.1$ to $0.0847$ at $\kappa = 0.9$ (MAPE from $42\%$ to $125\%$), whereas CF2's forest-based price model absorbs most of the shift, and its RMSE rises only from $0.0246$ to $0.0330$ with the short pre-period. Because the shape of the nonlinear shift is redrawn each iteration, T-GDID's growing RMSE reflects iteration-specific price modeling error whose sign and size depend on the drawn shape (i.e., produced by misspecified pre- and post-price models). Coverage responds less to the shift than accuracy does. Both estimators lose a few percentage points as $\kappa$ rises, but CF2 with the short pre-period stays 9 to 12 percentage points above T-GDID at every intensity in the $\geq 3$k subset ($89.8\%$ versus $77.7\%$ at $\kappa = 0.1$, and $85.1\%$ versus $77.2\%$ at $\kappa = 0.9$). 

Comparing the two arms at the same intensity, CF2's RMSE in the $\geq 3$k subset reaches similar levels under both forms at $\kappa = 0.9$ ($0.0288$ linear and $0.0330$ nonlinear), whereas T-GDID's ranges from $0.0258$ to $0.0847$. The form of the shift therefore matters mainly through T-GDID, which is exposed to large approximation error when the shift does not lie in its linear span (most likely in real-world scenarios). The evidence supports the mechanism of Appendix \ref{app:indshift-deriv} and the conclusion that the main-text comparison, which omits the treatment-induced shift, understates CF2's advantage over T-GDID when the induced shift is nonlinear.

\setcounter{table}{0}
\begin{table}[H]
\centering
\caption{Estimator Performance under a Treatment-Induced Equilibrium Shift (3--10~km control, Scenario 1)}
\label{tab:indshift}
\small
\resizebox{\textwidth}{!}{%
\begin{tabular}{@{}ll ccc @{\hspace{1.4em}} ccc@{}}
\toprule
& & \multicolumn{6}{c}{Shift intensity $\kappa$ (share of post $\ln P$ SD)} \\
\cmidrule(l){3-8}
& & \multicolumn{3}{c}{All iterations} & \multicolumn{3}{c}{No.\ Treated $\geq$ 3k} \\
\cmidrule(lr){3-5}\cmidrule(l){6-8}
Estimator & Metric & $\kappa=0.1$ & $0.5$ & $0.9$ & $\kappa=0.1$ & $0.5$ & $0.9$ \\
\midrule
\multicolumn{8}{l}{\textbf{Panel A: Linear shift (placebo; in T-GDID's span)}} \\
T-GDID (3--10~km)      & RMSE                  & 0.0314 & 0.0314 & 0.0314 & 0.0258 & 0.0258 & 0.0258 \\
                       & MAPE (\%)             & 47.69  & 47.69  & 47.69  & 39.33  & 39.33  & 39.33  \\
                       & True eff.\ in CI (\%) & 83.5   & 83.5   & 83.5   & 81.7   & 81.7   & 81.7   \\
CF2 (short pre-period) & RMSE                  & 0.0360 & 0.0379 & 0.0402 & 0.0245 & 0.0268 & 0.0288 \\
                       & MAPE (\%)             & 53.10  & 56.37  & 59.38  & 37.77  & 41.57  & 44.42  \\
                       & True eff.\ in CI (\%) & 91.2   & 89.5   & 88.8   & 89.5   & 87.2   & 86.5   \\
CF2 (long pre-period)  & RMSE                  & 0.0329 & 0.0355 & 0.0379 & 0.0251 & 0.0273 & 0.0289 \\
                       & MAPE (\%)             & 49.38  & 53.09  & 56.05  & 39.21  & 42.29  & 44.68  \\
                       & True eff.\ in CI (\%) & 89.2   & 87.6   & 87.7   & 86.2   & 83.9   & 84.2   \\
CF2 (short) $-$ T-GDID & $\Delta$RMSE          & $+0.0046$ & $+0.0065$ & $+0.0088$ & $-0.0012$ & $+0.0010$ & $+0.0030$ \\
\midrule
\multicolumn{8}{l}{\textbf{Panel B: Nonlinear shift (interactions / level indicators)}} \\
T-GDID (3--10~km)      & RMSE                  & 0.0326 & 0.0551 & 0.0878 & 0.0273 & 0.0517 & 0.0847 \\
                       & MAPE (\%)             & 49.81  & 83.59  & 130.71 & 42.09  & 77.90  & 125.22 \\
                       & True eff.\ in CI (\%) & 80.8   & 78.1   & 79.2   & 77.7   & 75.1   & 77.2   \\
CF2 (short pre-period) & RMSE                  & 0.0361 & 0.0422 & 0.0477 & 0.0246 & 0.0295 & 0.0330 \\
                       & MAPE (\%)             & 53.15  & 62.92  & 69.93  & 37.50  & 44.83  & 49.96  \\
                       & True eff.\ in CI (\%) & 90.8   & 87.6   & 86.3   & 89.8   & 86.4   & 85.1   \\
CF2 (long pre-period)  & RMSE                  & 0.0332 & 0.0400 & 0.0457 & 0.0254 & 0.0307 & 0.0345 \\
                       & MAPE (\%)             & 49.52  & 60.49  & 67.67  & 39.16  & 47.26  & 52.40  \\
                       & True eff.\ in CI (\%) & 88.7   & 83.6   & 81.9   & 85.1   & 80.0   & 78.0   \\
CF2 (short) $-$ T-GDID & $\Delta$RMSE          & $+0.0035$ & $-0.0129$ & $-0.0401$ & $-0.0027$ & $-0.0222$ & $-0.0517$ \\
\bottomrule
\end{tabular}
}
\end{table}
\begin{flushleft}
{\footnotesize
\noindent \textit{Note}: Columns index the shift intensity $\kappa$ (the standard deviation of the induced log-price shift as a share of the post-period log-price standard deviation); the corresponding realized shift standard deviations are approximately $0.07$, $0.37$, and $0.67$ log points (post-period $\ln P$ SD $\approx 0.72$). The left block uses all completed iterations; the right block restricts to $1107$ iterations with at least $3{,}000$ treated observations (the large-sample regime in which the main text establishes CF2's advantage). Panel A reports the linear placebo arm. Panel B reports the nonlinear arm. Both arms use the 3--10~km control specification under Scenario 1 (homogeneous $5\%$ direct effect). The no-shift baseline RMSEs on the same iterations are $0.0314$ (T-GDID), $0.0355$ (CF2 short), and $0.0324$ (CF2 long) across all iterations, and $0.0258$, $0.0241$, and $0.0248$ in the $\geq 3$k block. RMSE and MAPE are measured against the true average DUET; ``True eff.\ in CI'' is the per-iteration empirical coverage of the nominal $95\%$ interval. The last row of each panel is the difference in RMSE between CF2 (short pre-period) and T-GDID with negative values favoring CF2.
}
\end{flushleft}

\newpage
\bibliographystyle{apalike}
\bibliography{bibliography}
\clearpage